**A tale of two stories: astrocyte regulation of synaptic depression and facilitation**

Maurizio De Pittà[1], Vladislav Volman[2,3], Hugues Berry[4], and Eshel Ben-Jacob[1,2,*]

1. School of Physics and Astronomy, Tel Aviv University, 69978 Ramat Aviv, Israel

2. Center for Theoretical Biological Physics, UCSD, La Jolla, CA 92093-0319, USA

3. Computational Neurobiology Laboratory, The Salk Institute, La Jolla, CA 92037, USA

4. Project-Team BEAGLE, INRIA Rhône-Alpes, Université de Lyon, LIRIS, UMR5205, F-69603, France

*Corresponding author:

Eshel Ben-Jacob:

eshel@tamar.tau.ac.il,

Tel.: +972-3-640-7845

Fax: +972-3-642-5787

# Abstract

Short-term presynaptic plasticity designates variations of the amplitude of synaptic information transfer whereby the amount of neurotransmitter released upon presynaptic stimulation changes over seconds as a function of the neuronal firing activity. While a consensus has emerged that the resulting decrease (depression) and/or increase (facilitation) of the synapse strength are crucial to neuronal computations, their modes of expression in vivo remain unclear. Recent experimental studies have reported that glial cells, particularly astrocytes in the hippocampus, are able to modulate short-term plasticity but the mechanism of such a modulation is poorly understood. Here, we investigate the characteristics of short-term plasticity modulation by astrocytes using a biophysically realistic computational model. Mean-field analysis of the model, supported by intensive numerical simulations, unravels that astrocytes may mediate counterintuitive effects. Depending on the expressed presynaptic signaling pathways, astrocytes may globally inhibit or potentiate the synapse: the amount of released neurotransmitter in the presence of the astrocyte is transiently smaller or larger than in its absence. But this *global* effect usually coexists with the opposite *local* effect on paired pulses: with release-decreasing astrocytes most paired pulses become facilitated, namely the amount of neurotransmitter released upon spike $i+1$ is larger than that at spike $i$, while paired-pulse depression becomes prominent under release-increasing astrocytes. Moreover, we show that the frequency of astrocytic intracellular $Ca^{2+}$ oscillations controls the effects of the astrocyte on short-term synaptic plasticity. Our model explains several experimental observations yet unsolved, and uncovers astrocytic gliotransmission as a possible transient switch between short-term paired-pulse depression and facilitation. This possibility has deep implications on the processing of neuronal spikes and resulting information transfer at synapses.

# Authors' summary

Synaptic plasticity is the capacity of a preexisting connection between two neurons to change in strength as function of neuronal activity. Because it admittedly underlies learning and memory, the elucidation of its constituting mechanisms is of crucial importance in many aspects of normal and pathological brain function. Short-term presynaptic plasticity refers to changes occurring over short time scales (milliseconds to seconds) that are mediated by frequency-dependent modifications of the amount of neurotransmitter released by presynaptic stimulation. Recent experiments have reported that glial cells, especially hippocampal astrocytes, can modulate short-term plasticity, but the mechanism of such modulation is poorly understood. Here, we explore a plausible form of modulation of short-term plasticity by astrocytes using a biophysically realistic computational model. Our analysis indicates that astrocytes could simultaneously affect synaptic release in two ways. First, they either decrease or increase the overall synaptic release of neurotransmitter. Second, for stimuli that are delivered as pairs within short intervals, they systematically increase or decrease the synaptic response to the second one. Hence, our model suggests that astrocytes could transiently trigger switches between paired-pulse depression and facilitation. This property explains several challenging experimental observations and has deep impact on our understanding of synaptic information transfer.

## Introduction

Activity-dependent modification of synaptic transmission critically moulds the properties of synaptic information transfer with important implications for computation performed by neuronal circuitry [1-4]. Multiple mechanisms could coexist in the same synapse, regulating the strength or the efficacy of synaptic transmission therein in a way that depends on the timing and frequency of prior activity at that same synaptic terminal [5].

One widely studied mechanism responsible for the dependence of synaptic transmission on past activity has been dubbed presynaptic short-term plasticity [6]. Upon repetitive action potential stimulation, the response of a presynaptic terminal — usually assessed as the amount of neurotransmitter molecules released from this latter — will not follow with uniform strength but will be modified in a time- and activity-dependent manner, leading either to facilitation or to depression of synaptic release, or to a mixture of both [2]. Such stimulus-related variations of presynaptic response can span a time scale from few milliseconds to seconds from the stimulus onset [2, 7] and fade away after sufficiently prolonged synaptic inactivity [3, 5].

The ability of a presynaptic terminal to convey stimulus-related information is determined by the probability to release neurotransmitter-containing vesicles upon arrival of action potentials [3, 6]. The release probability depends on the number of vesicles that are ready to be released, i.e. the readily releasable pool, but also on the state of the calcium ($Ca^{2+}$) sensor for the exocytosis of synaptic vesicles [8]. On the mechanistic level, both the finite size and the slow post-stimulus recovery of the readily releasable pool, that is the reintegration of the content of synaptic vesicles, give rise to the phenomenon of short-term presynaptic depression, with the extent of depression being determined by the frequency of prior synaptic stimulation [9]. The dependence of short-term facilitation on the pattern of synaptic activation is likely determined either by the slow removal of free presynaptic residual $Ca^{2+}$ or by the slow unbinding of this latter from the $Ca^{2+}$ sensor [3], although these issues are still debatable [10, 11].

Given the important role assumed by presynaptic short-term plasticity in neural computation [6, 12] and the variety of plastic responses — depression, facilitation or both — exhibited by central synapses [13, 14], it is important to unravel the mechanisms that might govern dynamical transitions between depressing and facilitating synapses. The goal of the present work was to investigate one such candidate mechanism: modulation of presynaptic plasticity by glial cells and astrocytes in particular.

Recent years have witnessed mounting evidence on a possible role of glial cells in the dynamics of neuronal networks [15]. In particular, the specific association of synapses with processes of astrocytes – the main type of glial cells in the hippocampus and the cortex [16-18] – together with the discovery of two-way astrocyte-neuron communication [19, 20], suggest an active role of these cells in modulation of synaptic transmission and information processing in the brain [21].

Astrocytes could modulate synaptic transmission at nearby synapses by releasing neurotransmitter (or "gliotransmitter") in a $Ca^{2+}$-dependent fashion [22]. In the hippocampus in particular, several studies have shown that astrocyte-released glutamate modulates short-term plasticity at excitatory synapses either towards depression or facilitation [23-25]. This is achieved by activation of presynaptic glutamate receptors [26] (see also Figure 1 for a schematic presentation). Thus, astrocytes are equipped with means to modulate the extent to which presynaptic terminal exhibits short-term depression or facilitation in response to sustained rhythmic stimulation [27].

We devised a biophysically plausible computational model to investigate the characteristics of astrocyte modulation of presynaptic short-term plasticity. Using the model, we were able to identify the parametric regime in which the synaptic response to action potential stimulation can switch from facilitating to depressing and vice versa. This ability to switch synaptic *modus operandi* depended critically on the characteristics of astrocyte-to-synapse signaling. These findings highlight the new potential role played by astrocytes in defining synaptic short-term plasticity and could explain contradicting experimental evidences.

Although based on experimental results in the hippocampus, [28-34], our description could also be extended to model other recognized neuron-glia signaling pathways such as GABAergic gliotransmission on interneuron-to-pyramidal cell synapses in the hippocampus [35], glia-mediated ATP release both on hippocampal synapses [36, 37] or in the hypothalamus [38] as well as in the retina [39], and glial modulation of neuromuscular transmission [40-42].

## Methods

### The road map of astrocyte regulation of presynaptic short-term plasticity

Regulation of synaptic transmission by astrocyte-released gliotransmitter is supported by an elaborate signaling network schematized in Figure 1. Here, we consider the well-characterized experimental case of glutamate-mediated astrocyte regulation of synaptic transmission in the hippocampus [27, 43]. At excitatory synapses there, astrocytes can respond to synaptically-released glutamate by intracellular $Ca^{2+}$ elevations that in turn, may trigger the release of further glutamate from the astrocytes [22, 44]. This astrocyte-released glutamate ($G_A$) diffuses in the extrasynaptic space and binds to presynaptic metabotropic glutamate receptors (mGluRs) or NMDA receptors (NMDARs) on neighboring presynaptic terminals [21, 30]. Glutamate activation of these receptors can modulate $Ca^{2+}$ influx into the presynaptic terminal, affecting the release probability of glutamate-containing synaptic vesicles [26]. Thus, glutamate release from the presynaptic terminal is expected to increase the astrocytic intracellular $Ca^{2+}$, eventually leading to glutamate release from that astrocyte. In turn, astrocytic glutamate modulates presynaptic $Ca^{2+}$ and thus affects the amount of glutamate released from that same synapse in response to action potentials that will follow [27].

Astrocyte $Ca^{2+}$ dynamics may also not be modulated by glutamate originating from the very presynaptic terminal that is regulated by the astrocyte, but rather by an exogenous source [45]. This could correspond to the heterosynaptic case whereby two distinct synapses, **A** and **B**, are contacted by processes from the same astrocyte [21]. Glutamate released by the presynaptic terminal of synapse **A** modulates astrocytic $Ca^{2+}$, leading to modulation of glutamate release from the presynaptic terminal of synapse **B**. Alternatively astrocyte $Ca^{2+}$ dynamics could be modulated by intercellular $IP_3$ diffusion from neighboring astrocytes through gap junctions [46] or by exogenous stimulation of the astrocyte by different techniques or external stimuli [47, 48], or occur spontaneously [49, 50].

Although both homosynaptic and non-homosynaptic scenarios equally occur physiologically [21, 45], here we focus only on the latter. This approach, which is often adopted in the majority of experiments [30-32, 49], presents several advantages. First, it allows us to characterize the effect of astrocytic glutamate on short-term synaptic plasticity in general, that is, independently of the nature of synaptic inputs. Second, it uses $Ca^{2+}$ signals to merely trigger glutamate exocytosis from the astrocyte. Thus we can focus on the timing of glutamate release without considering the complexity of the underlying Ca$^{2+}$ dynamics [48] which can be ultimately modeled by simple stereotypical analytical functions (Text S1, Section I.2). Third, it can be used in the derivation of a mean-field description of synaptic transmission [51, 52] aimed at understanding regulation of short-term synaptic plasticity by a large variety of astrocytic glutamate signals impinging on the synapse, without the need to consider an equally large number of cases.

## Modeling of the astrocyte-to-synapses interaction

### The Tsodyks-Markram model of a dynamical synapse

To describe the kinetics of a synaptic terminal, we use the model of an activity-dependent synapse first introduced by Tsodyks and Markram [6]. This model assumes that neurotransmitter resources in the presynaptic bouton are limited and only a fraction $x(t)$ of them is available for release at time $t$. Upon arrival of a presynaptic spike at time $t_i$, a fraction $u$ of these latter is released into the cleft, thus reducing $x$ by the amount of "released resources" $RR = ux$. As $x(t)$ recovers to its original value at a rate $\Omega_d$, the process mimics neurotransmitter depletion and reintegration [9]. The dynamics of $x(t)$ thus reads:

$$\dot{x} = \Omega_d (1 - x) - \sum_i ux\delta(t - t_i) \tag{1}$$

On a par with the classical quantal model of synaptic transmission [53], $x(t)$ is analogous to the probability of a glutamate-containing vesicle to be available for release at any time $t$, whereas $u$ corresponds to the probability of release of a docked vesicle [54]. Accordingly, $u$ biophysically correlates with the state of occupancy of the Ca$^{2+}$ sensor of synaptic glutamate exocytosis and its value is incremented following incoming spikes, mimicking Ca$^{2+}$ influx into the presynaptic terminal and its effects on release probability [8]. In particular, at each spike a fraction $U_0$ of the (1-$u$) vacant states of the sensors is occupied by presynaptic Ca$^{2+}$ ions and later returns to be available at rate $\Omega_f$. Hence, the dynamics of $u$ follows the equation

$$\dot{u} = -\Omega_f u + U_0 \sum_i (1 - u)\delta(t - t_i) \tag{2}$$

The parameter $U_0$ coincides with the value of $u$ for very low frequencies of stimulation so that it can be regarded as the basal value of synaptic release probability (Text S1, Section I.1).

**Mechanisms of short-term presynaptic plasticity**

Despite its apparent simplicity, the Tsodyks-Markram (TM) model (equations 1-2) can generate surprisingly complex synaptic dynamics including multiple mechanisms of short-term plasticity among which are facilitation and depression. Nonetheless, the occurrence of each mechanism ultimately depends on specific values of synaptic parameters and the rate and the pattern of synaptic activation [55]. The biophysical correlates of different synaptic parameters (e.g., time of recovery from synaptic depression and per-spike usage of synaptic resource) have been extensively documented for central synapses [14, 56], but relatively little effort was done to understand in the TM framework, the nature of transitions between facilitating and depressing synaptic response. Accordingly, we performed thorough theoretical and computational analysis of the TM model.

In Figure 2A we show a sample response of the TM model to a train of a few input spikes (*top*). The low frequency of the first four spikes largely enables the recovery of available synaptic resource $x$ between spikes (*middle*) so that depletion of releasable resources is limited. This process is coupled with a progressive increase of per-spike resource usage $u$, so that the amount of released resources (*RRs*) per spike (*bottom*) increases and short-term potentiation (STP or facilitation) of synaptic response is observed [3]. On the contrary, stimulating the model synapse with a series of high frequency spikes at $t = 300$ ms, results in a rapid increase of $u$ but also in a larger depletion of $x$, so that from one spike to the next one, progressively less neurotransmitter is available for release. Consequently, the amount of released resources decreases after each input spike hallmarking the onset of short-term depression (STD) [9]. Finally, a relatively long quiescence before the occurrence of last input spike in the series allows for partial recovery of $x$ while $u$ hardly changes, which accounts for the increase of resources released by the last spike with respect to immediately preceding ones (compare response in state "3" to the last response in state "2"). Thus, the frequency and the temporal pattern of synaptic stimulation can modulate the synaptic response either transiently facilitating it or transiently depressing it (see also Figure S1).

While the precise pattern of synaptic response is shaped by the timing of input spikes and depends also on initial conditions [13] (Figures 2A, S1, S2; see also Text S1, Section II.1), it is of interest to be able to characterize synaptic release and related plasticity on "average", namely over different trials of inputs with shared statistics. With this aim, mean-field analysis can be deployed to show that, depending on the basal value $U_0$, two fundamentally different behaviors can be exhibited by the TM model [51, 57] (Text S1, Section II.3, Figures S3, S7A ). When $U_0$ is larger than the threshold value $U_{thr} = \Omega_d/(\Omega_d + \Omega_f)$, the amount of released resources in the steady state is roughly independent of the input frequency $f_{in}$ for low-frequency synaptic stimulation, and decreases only above some cut-off frequency (Figure 2B, *left*). Hence, if $U_0 > U_{thr}$, the synapse is depressing. On the other hand, when $U_0 < U_{thr}$ the amount of released resources first increases up to a peak input frequency so that the synapse is facilitating, then it decreases afterwards, marking the onset of depression (Figure 2B, *right*). Therefore, both the cut-off frequency in depressing synapses and the peak frequency in facilitating ones, set an upper limit for the range of input frequencies beyond which STD is observed [6]. For this reason both the cut-off frequency and the peak frequency can be regarded as the "limiting" frequency ($f_{lim}$) for the onset of STD for the specific synapse under consideration.

The steady-state frequency response $RR_\infty(f_{in})$ of a synapse can be computed using the mean-field analysis (Text S1, Section II.4) as

$$RR_\infty = U_\infty X_\infty = \frac{U_0 \Omega_d (\Omega_f + f_{in})}{\Omega_d \Omega_f + U_0 (\Omega_d + \Omega_f) f_{in} + U_0 f_{in}^2} \tag{3}$$

The above equation can then be solved to obtain the expression for $f_{lim}$ which reads

$$f_{lim} = \begin{cases} \Omega_f \left( \sqrt{\dfrac{\Omega_d}{\Omega_f}\left(\dfrac{1-U_0}{U_0}\right)} - 1 \right) & \text{if } U_0 < U_{thr} \\[4mm] \dfrac{\Omega_d}{(1+\sqrt{2})U_0} & \text{if } U_0 > U_{thr} \end{cases} \tag{4}$$

Thus, depending on the value of $U_0$ with respect to the threshold $U_{thr}$, $f_{lim}$ is described by different analytical functions with different dependencies on synaptic parameters. Furthermore, while a negative slope of $RR_\infty(f_{in})$ coincides with the onset of depression, a positive slope marks occurrence of facilitation. Accordingly, two conditions are necessary for the occurrence of facilitation in the TM model: (1) that $U_0 < U_{thr}$, which guarantees the existence of $f_{in}$ values for which $RR_\infty$ could have either positive or negative slope; and (2) that $f_{in} < f_{lim}$, which assures that the input stimulus effectively falls within the frequency range of positive slope values of $RR_\infty$.

## Modeling the action of the astrocytic glutamate on synaptic release

Glutamate release from astrocytes bears several similarities with synaptic exocytosis [27, 44]. Both processes are $Ca^{2+}$-dependent [8, 58]. Furthermore, glutamate is released from astrocytes in quanta consistently with vesicle exocytosis [44, 59]. A vesicular compartment competent for regulated exocytosis, is indeed present in astrocytes [60, 61] and synaptic-like vesicle fusion and recycling is observed in concomitance with astrocytic glutamate exocytosis [62].

Based on such arguments, we assumed that the dynamics of astrocytic glutamate resources could be modeled in a way that is mathematically similar to the TM description of the dynamics of synaptic neurotransmitter resources, although it should be kept in mind that the biological interpretation of the two mechanisms is different [62]. Accordingly, we assumed that a fraction $x_A(t)$ of the intracellular astrocytic glutamate is available for release at any time $t$. Any increase of intracellular $Ca^{2+}$ concentration beyond a threshold value $C_{thr}$ [59, 63] results in the release of a constant fraction $U_A$ of $x_A$ to the extrasynaptic space, and this released gliotransmitter is later reintegrated into the pool of available glutamate resources of the astrocyte at rate $\Omega_A$ (Text S1, Section I.3).

The effect of the astrocyte-released glutamate ($G_A$) on the release probability of synaptic neurotransmitter is mediated by the activation of presynaptic glutamate receptors [30, 31, 33]. Several experiments showed that activation of these receptors could modulate the magnitude of $Ca^{2+}$ influx into the presynaptic terminal, thus defining the levels of residual $Ca^{2+}$ therein (reviewed in [26]). Furthermore, activation of presynaptic glutamate receptors can modulate the synaptic response to an action potential via changes in residual synaptic $Ca^{2+}$ [64]. It is important to note that this kind of modulation does not require synaptic activation by action potentials and is observed even in basal conditions [3, 8, 13], likely reflecting changes of the occupancy of $Ca^{2+}$ sensors of exocytosis of synaptic vesicles.

We modeled the effect of astrocytic glutamate on synaptic neurotransmitter release assuming the modulation of synaptic basal release probability $U_0$ by astrocytic glutamate. In particular, we assumed that $U_0$ is not a constant (as it is in the original TM model), but rather is a function $U_0(\Gamma)$ of the fraction $\Gamma$ of presynaptic glutamate receptors that are activated by astrocyte-derived glutamate. In the absence of quantitative physiological data, we assumed that the function $U_0(\Gamma)$ is analytic around zero and we considered its first-order expansion, i.e. $U_0(\Gamma) \cong U_0(0) + U_0^{'}(0)\Gamma$. The 0-th order term $U_0(0) = \text{const} = U_0^{*}$ corresponds to the value of $U_0$ in absence of the astrocyte; hence, in the 0-th order approximation, the model of short-term presynaptic plasticity is just the classical TM model. To express $U_0^{'}(0)$, we note that both $U_0(\Gamma)$ and $\Gamma$ represent fractions and as such are constrained to the interval [0, 1] so that it must be $0 \leq U_0(\Gamma) \leq 1$. Hence, we define $U_0^{'}(0) = -U_0^{*} + \alpha$ (with $0 \leq \alpha \leq 1$), and accordingly (Text S1; Section I.5):

$$U_0(\Gamma) = (1 - \Gamma)U_0^{*} + \alpha\,\Gamma \qquad (5)$$

In the above equation, the parameter α lumps in a phenomenological way, all the information related to the activation properties of presynaptic glutamate receptors that mediate the effect of astrocyte on synaptic release (see "The road map of astrocyte regulation of presynaptic short-term plasticity" in "Methods"). Finally, the fraction Γ of presynaptic glutamate receptors that are occupied by astrocyte-released glutamate $G_A$ is modeled as (Text S1, Section I.5):

$$\dot{\Gamma} = O_G\,G_A(1 - \Gamma) - \Omega_G\,\Gamma \qquad (6)$$

The above parameters $O_G$ and $\Omega_G$ are rate constants that biophysically correlate with the rise and decay of the effect of astrocyte glutamate on synaptic neurotransmitter release.

Figure 3 illustrates how in our model, astrocytic $Ca^{2+}$ oscillations (*top*) modulate synaptic basal release probability (*bottom*) via presynaptic receptors activation (Γ) by astrocyte-released glutamate ($G_A$) (*middle panels*). The observed saw-shaped increase of Γ is due to the large difference between the rise and decay rates of the astrocyte effect on synaptic release, being $O_G G_A \ll \Omega_G$ (Text S1, Appendix C). Since in our approximation, $U_0$ is a linear function of Γ, the time evolution of Γ also determines $U_0$ according to equation (5). Depending on the value of the effect parameter α, $U_0(\Gamma)$ can either decrease as low as $U_0(\Gamma, \alpha = 0) = U_0^{*} - U_0^{*}\Gamma$ (*bottom panel*, *green line*) or increase as high as $U_0(\Gamma, \alpha = 1) = U_0^{*} + (1 - U_0^{*})\Gamma$ (*bottom panel*, *magenta line*).

## Results

### Astrocyte can either depress or facilitate synaptic neurotransmitter release

We first studied the effect of astrocytic glutamate release on the transfer properties of our model synaptic terminal. Because the response of a synapse to action potential critically depends on the value of $U_0$ (equation 3), which in turn could be modulated by astrocytic glutamate binding to presynaptic glutamate receptors (equation 5), we expected that the steady-state frequency response of a synapse ($RR_\infty$) could also be modulated by the astrocyte-synapse signaling. Since both geometry of synaptic bouton and diffusion of glutamate in the extracellular space are beyond the scope of the present work, we implicitly assumed, based on experimental evidence [31], that the release site of astrocytic glutamate apposes targeted presynaptic glutamate receptors. When the intracellular $Ca^{2+}$ in the astrocyte crossed over the threshold of glutamate exocytosis (Figure 4A, *top*, *dashed red line*), the extracellular concentration of glutamate in proximity of presynaptic receptors first increased rapidly and then decayed exponentially at rate $\Omega_c$, as a result of the concomitant uptake by astrocytic glutamate transporters and diffusion away from the site of exocytosis (Figure 4A, *middle*) (see also Text S1, Section I.4; Figure S5).

For $\alpha = 0$, equations (5-6) predict that this glutamate peak should lead to a sharp decay of $U_0$, followed by a slower recovery phase (Figure 4C, *left*). Using equation (5), we can also predict the resulting dependence of the steady-state synaptic response $RR_\infty$ on the input frequency (Figure 4C, *middle*). In the absence of astrocytic glutamate release (*thick dashed black line*), $RR_\infty$ monotonously decreases for increasing input frequency $f_{in}$ for the merely depressing synapse considered in this figure. At the release of astrocytic glutamate (Figure 4A, *middle*), the peak of bound presynaptic receptors (Figure 4A, *bottom*) and the resulting sharp drop of $U_0$ (Figure 4C, *left*, *black mark*) induce a strong decrease of the steady-state amount of released resources at low to intermediate input frequencies (0.1 – 10 Hz) (Figure 4C, middle, *thick red line*). In addition, the steady-state response loses its monotonicity and displays a peak frequency characteristic of facilitating synapses (see "Mechanisms of short-term presynaptic plasticity" in "Methods"). The $RR_\infty$ curve then slowly transforms back to its baseline form (*thin colored lines*) and the peak synaptic input frequency appears to progressively shift toward smaller input frequencies (*thick dashed arrow*). Hence, for $\alpha = 0$, the limiting frequency (equation 4) is predicted to sharply increase following astrocytic glutamate release and then to slowly relax back to smaller values (Figure 4C, *right*).

The exact opposite picture instead describes the scenario of α = 1 (Figure 4B). In this case, the parameter $U_0$ increases upon astrocytic glutamate release (Figure 4B, *left*) causing a dramatic increase of the steady-state response $RR_\infty$ for a range of frequencies within $0.1 - 10$ Hz (Figure 4B, *middle*). Accordingly, the limiting frequency of the synapse dramatically reduces following astrocytic glutamate release, and slowly recovers back to its baseline value (Figure 4B, *right*). Taken together, the above results of the mean-field analysis predict that, depending on the parametric scenario, astrocyte can either transiently decrease, when α = 0, or increase, if α = 1, the release of a model synapse.

To assess the validity of these predictions, we show in Figure 5 the responses of two different model synapses (A: depressing; B: facilitating) to Poisson spike trains delivered at frequency $f_{in}$ (Figure 5, *top panels* for specific realizations of such spike trains). To simplify the presentation, we considered the case in which a single $Ca^{2+}$ peak (Figure 5, *middle*) is sufficient to trigger the release of glutamate from the astrocyte. The synaptic response under different scenarios of astrocytic glutamate modulation (A: α = 0; B: α = 1) is then compared to the "Control" scenario obtained for the model synapses without astrocyte. In the case of α = 0 (Figure 5A, *bottom*) the amount of resources released by the model synapse steeply decreased at the onset of glutamate release from the astrocyte (*green area*) and slowly, i.e. tens of seconds, recovered to the levels comparable to those of the control scenario (*blue area*). The opposite effect was observed instead for α = 1 (Figure 5B). The synaptic response in this case was strongly augmented by astrocytic glutamate (*magenta area*) and then slowly decayed back to the levels obtained in control conditions (Figure 5B, *bottom*).

Collectively our mean field analysis (Figure 4) and simulations (Figure 5) suggest that glutamate release by the astrocyte can induce STD or STP of synaptic response to action potentials (Figure S6). Which one between these scenarios occurs depends on the value of the "effect" parameter α that lumps together both the density and the biophysical properties of presynaptic receptors targeted by astrocytic glutamate. These results are consistent with a large body of experimental observations in the hippocampus, where astrocyte-released glutamate could transiently decrease [33] or increase the synaptic response to stimulation [30-32, 34, 49].

# Astrocyte-synapse signaling mediates transitions between paired-pulse depression and facilitation at the same synapse

## Modulation of paired-pulse plasticity by astrocytic glutamate

The analysis presented above disclosed two independent routes to affect synaptic efficacy. (1) On one hand astrocyte-to-synapse signaling could either decrease ($\alpha = 0$) or increase ($\alpha = 1$) synaptic release. (2) On the other hand, the synapse itself, in the absence of the astrocyte, could exhibit STP if $U_0 < U_{thr}$ and $f_{in} < f_{lim}$, or STD otherwise. In principle, these two independent routes give rise to four possible scenarios to modulate the strength of synaptic response. For the sake of clarity, we restrict our attention in the rest of the paper to the intuitively simpler cases of "release-decreasing" astrocytes on otherwise depressing synapses, and of "release-increasing" astrocytes on otherwise facilitating synapses. The complementary cases – i.e. release-decreasing astrocytes on facilitating synapses and release-increasing astrocytes on depressing synapses – are addressed in the Supplementary Online Material (Figures S10 and S11 respectively).

Earlier studies suggested that variations of basal probability of synaptic release due to the activation of presynaptic glutamate receptors, are also expected to change synaptic plasticity as assessed by paired-pulse ratio (PPR) tests [5, 31]. Thus, we set to investigate how astrocytic glutamate modulated synaptic release in pairs of consecutive spikes for the realistic scenario of stochastic input trains such as those in Figure 5 (*top*) [65]. To this aim, we considered the synaptic response to Poisson-distributed spikes, computing the PPR value for each pair of consecutive spikes in the train (Text S1, Section II.2). The results are summarized in the histograms in Figures 5C,D in terms of ratio PPF/PPD of the number of facilitated pulse pairs (i.e. pairs for which PPR > 1) over the number of depressed ones (i.e. pairs such that PPR < 1), averaged over $n = 100$ Poisson spike trains with the same average frequency.

In the "Control" simulation (i.e. without astrocyte modulation), the depressing synapse of Figure 5A was expectedly characterized by PPF/PPD < 1 (Figure 5C, *blue bar*). By contrast, when a release-decreasing astrocyte was incorporated in this synapse, paired-pulse facilitation dominated, with PPF/PPD > 1 (*green bar*). The opposite picture was observed instead in the alternative scenario of release-increasing astrocyte modulating the facilitating synapse in Figure 5B. In this latter in fact, while in control simulations the synaptic response was consistently characterized by a ratio PPF/PPD > 1 (Figure 5D, *blue bar*), the release-increasing astrocyte shifted instead the balance between facilitated and depressed pairs in favor of these latter thus resulting in PPF/PPD < 1 (*magenta bar*).

To rule out the possibility that such increase of PPF (PPD) could have resulted out of the slow increase (decrease) of synaptic release during the recovery (decay) of the effect of release-decreasing (increasing) astrocytic glutamate, model synapses were also stimulated by pairs of spikes (Figure 6). For each spike pair we compared the amount of resources released after the first spike in the pair ($RR_1$) to those released after the second spike in the pair ($RR_2$). The averaged paired-pulse ratio, defined as $\mathrm{PPR} = \langle RR_2 / RR_1 \rangle$, was expected to be larger than 1 for a potentiating synapse, but less than 1 in the case of a depressing synapse (Text S1, Section II.2; Figure S2).

In absence of glutamate release from the astrocyte a depressing synapse responded to a spike pair releasing an amount of resources at the second spike that was *less* than the one at the first spike (Figure 6A, *middle*). Accordingly, the average paired-pulse ratio in this case was PPR < 1 (Figure 6C, *blue bar*). In presence of release-decreasing astrocyte however (equation 5-6, α = 0), the response of that same synapse to paired-pulse stimulation changed from depressing to facilitating – that is the amount of resources released upon the second spike in a pair was *larger* than that at the first spike (Figure 6A, *bottom*) –, and the average PPR became larger than 1 (Figure 6C, *green bar*). For the scenario of a release-increasing astrocyte on otherwise facilitating synapse, the exact opposite was observed (Figures 6B,D). Namely, glutamate release from the astrocyte transformed the model synapse from facilitating, i.e. PPR > 1 (Figure 6D, *blue bar*) to depressing, i.e. PPR < 1 (Figure 6D, *magenta bar*).

Taken together, the above results obtained both from Poisson input spike trains and paired-pulse stimulation hint that astrocytes could modulate short-term paired-pulse synaptic plasticity in a nontrivial way, triggering transitions from PPD to PPF or vice versa.

## Theoretical explanation of astrocyte-mediated transitions between PPD and PPF

Although the exact order of PPF and PPD for generic input spike trains, such as those in Figure 5, depends on the detail of spike timings of the stimulus, in light of the above observations, the effect of astrocytic glutamate on paired-pulse plasticity is expected to be similar for individual spike trains but all sharing the same statistics. We show this in the raster plots in Figures 7A,B (*left*) for the released resources of a depressing synapse modulated by a release-decreasing          astrocyte          (i.e. α = 0).          Facilitated (*green dots*)          vs. depressed          spike pairs (*magenta dots*) are displayed for 100 simulated Poisson spike trains sharing the same average frequency. In absence of the astrocyte ("Control"), the majority of pulse pairs is depressed given the depressing nature of the synapse, but as soon as the depressing astrocyte releases a glutamate pulse (Figure 7B, *black mark* at time $t = 10$ s), the number of facilitated pulse pairs becomes the majority, (*green dots* in the raster plot in Figure 7B, *left*). Notably, the alternation of facilitated and depressed pairs is different for each trial, but on average, the number of facilitated pairs increases for all trials right after astrocytic glutamate release.

The occurrence of this scenario can be explained by noting that the effect of the depressing astrocyte complies the two conditions required for PPF (see "Mechanisms of short-term presynaptic plasticity" in "Methods") namely: (1) that the baseline synaptic release probability $U_0$ is less than the switching threshold $U_{thr}$ and (2) that the frequency of presynaptic spikes is less than the limiting frequency of the synapse. In Figures 7A,B (*middle*) we show how $U_0$ (*black line*) changes during the stimulus with respect to $U_{thr}$ (*blue line*). In the absence of the astrocyte, $U_0$ is constant and because the synapse is depressing, it is larger than $U_{thr}$ (Figure 7A, *middle*). In presence of the astrocyte instead, $U_0$ changes, rapidly decreasing beyond $U_{thr}$ at the onset of glutamate exocytosis from the astrocyte (Figure 7B, *middle*), so that the first condition of facilitation is satisfied.

With regards to the second condition, in Figures 7A,B (*right*) we compare the instantaneous input frequency $f_{in}$ (*black line*), i.e. the inverse of the interspike interval averaged over trials, to the limiting frequency $f_{lim}$, as given by our mean-field analysis (Figure 4C, *right*, reported as the *dark red line* in Figure 7B, *right*). In control conditions, $f_{lim}$ is fixed (because $U_0$ is constant, see equation 4), and intersections of $f_{lim}$ with $f_{in}$ do not change synaptic plasticity, because in these conditions $U_0 > U_{thr}$ anyway. On the other hand, in presence of the astrocyte, $f_{lim}$ intersects $f_{in}$ at two points (Figure 7.B, *right*): first at the onset of glutamate release and then about 60 seconds later. Hence, at the first intersection of the two curves, the input frequency becomes smaller than the limiting frequency and the second condition for facilitation is also verified. In the raster plot of Figure 7B (*left*), this is marked by a dramatic increase at $t = 10$ s, of the number of green dots that mark facilitated pulse pairs. Conversely, after the second intersection, the input frequency becomes again larger than the limiting frequency, and the return to an essentially depressing regime can be noticed in the associated raster plot by an increasing occurrence of PPD towards the end of the considered time window.

The mirror reasoning also explains why a release-increasing astrocyte increases the chances of observing depressed paired-pulses in a facilitating synapse (Figures 7C,D). Here we start from a control case where both conditions for PPF are satisfied, that is $U_0 < U_{thr}$ (Figure 7C, *middle*) and $f_{in} < f_{lim}$ (Figure 7C, *right*). Upon glutamate release by the astrocyte this scenario changes instead because $U_0$ increases below $U_{thr}$ thus bringing forth predominant PPD as can be seen in the raster plot in Figure 7D (*left*). Depressed pulse pairs remain predominant also when $U_0$ recovers back to values below $U_{thr}$ towards the end of the considered time window, i.e. at $t = 100$ s. In this case in fact, $f_{lim}$ drops to zero by definition (equation 4) (results not shown) becoming less than $f_{in}$, which accounts for the predominance of PPD. Identical results can also be obtained by analysis of the slope of the frequency response curve as a function of $f_{in}$ (Figure S9).

In summary, our hitherto analysis shows that the effect of the astrocyte can be segregated into two components. First, the astrocyte modulates the overall amount of synaptic resources released after each input spike compared to the case without it. This imposes global decrease or increase of synaptic efficacy in terms of amount of released neurotransmitter. Second, because this effect shifts the location of the limiting frequency of the synapse, the astrocyte can also simultaneously modulate paired-pulse plasticity. Notably, these modulations are in the opposite direction with respect to the global depressing effect. That is, while a release-decreasing astrocyte is predicted to enhance PPF, a release-increasing one could instead reinforce PPD.

# Persistent $Ca^{2+}$ oscillations in astrocytes can regulate presynaptic short-term plasticity

## Different $Ca^{2+}$ dynamics correspond to different frequencies of glutamate release from the astrocyte

In many cases, astrocytic processes are found to display oscillating $Ca^{2+}$ dynamics. Because the $Ca^{2+}$ threshold for glutamate release is relatively low compared to the amplitude of $Ca^{2+}$ signal [47, 59, 63], one would expect persistent exocytosis of glutamate into the extrasynaptic space [47]. Thus, we proceeded to study the implications of such persistent glutamate release from the astrocyte on modulation of short-term presynaptic plasticity.

Generally speaking, the frequency of astrocytic $Ca^{2+}$ oscillations translates into the frequency at which astrocytic glutamate is released, implying that high rates of $Ca^{2+}$ oscillations would likely lead to stronger and faster depletion of releasable astrocytic glutamate [59]. Conversely, if the frequency of $Ca^{2+}$ oscillations is much smaller than the recovery rate $\Omega_A$ [44, 66], glutamate can recover in between the oscillation peaks and roughly the same amount of glutamate be released per oscillation. The rate at which the astrocytic glutamate pool is depleted is also likely to depend on the amplitude of $Ca^{2+}$ oscillations, with smaller-amplitude oscillations corresponding to lower probability of exocytosis [59].

The effects of amplitude and frequency of astrocytic $Ca^{2+}$ oscillations on the modulation of synaptic response properties are summarized in Figure 8. We considered three different stereotypical patterns of $Ca^{2+}$ oscillations modulated with time of their amplitude (AM), frequency (FM) or both (AFM) (Text S1, Section I.2; Figure S4). The $Ca^{2+}$ threshold $C_{thr}$ for glutamate exocytosis (*dashed red line*) was such that FM oscillations (Figure 8B) always crossed it, each triggering a single glutamate release event [67]. Conversely, AM or AFM oscillations (Figures 8A and 8C) did not always lead to the release of glutamate, as the $Ca^{2+}$ levels did not always reach $C_{thr}$. Thus, while the FM oscillations triggered glutamate exocytosis at the same frequency as their own (Figure 8B, *bottom*), the amplitude of AM and AFM oscillations selectively discriminated which oscillations triggered glutamate exocytosis, eventually dictating the frequency of "measured" glutamate release events. Hence, AM oscillations at *constant* frequency (Figure 8A, *top*) would generate sequences of glutamate release events of identical magnitude yet at *variable* frequency (Figure 8A, *bottom*). An implication of this mechanism is that different patterns of $Ca^{2+}$ oscillations could be encoded mainly by the frequency rather than the magnitude of astrocytic glutamate release.

## Astrocyte regulates transitions between facilitation and depression

Results presented in the previous section lead to the model prediction that if astrocytic $Ca^{2+}$ dynamics is encoded by the frequency $f_C$ of "measured" glutamate release events (GREs), then this frequency should critically shape the astrocytic modulatory effect on synaptic plasticity. We demonstrate this in Figure 9, where the effect of different GRE frequencies $f_C$ on facilitated (PPF) vs. depressed (PPD) pulse pairs is shown for $n = 100$ Poisson spike trains with the same average input rate. For a "release-decreasing" astrocyte acting on a depressing synapse, higher rates of GRE lead to stronger facilitation (Figure 9A). By contrast, increasing GRE frequency results in more synaptic depression when a "release-increasing" astrocyte modulates a facilitating synapse (Figure 9B).

We set to determine the effect that the rate $f_C$ of GREs might have on the basal synaptic release probability $U_0$. We employed mean-field analysis, assuming the existence of multiple GREs at different frequencies. The steady-state synaptic basal release probability $U_{0\infty}$ was computed as an explicit function of the GRE frequency $f_C$, the four rates $\Omega_A$, $\Omega_c$, $O_G$ and $\Omega_G$ (see Table S1 for an explanation), the total amount of releasable astrocytic glutamate $\beta$, and the effect parameter $\alpha$ (the detailed derivation can be found in the Text S1, Section II.5; see also Figures S7B,S8):

$$U_{0\infty} = \frac{\Omega_A \Omega_c \Omega_G U_0^* + \left(\Omega_c \Omega_G U_0^* + \alpha \beta \Omega_A O_G\right) U_A f_C}{\Omega_A \Omega_c \Omega_G + \left(\Omega_c \Omega_G + \beta \Omega_A O_G\right) U_A f_C} \qquad (7)$$

Experimental data [30-34] suggests that the rate-limiting step in $U_0$ dynamics is due to the slow astrocyte modulation of synaptic neurotransmitter release. Thus, we assumed that $\Omega_G << \Omega_A << \Omega_c, G_A O_G$ (Table S1). This implies that $U_{0\infty} \to \alpha$ in the high GRE rate regime (for $f_C >> \Omega_G$). In other words, in the presence of fast astrocytic Ca$^{2+}$ oscillations that would cause persistent release of glutamate at a high rate, the synaptic basal release probability would be stable and would be defined by the nature of presynaptic glutamate receptors.

This concept is further elucidated in Figures 10. At $f_C$ = 0.001 Hz, when glutamate release from astrocyte is sporadic, a facilitating synapse without astrocyte expectedly has basal release probability $U_0$ lower than the threshold value $U_{thr}$ at which the synapse switches from facilitating to depressing (Figure 10A, *left*, *blue line*). When a release-increasing astrocyte is added, our prediction suggests that $U_0$ should *increase* towards $\alpha$ for progressively higher GRE rates. Indeed, above a critical rate $f_{thr}$ of GREs, $U_0$ crosses over from facilitating regime to depressing one (*magenta area* in Figure 10A). For release-decreasing astrocytes acting on depressing synapses, the opposite holds instead, as shown in Figure 10B (*left*). In this latter case however the astrocyte effectively induces PPD-to-PPF transition only if the condition for facilitation on the limiting frequency condition is also satisfied, that is if the input rate of incoming spikes is such that $f_{in} < f_{lim}$ (see "Mechanisms of short-term presynaptic plasticity" in "Methods").

The synaptic limiting frequency $f_{lim}$ for the above-discussed cases as function of the GRE rate is shown in Figures 10A,B (*right*). Taking as an example the case of the release-decreasing astrocyte that modulates the depressing synapse in Figure 10B, we note that for incoming spikes at average frequency $f_{in}$ = 1.5 Hz, PPF is effectively expected to prevail on PPD for GRE rates close to $f_{thr} \approx 0.0015$ Hz (*dashed-dotted blue line*). On the other hand, for higher input frequencies, $f_{thr} \approx 0.0015$ Hz is not the effective threshold for the switch between PPD and PPF because for such input rates, the condition required for facilitation that $f_{in} < f_{lim}$ is verified only for $f_c > 0.0015$ Hz. Opposite dependence can be found for the release-increasing astrocyte on the facilitating synapse in Figure 10A.

## Discussion

The character of synaptic information transfer is shaped by several factors [2]. Synaptic strength at any given moment is determined by an earlier "activation history" of that same synapse [3, 5]. Structural and functional organization of presynaptic bouton affects the release and reintegration of neurotransmitter vesicles, ultimately defining the filtering feature (depressing or facilitating) of a synapse in response to spike train stimulation [3, 68]. Existing models of synaptic dynamics assume "fixed plasticity mode", in which the depression/facilitation properties of a synapse do not change with time. However, in biological synapses, plasticity itself seems to be a dynamic feature; for example, the filtering characteristics of a given synapse is not fixed, but rather can be adjusted by modulation of the initial release probability of docked vesicles [13]. Using a computational modeling approach, we showed here that astrocytes have the potential to modulate the flow of synaptic information via glutamate release pathway. In particular, astrocyte-mediated regulation of synaptic release could greatly increase paired-pulse facilitation (PPF) at otherwise depressing synapses (and thus switch the synapse from depressing to facilitating); conversely, it could reinforce paired-pulse depression (PPD) at otherwise facilitating synapses (therefore switching the synapse from facilitating to depressing). These findings imply that astrocytes could dynamically control the transition between different "plasticity modes". The present model also lends an explanation to several pieces of experimental data, as we detail below.

In agreement with experimental results [26, 27], our model suggests that the type of presynaptic glutamate receptors targeted by astrocytic glutamate critically determines the type of modulation that takes place. The modulatory action of an astrocyte is lumped in our model into the so-called "effect" parameter α: lower values of α make the action of an astrocyte depressing with respect to the overall synaptic release but also increase paired-pulse facilitation. On the contrary higher α values make the effect of an astrocyte facilitating but at the same time paired-pulse depression is enhanced. Interestingly, some recent experiments on perforant path-granule cell synapses in the hippocampal dentate gyrus, show that facilitation of synaptic release mediated by astrocyte-derived glutamate correlates with a decrease of paired-pulse ratio [31]. Our model provides a natural explanation of these experimental results.

Several lines of experimental evidence suggest that different types of glutamate receptors may be found at the same synaptic bouton [26]. The different types of receptors have different activation properties and hence could be recruited simultaneously or in a complex fashion [30, 41]. Thus it is likely that α could take intermediate values between 0 and 1. In one particular scenario, concurrence of astrocyte-mediated depression and facilitation could also lead these two effects to effectively cancel each other so that no apparent modulation of synaptic release is observed. Interestingly, in some recent studies, the $Ca^{2+}$-dependent release of glutamate from astrocytes was reported not to affect synaptic function [69, 70], thus questioning the vast body of earlier experimental evidence pointing to the contrary. In our model we posit that an apparent lack of astrocytic effect on synaptic function could arise when the "effect" parameter α exactly matches the basal release probability of that presynaptic terminal, that is when α = $U_0$* (in which case equation 5 becomes $U_0(\Gamma) = \alpha$, meaning that $U_0$ does not depend on Γ anymore). This scenario would lead to concurrence of astrocyte-mediated depression and facilitation with no net observable effect on synaptic transmission.

Whether *de facto* astrocytes decrease or increase synaptic release likely depends on the specific synapse under consideration and the functional implications that such different modulations could lead to [15, 21, 30]. In the former case for example, enhanced PPF could be not functionally relevant if release of neurotransmitter is strongly reduced by astrocyte glutamate signaling. In such situation the astrocyte would essentially shut down synaptic transmission, hindering the flow of information carried by presynaptic spikes [71]. On the other hand, for astrocyte-induced facilitation, an increase of released neurotransmitter could correspond to a similar increase of transmission probability [72]. However, the associated modulations of paired-pulse plasticity could also account for complex processing of specific – i.e. temporal vs. rate – features of input spike trains [2, 51, 57] that could not otherwise be transmitted by the single synapse, that is without the astrocyte.

In a recent line of experiments on frog neuromuscular junction, it was observed that glial cells could govern the outcome of synaptic plasticity based on their ability to bring forth variegated $Ca^{2+}$ dynamics [40, 41]. In other words, different patterns of $Ca^{2+}$ oscillations in perisynaptic glia were shown to activate different presynaptic receptors and thus to elicit different modulatory effects on neurotransmitter release [41]. This scenario would call for a future modification of our model to include a dependence on astrocytic $Ca^{2+}$ dynamics of the effect parameter α. Nevertheless such observations are generally bolstered by our study. Our model predicts the existence of a threshold frequency for $Ca^{2+}$ oscillations below which PPD (PPF) is predominant with respect to PPF (PPD) and above which the opposite occurs. This supports the idea that different spatiotemporal $Ca^{2+}$ dynamics in astrocytes, possibly due to different cellular properties [73-76], could provide specialized feedback to the synapse [40]. Moreover, our model suggests that different types of presynaptic glutamate receptors might not be necessary to trigger different modulations of synaptic transfer properties. The fact that the frequency of $Ca^{2+}$ oscillations could bias synaptic paired-pulse plasticity subtends the notion that not only the nature of receptors, but also the dynamics of their recruitment by gliotransmitter could be a further critical factor in the regulation of synaptic plasticity [27, 41]. This latter could eventually be dictated by the timing and the amount of released glutamate [27, 44] as well as by the ultrastructure of astrocytic process with respect to synaptic terminals which defines the geometry of extracellular space [18, 77] thus controlling the time course of glutamate therein [78].

Remarkably, the threshold frequency of $Ca^{2+}$ oscillations that discriminates between PPD and PPF falls, in our analysis, within the range $<2.5$ min$^{-1}$ of spontaneous $Ca^{2+}$ oscillations displayed by astrocytes in basal conditions independently of neuronal activity [32, 49, 50, 79, 80], hinting a possible role for these latter in the regulation of synaptic physiology. Spontaneous $Ca^{2+}$ oscillations can indeed trigger astrocytic glutamate release [32, 79-82] which could modulate ambient glutamate leading to tonic activation of presynaptic receptors [26, 83]. In this fashion, spontaneous glutamate gliotransmission could constitute a mechanism of regulation of basal synaptic release. Notably, in a line of recent experiments, selective metabolic arrest of astrocytes was observed to depress Schaffer collateral synaptic transmission towards increasing PPF, consistently with a reduction of the basal synaptic release probability as predicted by our analysis [49]. The latter could be also relevant in the homosynaptic case of astrocytic glutamate exocytosis evoked by basal activity of the same presynaptic terminal that it feeds back to [21, 27, 84]. In such conditions, the ensuing influence of astrocytic glutamate on synaptic release correlates with the incoming synaptic stimulus also through $Ca^{2+}$ dynamics in the astrocyte [85], unraveling potentially new mechanisms of modulation of synaptic transmission and plasticity.

Although we focused on regulation of astrocyte at single synapses, our analysis could also apply to synaptic ensembles [51, 54] that could be "contacted" either by the same astrocytic process [30, 32, 80] or by different ones with locally synchronized $Ca^{2+}$ dynamics [81]. In this case, modulation of the release probability by the astrocyte would support the existence of "functional synaptic islands" [86], namely groups of synapses, intermittently established by different spatiotemporal $Ca^{2+}$ dynamics, whose transmission mode and plasticity share common features. The implications that such dynamic astrocyte-synapse interaction might have with regard to information flow in neural circuitry remain to be investigated.

Due to their capacity to modulate the characteristics of synaptic transmission, astrocytes could also alter the temporal order of correlated pre- and postsynaptic spiking that critically dictates spike-timing dependent plasticity (STDP) at the synapse [87]. Thus, astrocyte modulation of short-term plasticity could potentially contribute to ultimately shape persistent modifications of synaptic strength [49, 88, 89] underlying processing, memory formation and storage that provides the exquisite balance, subtlety and smoothness of operation for which nervous systems are held in awe [90]. Future combined physiological and computational studies will determine whether or not this is the case.

## Acknowledgements

The authors wish to thank Misha Tsodyks, Andrea Volterra and Richard Robitaille for illuminating discussions. M. D. P. also thanks the Center for Theoretical Biological Physics (CTBP) at University of California at San Diego for hospitality while part of this research was carried out.

## Funding

M. D. P. and V. V. acknowledge the support of the U.S. National Science Foundation I2CAM International Materials Institute Award, Grants DMR-0844115 and DMR-0645461. This research was supported by the Tauber Family Foundation, by the Maguy-Glass Chair in Physics of Complex Systems at Tel Aviv University, by the NSF-sponsored Center for Theoretical Biological Physics (CTBP), PHY grants 0216576 and 0225630 and by the University of California at San Diego.

## Author contributions

Conceived and designed the study: MDP VV HB EBJ. Performed the simulations: MDP. Analyzed the data and wrote the paper: MDP VV HB EBJ.

# References

1.      Barak O, Tsodyks M (2007) Persistent activity in neural networks with dynamic synapses. PLoS Comput Biol 32: e35.

2.      Abbott LF, Regehr WG (2004) Synaptic computation. Nature 431: 796-803.

3.      Zucker RS, Regehr WG (2002) Short-term synaptic plasticity. Annual Rev Physiol 64: 355-405.

4.      Mongillo M, Barak O, Tsodyks M (2008) Synaptic theory of working memory. Science 319: 1543-1546.

5.      Citri A, Malenka RC (2008) Synaptic plasticity: multiple forms, functions and mechanisms. Neuropsychopharmacology Rev 33: 18-41.

6.      Tsodyks MV, Markram H (1997) The neural code between neocortical pyramidal neurons depends on neurotransmitter release probability. Proc Natl Acad Sci USA 94: 719-723.

7.      Dobrunz LE, Stevens CF (1997) Heterogeneity of release probability, facilitation, and depletion at central synapses. Neuron 18: 995-1008.

8.      Südhof TC (2004) The synaptic vesicle cycle. Annu Rev Neurosci 27: 509-547.

9.      Schneggenburger R, Sakaba T, Neher E (2002) Vesicle pools and short-term synaptic depression: lessons form a large synapse. Trends Neurosci 25: 206-212.

10.     Nadkarni S, Bartol TM, Sejnowski TJ, Levine H (2010) Modelling vesicular release at hippocampal synapses. PLoS Comput Biol 6: e1000983.

11.     Sun J, Pang ZP, Qin D, Fahim AT, Adachi R, Südhof TC (2007) A dual-$Ca^{2+}$-sensor model for neurotransmitter release in a central synapse. Nature 450: 676-682.

12.     Abbott LF, Varela JA, Sen K, Nelson SB (1997) Synaptic depression and cortical gain control. Science 25: 220-224.

13.     Dittman JS, Kreitzer AC, Regehr WG (2000) Interplay between facilitation, depression, and residual calcium at three presynaptic terminals. J Neurosci 20: 1374-1385.

14.     Debanne D, Guerineau NC, Giihwiler BH, Thompson SM (1996) Paired-pulse facilitation and depression at unitary synapses in rat hippocampus: quantal fluctuation affects subsequent release. J Physiol 491: 163-176.

15.     Haydon PG, Carmignoto G (2006) Astrocyte control of synaptic transmission and neurovascular coupling. Physiol Rev 86: 1009-1031.

16.     Herculano-Houzel S (2009) The human brain in numbers: a linearly scaled-up primate brain. Frontiers Human Neurosci 3: 1-11.

17.     Savchenko VL, McKanna JA, Nikonenko IR, Skibo GG (2000) Microglia and astrocytes in the adult rat brain: comparative immunocytochemical analysis demonstrates the efficacy of lipocortin 1 immunoreactivity. Neuroscience 96: 196-203.

18.     Ventura R, Harris KM (1999) Three-dimensional relationships between hippocampal synapses and astrocytes. J Neurosci 19: 6897-6906.

19.     Haydon PG (2001) Glia: listening and talking to the synapse. Nature Rev Neurosci 2: 185-193.

20.     Araque A, Parpura V, Sanzgiri RP, Haydon PG (1999) Tripartite synapses: glia, the unacknowledged partner. Trends Neurosci 23: 208-215.

21.     Volterra A, Meldolesi J (2005) Astrocytes, from brain glue to communication elements: the revolution continues. Nature Rev Neurosci 6: 626-640.

22.     Parpura V, Zorec R (2010) Gliotransmission: exocytotic release from astrocytes. Brain Res Rev 63: 83-92.

23.     Perea G, Navarrete M, Araque A (2009) Tripartite synapse: astrocytes process and control synaptic information. Trends Neurosci 32: 421-431.

24.     Agulhon C, Petravicz J, McMullen AB, Sweger EJ, Minton SK, et al. (2008) What is the role of astrocyte calcium in neurophysiology? Neuron 59: 932-946.

25.     Barnes BA (2008) The mystery and magic of glia: a perspective on their roles in health and disease. Neuron 60: 430-440.

26.     Pinheiro PS, Mulle C (2008) Presynaptic glutamate receptors: physiological functions and mechanisms of action. Nature Rev 9: 423-436.

27.     Santello M, Volterra A (2009) Synaptic modulation by astrocytes via $Ca^{2+}$-dependent glutamate release. Neuroscience 158: 253-259.

28.     Andersson M, Hanse E (2010) Astrocytes impose postburst depression of release probability at hippocampal glutamate synapses. J Neurosci 30: 5776-5780.

29.     Andersson M, Blomstrand F, Hanse E (2007) Astrocytes play a critical role in transient heterosynaptic depression in the rat hippocampal CA1 region. J Physiol 585: 843-852.

30.     Perea G, Araque A (2007) Astrocytes potentiate transmitter release at single hippocampal synapses. Science 317: 1083-1086.

31.     Jourdain P, Bergersen LH, Bhaukaurally K, Bezzi P, Santello M, Domercq M, et al. (2007) Glutamate exocytosis from astrocytes controls synaptic strength. Nature Neurosci 10: 331-339.

32.     Fiacco TA, McCarthy KD (2004) Intracellular astrocyte calcium waves *in situ* increase the frequency of spontaneous AMPA receptor currents in CA1 pyramidal neurons. J Neurosci 24: 722-732.

33.     Araque A, Parpura V, Sanzgiri RP, Haydon PG (1998) Glutamate-dependent astrocyte modulation of synaptic transmission between cultured hippocampal neurons. Eur J Neurosci 10: 2129-2142.

34.     Araque A, Sanzgiri RP, Parpura V, Haydon PG (1998) Calcium elevation in astrocytes causes an NMDA receptor-dependent increase in the frequency of miniature synaptic currents in cultured hippocampal neurons. J Neurosci 18: 6822-6829.

35.     Kang J, Jiang L, Goldman SA, Nedergaard M (1998) Astrocyte-mediated potentiation of inhibitory synaptic transmission. Nat Neurosci 1: 683-692.

36.     Serrano A, Haddjeri N, Lacaille J, Robitaille R (2006) GABAergic network activation of glial cells underlies heterosynaptic depression. J Neurosci 26: 5370-5382.

37.     Pascual O, Casper KB, Kubera C, Zhang J, Revilla-Sanchez R, et al. (2005) Astrocytic purinergic signaling coordinates synaptic networks. Science 310: 113-116.

38.     Gordon GRJ, Iremonger KJ, Kantevari S, Ellis-Davies GCR, MacVicar BA, et al. (2009) Astrocyte-mediated distributed plasticity at hypothalamic glutamate synapses. Neuron 64: 391-403.

39.     Newman EA (2003) Glial cell inhibition of neurons by release of ATP. J Neurosci 23: 1659-1666.

40.     Rousse I, St-Amour A, Darabid H, Robitaille R (2010) Synapse-glia interactions are governed by synaptic and intrinsic glial properties. Neuroscience 167: 621-632.

41.     Todd KJ, Darabid H, Robitaille R (2010) Perisynaptic glia discriminate patterns of motor nerve activity and influence plasticity at the neuromuscular junction. J Neurosci 30: 11870-11882.

42.     Robinson R (1998) Modulation of synaptic efficacy and synaptic depression by glial cells at the frog neuromuscular junction. Neuron 21: 847-855.

43.     Fellin T (2009) Communication between neurons and astrocytes: relevance to the modulation of synaptic and network activity. J Neurochem 108: 533-544.

44.     Montana V, Malarkey EB, Verderio C, Matteoli M, Parpura V (2006) Vesicular transmitter release from astrocytes. Glia 54: 700-715.

45.     Giaume C, Koulakoff A, Roux L, Holcman D, Rouach N (2010) Astroglial networks: a step further in neuroglial and gliovascular interactions. Nature Reviews Neuroscience 11: 87-99.

46.     Kang N, Xu J, Xu Q, Nedergaard M, Kang J (2005) Astrocytic glutamate release-induced transient depolarization and epileptiform discharges in hippocampal CA1 pyramidal neurons. J Neurophysiol 94: 4121-4130.

47.     Shigetomi E, Kracun S, Sovfroniew MS, Khakh BS (2010) A genetically targeted optical sensor to monitor calcium signals in astrocyte processes. Nature Neurosci 13: 759-766.

48.     Nimmerjahn A (2009) Astrocytes going live: advances and challenges. J Physiol 587: 1639-1647.

49.     Bonansco C, Couve A, Perea G, Ferradas CA, Roncagliolo M, et al. (2011) Glutamate released spontaneously from astrocytes sets the threshold for synaptic plasticity. Eur J Neurosci 33: 1483-1492.

50.     Nett WJ, Oloff SH, McCarthy KD (2002) Hippocampal astrocytes in situ exhibit calcium oscillations that occur independent of neuronal activity. J Neurophysiol 87: 528-537.

51.     Tsodyks M, Pawelzik K, Markram H (1998) Neural networks with dynamic synapses. Neural Computation 10: 821-835.

52.     Amit DJ, Tsodyks MV (1991) Quantitative study of attractor neural network retrieving at low spike rates: I. Substrate-spikes, rates and neuronal gain. Network 2: 259-273.

53.     Del Castillo J, Katz B (1954) Quantal components of the end-plate potential. J Physiol 124: 560-573.

54.     Fuhrmann G, Segev I, Markram H, Tsodyks M (2002) Coding of temporal information by activity-dependent synapses. J Neurophysiol 87: 140-148.

55.     Markram H, Pikus D, Gupta A, Tsodyks M (1998) Potential for multiple mechanisms, phenomena and algorithms for synaptic plasticity at single synapses. Neuropharmacology 37: 489-500.

56.     Dittman JS, Regehr WG (1998) Calcium dependence and recovery kinetics of presynaptic depression at the climbing fiber to Purkinje cell synapse. J Neurosci 18: 6147-6162.

57.     Tsodyks M (2005) Activity-dependent transmission in neocortical synapses. In: Chow CC, Gutkin B, D H, C M, Dalibard J, editors, Methods and Models in Neurophysics, Elsevier. pp. 245-266.

58.    Lee W, Parpura V (2007) Exocytotic release of glutamate from astrocytes: comparison to neurons. In: Bean A, editor, Protein trafficking in neurons, Amsterdam, The Netherlands: Elsevier. pp. 329-365.

59.    Pasti L, Zonta M, Pozzan T, Vicini S, Carmignoto G (2001) Cytosolic calcium oscillations in astrocytes may regulate exocytotic release of glutamate. J Neurosci 21: 477-484.

60.    Bergersen LH, Gundersen V (2009) Morphological evidence for vesicular glutamate release from astrocytes. Neuroscience 158: 260-265.

61.    Zhang Q, Fukuda M, Van Bockstaele E, Pascual O, Haydon PG (2004) Synaptotagmin IV regulates glial glutamate release. Proc Natl Acad Sci USA 101: 9441–9446.

62.    Marchaland J, Calì C, Voglmaier SM, Li H, Regazzi R, et al. (2008) Fast subplasma membrane $Ca^{2+}$ transients control exo-endocytosis of synaptic-like microvesicles in astrocytes. J Neurosci 28: 9122-9132.

63.    Parpura V, Haydon PG (2000) Physiological astrocytic calcium levels stimulate glutamate release to modulate adjacent neurons. Proc Natl Acad Sci USA 97: 8629-8634.

64.    Hori T, Takahashi T (2009) Mechanisms underlying short-term modulation of transmitter release by presynaptic depolarization. J Physiol 587: 2987-3000.

65.    Softky W, Koch C (1993) The highly irregular firing pattern of cortical cells is inconsistent with temporal integration of random EPSPs. J Neurosci 13: 334-350.

66.    Bowser DN, Khakh BS (2007) Two forms of single-vesicle astrocyte exocytosis imaged with total internal reflection fluorescence microscopy. Proc Natl Acad Sci USA 104: 4212-4217.

67.    Pasti L, Volterra A, Pozzan T, Carmignoto G (1997) Intracellular calcium oscillations in astrocytes: a highly plastic, bidirectional form of communication between neurons and astrocytes *in situ*. J Neurosci 17: 7817-7830.

68.    Markram H, Wang Y, Tsodyks M (1998) Differential signaling via the same axon of neocortical pyramidal neurons. Proc Natl Acad Sci USA 95: 5323–5328.

69.    Agulhon C, Fiacco TA, McCarthy KD (2010) Hippocampal short- and long-term plasticity are not modulated by astrocyte $Ca^{2+}$ signalling. Science 327: 1250-1254.

70.    Fiacco TA, Agulhon C, Taves SR, Petravicz J, Casper KB, et al. (2007) Selective stimulation of astrocyte calcium in situ does not affect neuronal excitatory synaptic activity. Neuron 54: 611-626.

71.    Volman V, Ben-Jacob E, Levine H (2007) The astrocyte as a gatekeeper of synaptic information transfer. Neur Comput 19: 303-326.

72.     Nadkarni S, Jung P, Levine H (2008) Astrocytes optimize the synaptic transmission of information. PLoS Comput Biol 4: e1000088.

73.     Goldberg M, De Pittà M, Volman V, Berry H, Ben-Jacob E (2010) Nonlinear gap junctions enable long-distance propagation of pulsating calcium waves in astrocyte networks. PLoS Comput Biol 6: e1000909.

74.     De Pittà M, Goldberg M, Volman V, Berry H, Ben-Jacob E (2009) Glutamate-dependent intracellular calcium and $IP_3$ oscillating and pulsating dynamics in astrocytes. J Biol Phys 35: 383-411.

75.     De Pittà M, Volman V, Levine H, Ben-Jacob E (2009) Multimodal encoding in a simplified model of intracellular calcium signaling. Cogn Proc 10: 55-70.

76.     De Pittà M, Volman V, Levine H, Pioggia G, De Rossi D, et al. (2008) Coexistence of amplitude and frequency modulations in intracellular calcium dynamics. Phys Rev E 77: 030903(R).

77.     Haber M, Zhou L, Murai KK (2006) Cooperative astrocyte and dendritic spine dynamics at hippocampal excitatory synapses. J Neurosci 26: 8881-8891.

78.     Rusakov DA, Kullmann DM (1998) Extrasynaptic glutamate diffusion in the hippocampus: ultrastructural constraints, uptake, and receptor activation. J Neurosci 18: 3158-3170.

79.     Angulo MC, Kozlov AS, Charpak S, Audinat E (2004) Glutamate released from glial cells synchronizes neuronal activity in the hippocampus. J Neurosci 24: 6920-6927.

80.     Fellin T, Pascual O, Gobbo S, Pozzan T, Haydon PG, et al. (2004) Neuronal synchrony mediated by astrocytic glutamate through activation of extrasynaptic NMDA receptors. Neuron 43: 729-743.

81.     Sasaki T, Kuga T, Namiki S, Matsuki N, Ikegaya Y (2011) Locally synchronized astrocytes. Cerebral Cortex : In press.

82.     Tian GF, Azmi H, Takahiro T, Xu Q, Peng W, et al. (2005) An astrocytic basis of epilepsy. Nature Med 11: 973-981.

83.     Oliet SHR, Piet R, Poulain DA (2001) Control of glutamate clearance and synaptic efficacy by glial coverage of neurons. Science 292: 923-926.

84.     Bezzi P, Volterra A (2001) A neuron-glia signalling network in the active brain. Curr Opinion Neurobiol 11: 387-394.

85.     Aguado F, Espinosa-Parrilla JF, Carmona MA, Soriano E (2002) Neuronal activity regulates correlated network properties of spontaneous calcium transients in astrocytes *in situ*. J Neurosci 22: 9430-9444.

86.     Halassa MM, Fellin T, Takano H, Dong JH, Haydon PG (2007) Synaptic islands defined by the territory of a single astrocyte. J Neurosci 27: 6473-6477.

87.     Dan Y, Poo MM (2004) Spike timing-dependent plasticity of neural circuits. Neuron 44: 23-30.

88.     Henneberger C, Papouin T, Oliet SHR, Rusakov DA (2010) Long-term potentiation depends on release of D-serine from astrocytes. Nature 463: 232-237.

89.     Santello M, Volterra A (2010) Astrocytes as aide-mémoires. Nature 463: 169-170.

90.     Abbott LF, Nelson SB (2000) Synaptic plasticity: taming the beast. Nature 3: 1178-1183.

91.     Li Y, Rinzel J (1994) Equations for InsP$_3$ receptor-mediated $[Ca^{2+}]_i$ oscillations derived from a detailed kinetic model: A Hodgkin-Huxley like formalism. J Theor Biol 166: 461-473.

92.     Diamond JS, Jahr CE (2000) Synaptically released glutamate does not overwhelm transporters on hippocampal astrocytes during high-frequency stimulation. J Neurophysiol 83: 2835-2843.

93.     Cochilla AJ, Alford S (1998) Metabotropic glutamate receptor-mediated control of neurotransmitter release. Neuron 20: 1007-1016.

94.     Gereau RW IV, Conn J (1995) Multiple presynaptic metabotropic glutamate receptors modulate excitatory and inhibitory synaptic transmission in hippocampal area CA1. J Neurosci 15: 6879-6889.

95.     Baskys A, Malenka RC (1991) Agonist at metabotropic glutamate receptors presynaptically inhibit EPSCs in neonatal rat hippocampus. J Physiol 444: 687-701.

**A tale of two stories: astrocyte regulation of synaptic depression and facilitation**

Maurizio De Pittà, Vladislav Volman, Hugues Berry, and Eshel Ben-Jacob

# SUPPLEMENTARY TEXT S1

The organization of the present Supporting Text is as following.

# I.    Model description

## 1.    The Tsodyks-Markram model of synaptic release

Mechanisms of short-term synaptic depression and facilitation at excitatory hippocampal synapses can be realistically mimicked by the Tsodyks-Markram (TM) model of activity-dependent synapse [68, 6]. The model considers two variables $u$ and $x$, which respectively correlate with the state of occupancy of the calcium ($Ca^{2+}$) sensor of synaptic glutamate exocytosis and the fraction of glutamate available for release at any time [3].

At resting (basal) conditions, the occupancy of the sensor is minimal so that $u = 0$. Each presynaptic spike occurring at time $t_i$ (and modeled by a Dirac delta) triggers $Ca^{2+}$ influx into the presynaptic terminal thus increasing $u$. In particular, the model assumes that a fraction $U_0$ of $1-u$ vacant states of the sensor is first occupied by incoming $Ca^{2+}$ ions [4, 5], and is following recovered at rate $\Omega_f$. Hence, $u(t)$ evolves according to the equation

$$\dot{u} = -\Omega_f u + U_0 \sum_i (1-u)\delta(t-t_i) \qquad (1)$$

Following the increase of $u$ upon action potential arrival, an amount $ux$ of presynaptic glutamate is released into the cleft while the pool of synaptic glutamate (assumed to be constant in size) is replenished at rate $\Omega_d$. The equation for $x(t)$ then reads

$$\dot{x} = \Omega_d (1-x) - \sum_i ux\delta(t-t_i) \qquad (2)$$

The fraction of released glutamate resources $RR$, upon arrival of a presynaptic spike at $t = t_i$ is given by

$$RR(t_i) = u(t_i^+) \cdot x(t_i^-) \qquad (3)$$

where $u(t_i^+)$ and $x(t_i^-)$ denote the values of $u$ and $x$ immediately respectively *after* and *before* the spike at $t = t_i$.

On a par with the classical quantal model of synaptic transmission [53], $x(t)$ is analogous to the probability of a glutamate-containing vesicle to be available for release at any time $t$; $u(t)$ corresponds instead to the probability of release of a docked vesicle; and finally $RR(t_i)$ represents the probability of release (for every release site) at the time $t_i$ of the spike [54]. Then the parameter $U_0$ in the above equation (1), coincides with the value reached by $u$ immediately after the first spike of a train, starting from resting conditions (i.e. $u(0) = 0$, $x(0) = 1$). Since this situation also corresponds to the case of basal stimulation, that is of a stimulus at very low frequency, $U_0$ can be regarded as the basal value of synaptic release probability too [51]. The TM formulation ignores the stochastic nature of synaptic release and reproduces the average synaptic release event generated by any presynaptic spike train [54].

## 2.     Astrocytic calcium dynamics

Intracellular $Ca^{2+}$ concentration in astrocytes can be modulated by several mechanisms [24]. These include $Ca^{2+}$ influx from the extracellular space or controlled release from intracellular $Ca^{2+}$ stores such as the endoplasmic reticulum (ER) and mitochondria [10]. Inositol trisphosphate- ($IP_3$-) dependent $Ca^{2+}$-induced $Ca^{2+}$ release (CICR) from the ER is considered though the primary mechanism responsible of intracellular $Ca^{2+}$ dynamics in astrocytes [48]. In this latter, $IP_3$ second messenger binds to receptors localized on the cytoplasmic side of the ER which open releasing $Ca^{2+}$ from the ER in an autocatalytic fashion [10]. Due to the nonlinear properties of such receptor/channels, CICR is essentially oscillatory [12]. The pattern of $Ca^{2+}$ oscillations depends on the intracellular $IP_3$ concentration, hence one can think of the $Ca^{2+}$ signal as being an encoding of information on this latter [76]. Notably, this information encoding could use amplitude modulations (AM), frequency modulations (FM) or both modulations (AFM) of $Ca^{2+}$ oscillations [76-17]. Accordingly, we consider stereotypical functions that reproduce all these possible encoding modes (Figure S4). Let $C(t)$ denote the $Ca^{2+}$ signal and $m_i(t)$ (with $i =$ AM, FM) the modulating signal related to the $IP_3$ concentration. Then, AM-encoding $Ca^{2+}$ dynamics can be modeled by:

$$C(t) = C_0 + m_{AM}(t)\sin^w\!\left(2\pi f_C t + \varphi_C\right) \qquad (4)$$

FM-encoding $Ca^{2+}$ dynamics instead can be mimicked by the equation:

$$C(t) = C_0 + \sin^w\!\left(2\pi \cdot m_{FM}(t) \cdot f_C t + \varphi_C\right) \qquad (5)$$

Eventually, AFM-encoding comprises both the above in the single generic equation:

$$C(t) = C_0 + m_{AM}(t)\sin^w\!\left(2\pi \cdot m_{FM}(t) \cdot f_C t + \varphi_C\right) \qquad (6)$$

In the above equations, $f_C$ and $\varphi_C$ denote the frequency and the phase of $Ca^{2+}$ oscillations respectively. Moreover, the exponent $w$ is taken as positive even integer to adjust the shape of $Ca^{2+}$ oscillations, i.e. from sinusoidal to more pulse-like oscillations (namely pulses of width much smaller than their wavelength) [74, 18].

The exact functional form of $m_i(t)$ depends on inherent cellular properties of the astrocyte [76, 75, 19]. Notwithstanding, several theoretical studies showed that for increasing $IP_3$ concentrations, $Ca^{2+}$ oscillations are born via some characteristic bifurcation pathways [20]. In particular, while AM $Ca^{2+}$ dynamics could be explained by a supercritical Hopf bifurcation, FM features are born via saddle-node on homoclinic bifurcation [76-75]. Notably, both these bifurcations are characterized by similar functional dependence on the bifurcation parameter, i.e. the $IP_3$ concentration in our case, respectively for amplitude and period of oscillations at their onset [21, 22]. This scenario thus allows considering analogous $m_i(t)$ in equations (4-6) of the form $m_i(t) = k_i \sqrt{(IP_3(t) - I_\mathrm{b})}$, where $I_\mathrm{b}$ is the threshold $IP_3$ concentration that triggers CICR and $k_i$ is a scaling factor [21]. We assume that $IP_3(t)$ is externally driven either by gap junction-mediated intercellular diffusion from neighboring astrocytes [23-25], or by external stimulation of the cell [18, 32-59]. In this fashion we can control the pattern of $Ca^{2+}$ oscillations yet preserving the essence of the complex network of chemical reactions underlying $IP_3/Ca^{2+}$-coupled signals [74, 73].

## 3.     Glutamate exocytosis from the astrocyte

Calcium induced $Ca^{2+}$ release triggered by $IP_3$ is observed to induce glutamate exocytosis from astrocytes [22, 30]. Additional data also suggest an involvement of ryanodine/caffeine-sensitive internal $Ca^{2+}$ stores [31], notwithstanding evidence for the existence of RyR-mediated $Ca^{2+}$ signaling in astrocytes are contradictory [32] and this possibility is not considered in this study.

A large amount of evidence suggests that glutamate exocytosis from astrocytes resembles its synaptic homologous [32, 27]. Astrocytes indeed possess a vesicular compartment that is competent for regulated exocytosis [60]. Glutamate-filled vesicles in astrocytic processes in rodents' dentate gyrus closely resemble synaptic vesicles in excitatory nerve terminals [30, 31]. Similarly to synapses, astrocytes also express SNARE proteins necessary for exocytosis [36] as well as the proteins responsible for sequestering glutamate into vesicles [44]. Indeed synaptic-like plasma-membrane fusion, trafficking and recycling of astrocytic glutamate vesicles were observed [38-40] and quantal glutamate release hallmarking vesicle exocytosis [53] was measured accordingly [59, 40, 66], [62]]. Moreover, experiments suggest that release of glutamate is likely much faster than its reintegration [38, 40] in a fashion akin to that of synaptic exocytosis [8].

Based on these arguments, we model astrocytic glutamate exocytosis similarly to synaptic release. Thus, we postulate the existence of an astrocytic pool $x_A$ of releasable glutamate resources that is limited and constant in size. Then, upon any "proper" intracellular $Ca^{2+}$ increase at time $\tau_i$, an amount $U_A x_A$ of such resources is released into the extrasynaptic space and is later reintegrated into the pool at rate $\Omega_A$. Hence, the equation for $x_A$ reads

$$\dot{x}_A = \Omega_A \left(1 - x_A\right) - U_A \sum_i x_A \, \delta(t - \tau_i) \tag{7}$$

where the parameter $U_A$ is the astrocytic analogous of synaptic basal release probability $U_0$ (equation 1). The instants $\tau_i$ at which glutamate is released from the astrocyte are dictated by the $Ca^{2+}$ dynamics therein. While $Ca^{2+}$ oscillations trigger synchronous release, sustained nonoscillatory $Ca^{2+}$ increases were observed to induce glutamate exocytosis only during their initial rising phase [59, 44]. Furthermore, glutamate release occurs only if intracellular $Ca^{2+}$ concentration exceeds a threshold value [63, 59]. Therefore, in agreement with these experimental observations, we assume that astrocytic glutamate release occurs at any time $t = \tau_i$ such that $C(\tau_i) = C_{thr}$ and $dC/dt\big|_{\tau_i} > 0$, where $C(t)$ is described by equations (4-6) and $C_{thr}$ stands for the $Ca^{2+}$ threshold of glutamate exocytosis.

This description lumps into a single release event the *overall* amount of glutamate released by a $Ca^{2+}$ increases beyond the threshold, independently of the underlying mechanism of exocytosis, which could involve either a single or multiple vesicles [30, 39, 45]. In this latter scenario, the error introduced by our description might be conspicuous if asynchronous release occurs in presence of fast clearance of glutamate in the extrasynaptic space. Nonetheless, further experiments are needed to support this possibility [22].

## 4.  Glutamate time course in the extrasynaptic space

A detailed modeling of glutamate time course in the extrasynaptic space (ESS) is beyond the scope of this study. Accordingly, we simply assume that the ESS concentration of glutamate $G_A$ is mainly dictated by: (1) the frequency and the amount of its release from the astrocyte; and (2) its clearance rate due to astrocytic glutamate transporters along with diffusion away from the release site [46-48]. In other words, $G_A$ evolves according to the generic equation

$$\dot{G}_A = v_{release} - v_{diffusion} - v_{uptake} \tag{8}$$

where $v_{release}$, $v_{diffusion}$ and $v_{uptake}$ respectively denote the rates of glutamate release, diffusion and uptake and are following detailed. Binding of astrocyte-released glutamate by astrocytic glutamate receptors [24, 30] is also not included in equation (8) for simplicity. Activation of such receptors by excess glutamate in the ESS is in fact strongly limited by fast glutamate buffering by astrocytic transporters [49] and is further supported by the experimental evidence that autocrine receptor activation does not essentially affect intracellular $Ca^{2+}$ dynamics in the astrocyte [85, 51].

Glutamate release in our description occurs instantaneously by exocytosis from the astrocyte. In particular, the fraction $RR_A$ of astrocytic glutamate resources released at $t = \tau_i$ by the *i*-th $Ca^{2+}$ increase beyond the threshold $C_{thr}$ (Section I.3), is given by (equation 7):

$$RR_A\left(\tau_i\right) = U_A x_A\left(\tau_i^-\right) \tag{9}$$

If we suppose that the total number of releasable glutamate molecules in the astrocyte is $M$ then, the maximal contribution to glutamate concentration in the ESS, that is for $RR_A\left(\tau_i\right) = 1$, equals to

$$G_{max} = M / N_A V_e \tag{10}$$

where $N_A$ is the Avogadro constant and $V_e$ is the volume of the ESS "of interest", namely the ESS comprised between the astrocytic process where release occurs, and the presynaptic receptors targeted by astrocytic glutamate. In general though, the contribution $G_{rel}$ to glutamate in the ESS is only a fraction $RR_A(\tau_i)$ of the maximal contribution, that is

$$G_{rel}\left(\tau_i\right) = G_{max} \cdot RR_A\left(\tau_i\right) \tag{11}$$

We can further express then the number $M$ of glutamate molecules in the astrocyte in terms of parameters that can be experimentally estimated noting that, $M$ equals the number of molecules per vesicle $M_v$, times the number of vesicles available for release $n_v$. In turn, the number of molecules per synaptic vesicle can be estimated to be proportional to the product of the vesicular glutamate concentration $G_v$, times the vesicular volume $V_v$, being [52]

$$M_v = N_A G_v V_v \tag{12}$$

Under the hypothesis that all glutamate-containing vesicles in the astrocyte are identical both in size and content, the overall number of glutamate molecules in the astrocyte can be estimated as

$$M = n_v M_v = n_v N_A G_v V_v \tag{13}$$

Accordingly, replacing equations (10) and (13) in (11) provides the generic exocytosis contribution to glutamate concentration in the ESS, which can be written as

$$G_{rel}(\tau_i) = \rho_A n_v G_v \cdot RR_A(\tau_i) \tag{14}$$

where $\rho_A = V_v / V_e$ is the ratio between vesicular volume and the volume of the ESS of interest.

The total contribution of astrocytic glutamate exocytosis to the time course of glutamate in the ESS, is therefore the sum of all contributions by each single release event (given by equation 14). Hence, the rate of glutamate release in equation (8) can be written as

$$v_{release} = \sum_i G_{rel}(\tau_i) = \rho_A n_v G_v \sum_i RR_A(\tau_i) \tag{15}$$

Let us consider now the glutamate degradation terms in equation (8). Glutamate clearance due to lateral diffusion out of the ESS volume of interest follows Fick's first law of diffusion [53]. Then, assuming ESS isotropy, the rate of decrease of ESS glutamate concentration can be taken proportional to the concentration of astrocyte-released glutamate ($G_A$) by a factor $r_d$ which stands for the total rate of glutamate diffusion, that is

$$v_{diffusion} = r_d G_A \tag{16}$$

Glutamate uptake by astrocytic transporters can instead be approximated by Michaelis-Menten kinetics [78]. Accordingly, the uptake rate reads

$$v_{uptake} = v_u \frac{G_A}{G_A + K_u} \tag{17}$$

where $v_u$ is the maximal uptake rate and $K_u$ is the transporters' affinity for glutamate.

Equations (15), (16) and (17) replaced in equation (8) provide a generic concise description of glutamate time course in the ESS. A further simplification though can be made based on the experimental evidence that astrocytic glutamate transporters are not saturated in physiological conditions [92]. This scenario in fact is consistent with an ESS glutamate concentration $G_A$ such that $G_A \leq K_u$. Accordingly, the uptake rate can be taken roughly linear in $G_A$, that is $v_{uptake} = (v_u / K_u) G_A$ [46], and equation (8) reduces to

$$\dot{G}_A \approx \sum_i G_{rel}(\tau_i) - r_d G_A - \frac{v_u}{K_u} G_A = \sum_i G_{rel}(\tau_i) - \Omega_c G_A \tag{18}$$

where $\Omega_c = r_d + v_u / K_u$ denotes the overall rate of glutamate clearance (Figure S5). That is, under the hypothesis that astrocytic glutamate transporters are not saturated, the time course of ESS glutamate is monoexponentially decaying roughly in agreement with experimental observations [46, 56, 57].

Equations (7) and (18) provide a description of astrocyte glutamate exocytosis and control by the astrocyte of glutamate concentration in the ESS. A key assumption in their derivation is that despite part of the released glutamate re-enters the astrocyte by uptake, the contribution of this latter to the reintegration of releasable glutamate resources therein can be neglected at first instance [58]. In other words transporters merely function as glutamate "sinks", so that the supply of new glutamate needed to reintegrate releasable astrocytic resources must occur through a different route, independently of extracellular glutamate [22]. While this could constitute a drastic simplification, consistency of such assumption with experimental evidence can be based on the following arguments.

Differently from nerve terminals where its reuse as transmitter is straight forward [59], uptaken glutamate in astrocytes seems to be mostly involved in the metabolic coupling with neurons [60, 61], so that glutamate supply for astrocytic exocytosis is mainly provided by other routes [58]. Glutamate sequestered by astrocytic transporters is in fact metabolized either into glutamine or α-ketoglutarate, and this latter further into lactate [58, 61]. Both glutamine and lactate are eventually exported from astrocytes to the ESS from which they may enter neurons and be reused therein as precursors for synaptic transmitter glutamate [62].

Vesicular glutamate required for astrocytic glutamate exocytosis, can be synthetized *ex novo* instead mainly from glucose imported either from intracellular glycogen, circulation [63] or from neighboring astrocytes [64] by tricarboxylic acid cycle [22]. Alternatively it can also be obtained by transamination of amino acids such as alanine, leucine or isoleucine that could be made available intracellularly [58, 65]. Although α-ketoglutarate originated from glutamate uptake could also enter the tricarboxylic acid cycle, its role in astrocyte glutamate exocytosis however remains to be elucidated [22, 49, 67].

## 5.     Astrocyte modulation of synaptic release

Glutamate released from astrocytes can modulate synaptic transmission at nearby synapses. In the hippocampus in particular, several studies have shown that astrocyte-released glutamate modulates neurotransmitter release at excitatory synapses either towards a decrease [28-33] or an increase of it [32, 31, 30, 34]. This is likely achieved by specific activation of pre-terminal receptors, namely presynaptic glutamate receptors located far from the active zone [26]. Ultrastructural evidence indeed hints that glutamate-containing vesicles could colocalize with these receptors suggesting a focal action of astrocytic glutamate on pre-terminal receptors [31]. Such action likely occurs with a spatial precision similar to that observed at neuronal synapses [27] and is not affected by synaptic glutamate [32].

While astrocyte-induced presynaptic depression links to activation of metabotropic glutamate receptors (mGluRs) [29, 33, 74], in the case of astrocyte-induced presynaptic facilitation, ionotropic NR2B-containing NMDA receptors could also play a role [31, 34]. The precise mechanism of action in each case remains yet unknown. The inhibitory action of presynaptic mGluRs (group II and group III) might involve a direct regulation of the synaptic release/exocytosis machinery reducing $Ca^{2+}$ influx by inhibition of P/Q-type $Ca^{2+}$ channels [26, 75]. Conversely, the high $Ca^{2+}$ permeability of NR2B-containing NMDAR channels could be consistent with an increase of $Ca^{2+}$ influx that in turn would justify facilitation of glutamate release [76, 77]. Facilitation by group I mGluRs instead [32, 30] could be triggered by ryanodine-sensitive $Ca^{2+}$-induced $Ca^{2+}$ release from intracellular stores which eventually modulates presynaptic residual $Ca^{2+}$ levels [93].

Despite the large variety of possible cellular and molecular elements involved by different receptor types, all receptors ultimately modulate the likelihood of release of synaptic vesicles [26]. From a modeling perspective equations (1-2) can be modified to include such modulation in the fraction of released glutamate (equation 3). Accordingly, three scenarios can be drawn *a priori*. (1) Activation of presynaptic glutamate receptors could modulate one or both synaptic rates $\Omega_d$ and $\Omega_f$. (2) Modulations of release probability could be consistent instead with modulation of synaptic basal release probability $U_0$. (3) Alternatively release probability could be modulated making $x$ and/or $u$ in equations (1-2) − respectively the release probabilities of available-for-release and docked vesicles − explicitly depend on astrocyte-released glutamate (equation 18). This could be implemented for example including additional terms $\beta(G_A)$, $\gamma(G_A)$ in equations (1-2) respectively, that could mimic experimental observations.

Synaptic recovery rates $\Omega_d$ and $\Omega_f$ could indeed be modulated by presynaptic $Ca^{2+}$ [3] and thus by modulations of $Ca^{2+}$ concentration at the release site mediated by presynaptic glutamate receptors. Incoming action potentials though transiently affect presynaptic $Ca^{2+}$ levels, so that this process would depend on synaptic activity. On the contrary astrocyte modulation of synaptic release is activity independent [31, 49], and this first scenario does not seem to be realistic.

Modulation of release probability of available-for-release vesicles in the third scenario would occur for example, if the pool size of readily releasable vesicles changes [79]. Although this possibility cannot be ruled out, there is no evidence that such mechanism could be mediated by presynaptic glutamate receptors [26]. Furthermore, recordings of stratum radiatum CA1 synaptic responses to Schaffer collaterals paired-pulse stimulation showed paired-pulse ratios highly stable in time during astrocyte modulation [29]. Accordingly, it could be speculated that if the interpulse interval of delivered pulse pairs in such experiments was long enough to allow replenishment of the readily-releasable pool at those synapses, then the constancy of paired-pulse ratio would also require that the size of the readily-releasable pool is preserved. Therefore direct modulation of $x$ is unlikely.

We thus assume that astrocytic regulation of synaptic release could be brought forth by modulations of $u$, the release probability of docked vesicles. In this case then, either the modulation of $U_0$, i.e. the synaptic basal release probability, or the addition of a supplementary term to the right hand side of equation (1) could be implemented with likely similar effects. The former scenario though seems more plausible based on the following experimental facts. First, presynaptic receptors can modulate presynaptic residual $Ca^{2+}$ concentration by modulations of $Ca^{2+}$ influx thereinto [26]. Second, basal residual $Ca^{2+}$ could sensibly affect evoked synaptic release of neurotransmitter [64, 81]. Finally, third, astrocyte-modulation of synaptic release is independent of synaptic activity [31, 49, 30, 74], and so is the modulation of $U_0$ rather than the addition of a supplementary term $\gamma(G_A)$. Accordingly, we lump the effect of astrocytic glutamate $G_A$ into a functional dependence of $U_0$ such that $U_0 = U_0(G_A)$.

The time course of extracellular glutamate is estimated in the range of few seconds [56], notwithstanding the effect of astrocytic glutamate on synaptic release could last much longer, from tens of seconds [31, 30, 34] up to minutes [33]. This hints that the functional dependence of $U_0$ on $G_A$ mediated by presynaptic receptors is nontrivial. However, rather than attempting a detailed biophysical description of the complex chain of events leading from astrocytic glutamate binding of presynaptic receptors to modulation of resting presynaptic $Ca^{2+}$ levels, we proceed in a phenomenological fashion [71]. Accordingly, we define a dynamical variable $\Gamma$ that phenomenologically captures this interaction so that $U_0(G_A) = U_0(\Gamma(G_A))$. The variable $\Gamma$ exponentially decays at rate $\Omega_G$ which must be small in order to mimic the long-lasting effect of astrocytic glutamate on synaptic release. On the other hand, in presence of extracellular glutamate following astrocyte exocytosis, $\Gamma$ increases by $O_G G_A (1 - \Gamma)$ which includes the possible saturation of presynaptic receptors by astrocytic glutamate. Accordingly, the equation for $\Gamma$ reads

$$\dot{\Gamma} = O_G G_A (1 - \Gamma) - \Omega_G \Gamma \qquad (19)$$

Under the hypothesis that presynaptic $Ca^{2+}$ levels are proportional to presynaptic receptor occupancy [26], then $\Gamma$ biophysically correlates with the fraction of receptor bound to astrocyte-released glutamate. The total amount of receptors that could be potentially targeted by astrocytic glutamate is assumed to be preserved in time and so are the two rate constants in equation (19). Experimental evidence however hints a more complex reality. The coverage of synapses by astrocytic processes in fact could be highly dynamic [77] and trigger repositioning of the astrocytic sites for gliotransmission release [84]. It has been argued that this mechanism could further regulate the onset and duration of astrocyte modulation of synaptic release [27]. Nevertheless the lack of evidence in this direction allows the approximation of our description in equation (19).

The exact functional form of $U_0(\Gamma)$ depends on the nature of presynaptic receptors targeted by astrocytic glutamate. In the absence of more detailed data, we just assume that the related function $U_0(\Gamma)$ is analytic around zero and consider its first-order Taylor expansion

$$U_0(\Gamma) \cong U_0(0) + U_0^{'}(0)\Gamma \qquad (20)$$

In this framework, the expansion of order zero coincides with the synaptic basal release probability in the genuine TM model (equation 1), that is $U_0(0) = \text{const} = U_0^*$. On the contrary, the first-order term must be such that: (1) $U_0(\Gamma) \cong U_0^* + U_0^{'}(0)\Gamma$ is comprised between $[0,1]$ (being $U_0$ a probability); (2) for astrocyte-induced presynaptic depression, $U_0(\Gamma)$ must *decrease* with increasing $\Gamma$ in agreement with the experimental observation that the more the bound receptors the stronger the inhibition of synaptic release [33]; (3) for receptor-mediated facilitation of synaptic release instead, $U_0(\Gamma)$ must *increase* with the fraction of bound receptors. Accordingly, we consider the generic expression $U_0^{'}(0) = -U_0^* + \alpha$ with $\alpha \in [0,1]$ being a parameter that lumps information on the nature of presynaptic receptors targeted by astrocytic glutamate. The resulting expression for $U_0(\Gamma)$ is then

$$U_0(\Gamma) = (1 - \Gamma)U_0^* + \alpha\,\Gamma \qquad (21)$$

It is easy to show that all the above constraints are satisfied. The fact that $\Gamma \in [0,1]$ and $\alpha \in [0,1]$ also assures that $0 \leq U_0(\Gamma) \leq 1$, in agreement with the first condition. For α = 0, it is $U_0(\Gamma) = U_0^* - \Gamma U_0^*$, so that while in absence of the astrocyte (i.e. Γ = 0), $U_0(\Gamma)$ is maximal and equals $U_0$*, in presence of astrocytic glutamate (i.e. Γ > 0), $U_0(\Gamma)$ decreases by a factor $\Gamma U_0^*$ consistently with a *release-decreasing* action of the astrocyte on synaptic release (second condition) (Figure S6). Conversely, if α = 1, it is $U_0(\Gamma) = U_0^* + \Gamma(1 - U_0^*)$, and in absence of the astrocyte $U_0(\Gamma = 0)$ coincides with $U_0$* and is minimal while in presence of the astrocyte, $U_0$ increases by a factor $\Gamma(1 - U_0^*)$, as expected by a *release-increasing* action of astrocytic glutamate on synaptic release (third condition) (Figure S6).

In general, for intermediate values of $0 < \alpha < 1$ astrocyte-induced decrease and increase of synaptic release coexist, mirroring the activation of different receptor types at the same synaptic terminal by astrocytic glutamate [26]. However the distinction between release-decreasing vs. release-increasing action of the astrocyte is still possible. For 0 < α < $U_0$*, it is in fact $U_0(\Gamma) \leq U_0$*, so that a decrease of synaptic release prevails on an increase of this latter. On the contrary, when $U_0$* < α < 1, it is $U_0(\Gamma) \geq U_0$* and increase is predominant over decrease.

Our description can be adopted in principle to study modulation of synaptic release by astrocyte-released glutamate, independently of the specific type of presynaptic receptor that is involved. This especially holds true for the case of NMDAR-mediated astrocyte-induced increase of synaptic release. While activation of such receptors at postsynaptic terminals depends on the membrane voltage for the existence of a voltage-dependent $Mg^{2+}$ block [85, 86], this is apparently not the case for NR2B-containing presynaptic NMDA receptors targeted by the astrocyte [27]. Although the mechanism remains unknown [26], this scenario allows to use equations (19) and (21) for different receptor types which are then characterized on the mere basis of their different rates $O_G$ and $\Omega_G$ [87].

## II.      Model analysis

## 1.      Mechanisms of short-term depression and facilitation in the TM model of synaptic release

Depending on the frequency $f_{in}$ of presynaptic spikes and the choice of values of the three synaptic parameters $\Omega_d$, $\Omega_f$ and $U_0$, the TM model can mimic dynamics of both depressing and facilitating synapses [68, 51].

Presynaptic depression correlates with a decrease of probability of neurotransmitter release. Although this latter could be put forth by multiple mechanisms, the most widespread one appears to be a decrease in the release of neurotransmitter that likely reflects a depletion of the pool of ready-releasable vesicles [3]. In parallel, presynaptic depression could also be observed in concomitance of a reduction of $Ca^{2+}$ influx into the presynaptic terminal. Such reduction is consistent with subnormal residual $Ca^{2+}$ and thus with a reduced probability of release of docked vesicles [9]. On the contrary, synaptic facilitation is consistent with a short-term enhancement of release that correlates with increased residual $Ca^{2+}$ concentration in the presynaptic terminal. Because, increases of residual $Ca^{2+}$ correlate with increases of the released probability of docked vesicles [89], synaptic facilitation is therefore associated to higher release probability [3].

In the TM model, each presynaptic spike decreases the amount of glutamate available for release by *RR* (equation 3). The released glutamate by one spike is subsequently recovered at a rate $\Omega_d$. Yet, if the spike rate $f_{in}$ is larger than the recovery rate, namely $f_{in} \geq \Omega_d$, progressive depletion of the pool of releasable glutamate occurs. Thus each spike will release less glutamate than the preceding one and synaptic release is progressively depressed (Figure S1A). Clearly the onset of depression is more pronounced the larger the basal release probability $U_0$ since, in these conditions, depletion is deeper (Figure S1B).

Immediately after each presynaptic spike, the release probability is augmented by a factor $U_0(1-u)$ and following recovers to its original baseline value $U_0$ at rate $\Omega_f$. If the spike rate is larger than $\Omega_f$ though, i.e. $f_{in} \geq \Omega_f$, the release probability progressively grows with incoming spikes, and facilitation occurs (Figure S1C). One though should keep in mind that facilitating and depressing mechanisms are intricately interconnected and stronger facilitation leads to higher *u* values which in turn leads to stronger depression. Accordingly, facilitating presynaptic spike trains, namely spike trains that are characterized by $f_{in} \geq \Omega_f$, eventually bring forth depression of synaptic release if they last sufficiently long (Figure S1D).

## 2.    Characterization of paired-pulse depression and facilitation

Short-term depression and facilitation can be characterized by paired-pulse stimulation [14]. Consider the pair of neurotransmitter release events, labeled by $RR_1$ and $RR_2$, triggered by a pair of presynaptic spikes. Then, the paired-pulse ratio (PPR) is defined as $\mathrm{PPR} = RR_2/RR_1$ and can be used to discriminate between short-term paired-pulse depression (PPD) and/or facilitation (PPF) displayed by the synapse under consideration [3]. Indeed, if the paired-pulse stimulus is delivered to the synapse at rest, then PPR values larger than 1 imply that the amount of resources released by the second spike in the pair is larger than the one due to the first spike, i.e. $RR_2 > RR_1$, thus synaptic release is facilitated or equivalently, PPF occurs. Conversely, if PPR < 1 then $RR_2 < RR_1$ which marks the occurrence of PPD (Figure S2A).

When trains of presynaptic spikes in any sequence are considered instead, the above scenario is complicated by the fact that for each *i*-th pair of consecutive spikes in the train, the value of $\mathrm{PPR}_i = RR_i/RR_{i-1}$ depends on the past activity of the synapse. In this context in fact the released resources $RR_{i-1}$ at the (*i*-1)-th spike are dictated by the state of the synapse upon arrival of the (*i*-1)-th spike which in turn depends on the previous spikes in the train. Accordingly, values of $\mathrm{PPR}_i > 1$ ($\mathrm{PPR}_i < 1$) are not any longer a sufficient condition to discriminate between PPD and PPF. This concept can be elucidated considering the difference of released resources $\Delta RR_i$ associated with $\mathrm{PPR}_i$, namely:

$$
\begin{aligned}
\Delta RR_i &= RR_i - RR_{i-1} \\
&= u_i x_i - u_{i-1} x_{i-1} \\
&= (u_i - u_{i-1})x_i + u_{i-1}(x_i - x_{i-1}) \\
&= x_i \Delta u_i + u_{i-1} \Delta x_i
\end{aligned}
\tag{22}
$$

where $u_i = u(t_i^+)$ and $x_i = x(t_i^-)$ (equation 3). According to definition of PPF, one would expect to measure $\mathrm{PPR}_i > 1$ and thus $\Delta RR_i > 0$ merely when the probability of release of docked vesicles has increased from one spike to the following one, that is when $\Delta u_i > 0$ [3]. Nonetheless, equation (22) predicts that $\Delta RR_i > 0$ (thus $\mathrm{PPR}_i > 1$) can also be found when $\Delta u_i < 0$, if in between spikes, sufficient synaptic resources are *recovered*, that is if $\Delta x_i > 0$. This situation occurs whenever $\left|\Delta u_i\right|/u_{i-1} < \Delta x_i/x_i$ (equation 22) and corresponds to the mechanism of synaptic plasticity dubbed as "recovery from depletion", a further mechanisms of short-term synaptic plasticity that is different from both PPF and PPD [13] (Figure S2B-C).

For the purpose of our analysis nevertheless, distinction between facilitation (or PPF) and recovery was observed to be redundant in the characterization of astrocyte modulations of paired-pulse plasticity in most of the studied cases (results not shown). Accordingly, PPD and PPF were distinguished on the mere basis of their associated PPR value: PPD when PPR < 1 and PPF when PPR > 1 (Figures 5-7, 9, S10, S11). The only exception to this rationale was found when release-increasing astrocytes regulate neurotransmitter release from depressing synapses. Here, subtler changes of paired-pulse plasticity by astrocytic glutamate required the mechanism of recovery to be taken into account too (Figure S11).

## 3.    Mean-field description of synaptic release

An advantage of the TM model is that it can be used to derive a mean-field description of the average synaptic dynamics in responses to many different inputs sharing the same statistics without having to solve an equally large number of equations [51, 57]. The derivation of such description, originally developed for the mean-field dynamics of large neural populations [51], is following outlined.

The first step in the derivation is to rewrite the equation for $u$ (equation 1) in terms of $u(t_i^+)$, since this latter value, namely the value of $u$ immediately after the arrival of a spike at $t_i$, is the one that appears in equation (2). With this regard, we note that [57]:

$$u\left(t_i^+\right) = u + U_0\left(1-u\right)$$
$$\Rightarrow u = \frac{u\left(t_i^+\right) - U_0}{1 - U_0} \tag{23}$$

Accordingly, substituting this latter into equation (1) and redefining $u$ hereafter as $u \leftarrow u\left(t_i^+\right)$, we obtain:

$$\dot{u} = \Omega_f\left(U_0 - u\right) + U_0 \sum_i \left(1-u\right)\delta(t-t_i) \tag{24}$$

In this fashion, we can update $x$ and $u$ simultaneously at each spike, rather than first compute $u$ and then update $x$ as otherwise required by equations (1-2).

Consider then *N* presynaptic spike trains of same duration delivered to the synapse at identical initial conditions. The trial-averaged synaptic dynamics is described by equations (1-2), in terms of the mean quantities $\bar{u} = 1/N \sum_k^N u_k$ and $\bar{x} = 1/N \sum_k^N x_k$. That is

$$\dot{\bar{x}} = \Omega_d(1 - \bar{x}) - \frac{1}{N}\sum_k^N \sum_i ux\delta(t - t_{ik}) \qquad (25)$$

$$\dot{\bar{u}} = \Omega_f(U_0 - \bar{u}) + \frac{U_0}{N}\sum_k^N \sum_i (1 - u)\delta(t - t_{ik}) \qquad (26)$$

Focusing on a time interval Δ*t*, the above equations can be rewritten by their equivalent difference form

$$\bar{x}(t + \Delta t) - \bar{x}(t) = \Omega_d(1 - \bar{x}(t))\Delta t - \frac{1}{N}\sum_k^N u(t)x(t)\Delta_k(\Delta t) \qquad (27)$$

$$\bar{u}(t + \Delta t) - \bar{u}(t) = \Omega_f(U_0 - \bar{u}(t))\Delta t + \frac{U_0}{N}\sum_k^N (1 - u(t))\Delta_k(\Delta t) \qquad (28)$$

where $\Delta_k(\Delta t)$ is the number of spikes per time interval Δ*t* for the *k*-th trial and is a strongly fluctuating (stochastic) quantity itself.

Analysis of neurophysiological data revealed that individual neurons in vivo fire irregularly at all rates reminiscent of the so-called Poisson process [65]. Mathematically, the Poisson assumption means that at each moment, the probability that a neuron will fire is given by the value of the instantaneous firing rate and is independent of the timing of previous spikes. We thus take the *N* spike trains under consideration to be different realizations of the same Poisson process of average frequency $f_{in}(t)$ and we average in time over Δ*t* equations (27-28) (denoted by "$\langle \ \rangle$").

Thanks to the Poisson hypothesis, the variables $u(t)$, $u(t)x(t)$ and $\Delta_k(\Delta t)$ can be considered independent and thus averaged independently. Therefore, taking Δ*t* of the order of several interspike intervals and shorter than the longest time scale in the system between $1/\Omega_d$ and $1/\Omega_f$ [52], we note that the time average of $\Delta_k(\Delta t)$ can be estimated by $\langle \Delta_k(\Delta t) \rangle = f_{in}\Delta t$. Accordingly,

$$\langle \bar{x}(t + \Delta t) \rangle - \langle \bar{x}(t) \rangle = \Omega_d(1 - \langle \bar{x}(t) \rangle)\Delta t - \frac{1}{N}\sum_k^N \langle u(t)x(t) \rangle f_{in}\Delta t \qquad (29)$$

$$\langle \bar{u}(t + \Delta t) \rangle - \langle \bar{u}(t) \rangle = \Omega_f(U_0 - \bar{u}(t))\Delta t + \frac{U_0}{N}\sum_k^N (1 - \langle u(t) \rangle)f_{in}\Delta t \qquad (30)$$

Finally, dividing by Δ*t* the two equations above yields

$$\langle \dot{\bar{x}} \rangle = \Omega_d \left(1 - \langle \bar{x} \rangle\right) - \langle \bar{u} \rangle \langle \bar{x} \rangle f_{in} \tag{31}$$

$$\langle \dot{\bar{u}} \rangle = \Omega_f \left(U_0 - \langle \bar{u} \rangle\right) + U_0 \left(1 - \langle \bar{u} \rangle\right) f_{in} \tag{32}$$

where we made the approximation that $1/N \sum_{k}^{N} \langle ux \rangle = \langle \bar{u} \rangle \langle \bar{x} \rangle$. This would be possible only if $u(t)$ and $x(t)$ were statistically independent while they are not, as both are functions of the same presynaptic spikes. The relative error of this approximation can be estimated using the Cauchy-Schwarz inequality of the probability theory [51]:

$$\frac{\left| \langle ux \rangle - \langle u \rangle \langle x \rangle \right|}{\langle u \rangle \langle x \rangle} \leq c_u c_x \tag{33}$$

where $c_u$ and $c_x$ stand for the coefficient of variation of the random variables $u$ and $x$ respectively and satisfy the following

$$c_u^2 = \frac{(1 - U_0) f_{in}}{(\Omega_f + f_{in})^2 (2\Omega_f + U_0 (2 - U_0) f_{in})} \tag{34}$$

$$c_x^2 = \frac{(1 + c_u^2) \langle u \rangle f_{in}}{2\Omega_d + \langle u \rangle (1 - \langle u \rangle (1 + c_u^2)) f_{in}} \tag{35}$$

Accordingly, the self-consistency of the mean-field theory can be checked plotting the product $c_u c_x$ as a function of the frequency $f_{in}$ of the presynaptic spikes and the synaptic basal release probability $U_0$ (Figure S7A). For the cases considered in our study, the error does not exceed 10%.

## 4.    Frequency response and limiting frequency of a synapse

For the sake of clarity we will following denote by capital letters the mean quantities, hence $U = \langle \bar{u} \rangle$ and $X = \langle \bar{x} \rangle$. One of the advantages of the mean-field description derived above is the possibility to obtain an analytical expression for the mean amount of released resources as a function of the average input frequency. At steady state in fact, that is for $\dot{U} = \dot{X} = 0$, equations (31-32) can be solved for the equivalent steady state values (denoted by the subscript '$\infty$'):

$$U_\infty = \frac{U_0 (\Omega_f + f_{in})}{(\Omega_f + U_0 f_{in})} \tag{36}$$

$$X_\infty = \frac{\Omega_d}{\Omega_d + U_\infty f_{in}}$$

(37)

Accordingly, the mean steady-state released synaptic resources $RR_\infty$ are given by (equation 3):

$$RR_\infty = U_\infty X_\infty = \frac{U_0 \Omega_d (\Omega_f + f_{in})}{\Omega_d \Omega_f + U_0 (\Omega_d + \Omega_f) f_{in} + U_0 f_{in}^2}$$

(38)

The above equation provides the *frequency response* of the synapse in steady-state conditions.

The slope of the frequency response curve, that is $RR_\infty^{'} = \lim\limits_{\Delta f_{in} \to 0} \Delta RR_\infty / \Delta f_{in}$, can be used to distinguish between facilitating and depressing synapses. Indeed a negative slope implies that $\Delta RR_\infty < 0$ for increasing frequencies, which marks the occurrence of depression. Conversely, a positive slope value is when $\Delta RR_\infty > 0$ for $\Delta f_{in} > 0$, which reflects ongoing facilitation (Section II.2). Notably, for vanishing input frequencies, that is for $f_{in} \to 0$, the slope of the frequency response is $RR_\infty^{'}(f_{in} \to 0) = \left( (1 - U_0) \Omega_d - U_0 \Omega_f \right) / (\Omega_d \Omega_f)^2$, which can be either positive or negative depending on the sign of the numerator. With this regard, a threshold value of $U_0$, i.e. $U_{thr}$, can be defined as

$$U_{thr} = \frac{\Omega_d}{\Omega_d + \Omega_f}$$

(39)

so that if $U_0 < U_{thr}$ a synapse can be facilitating, if not, that is if $U_0 > U_{thr}$, the synapse is depressing (Figure S3A).

The frequency response of a depressing synapse is thus maximal for $f_{in} \to 0$, for which $RR_\infty \cong U_0$, and then monotonously decreases towards zero for increasing frequency. In this case, a cut-off frequency can be defined, beyond which the onset of depression is marked by a strong attenuation with respect to the maximal $RR_\infty$, that is an attenuation larger than 3 dB. For a facilitating synapse instead, the frequency response is nonmonotonic and bell-shaped and a peak frequency $f_{peak}$ can be recognized, at which $RR_\infty$ is maximal. This follows from the coexistence of facilitation and depression by means of which the stronger facilitation, the stronger depression. That is, for $0 < f_{in} < f_{peak}$, the slope of the frequency response is positive and thus facilitation occurs. For increasing frequencies on the other hand, the increase of facilitation is accompanied by growing depression, up to $f_{in} = f_{peak}$ when the two compensate, and afterwards depression takes over facilitation for $f_{in} > f_{peak}$.

Both the cut-off frequency and the peak frequency can be regarded as the *limiting frequency* $f_{\text{lim}}$ of the synapse for the onset of depression. Accordingly, $f_{\text{lim}}$ can be obtained from equation (38) and reads

$$f_{\text{lim}} = \begin{cases} \Omega_f \left( \sqrt{\dfrac{\Omega_d}{\Omega_f}\left(\dfrac{1-U_0}{U_0}\right)} - 1 \right) & \text{if } U_0 < U_{\text{thr}} \\ \dfrac{\Omega_d}{\left(1+\sqrt{2}\right)U_0} & \text{if } U_0 > U_{\text{thr}} \end{cases} \tag{40}$$

The corresponding value $RR_{\text{lim}}$ of steady-state average released synaptic resources is obtained replacing $f_{\text{in}}$ by $f_{\text{lim}}$ in equation (38) (Figure S3B):

$$RR_{\text{lim}} = \begin{cases} \dfrac{\Omega_d U_0}{U_0\left(\Omega_d - \Omega_f\right) + 2U_0\left(1 - U_0\right)\Omega_d\Omega_f} & \text{if } U_0 < U_{\text{thr}} \\ \dfrac{\Omega_d U_0}{\sqrt{2}} & \text{if } U_0 > U_{\text{thr}} \end{cases} \tag{41}$$

From the above analysis therefore, it follows that two are the conditions needed for a synapse to exhibit facilitation. These are: (1) $U_0 < U_{\text{thr}}$, and (2) $f_{\text{in}} < f_{\text{lim}}$. Alternatively, if we are able to estimate the slope of the frequency response curve for a given input frequency, then it is necessary and sufficient for the occurrence of facilitation that $\lim\limits_{\Delta f_{\text{in}} \to 0} \Delta RR_\infty / \Delta f_{\text{in}} > 0$ (Figure S9).

## 5.   Mean-field description of astrocyte-to-synapse interaction

We can extend the mean-field description of synaptic release to include modulation of this latter by astrocytic glutamate. The difference with respect to the mean-field description of synaptic release (equations 31-32) is that, when the astrocyte is taken into account, the synaptic basal release probability $U_0$ changes in time according to equation (21). Notwithstanding, in the case that astrocytic $Ca^{2+}$ dynamics and synaptic glutamate release are statistically independent (see "The road map of astrocyte regulation of presynaptic short-term plasticity" in "Methods" in the main text), equation (32) can be rewritten as following:

$$\left\langle \dot{\bar{u}} \right\rangle = \Omega_f \left( \left\langle \overline{U}_0 \right\rangle - \left\langle \bar{u} \right\rangle \right) + \left\langle \overline{U}_0 \right\rangle \left(1 - \left\langle \bar{u} \right\rangle \right) f_{\text{in}} \tag{42}$$

The mean basal probability of synaptic release ultimately depends on the frequency of glutamate exocytosis from the astrocyte. In order to seek an average description of this latter though, we need that at each moment the probability of glutamate exocytosis is independent of the timing of the previous release event, namely that glutamate exocytosis from the astrocyte is a Poisson process [52]. Indeed recent studies provide support to this scenario, hinting that the period of spontaneous astrocytic $Ca^{2+}$ oscillations could be reminiscent of a Poisson process [95, 96]. Moreover, glutamate exocytosis from astrocytes is reported to occur for $Ca^{2+}$ concentrations as low as 200 nM [63], that is lower than the average reported minimal peak $Ca^{2+}$ concentration of $200 - 250$ nM [59, 47]. This allows to assume that the majority of $Ca^{2+}$ oscillations triggers glutamate exocytosis from the astrocyte. Accordingly, the inter-event intervals between two consecutive glutamate release events, can be also assumed to be Poisson distributed. This allows averaging of equation (7) to obtain

$$\left\langle \dot{\overline{x}}_A \right\rangle = \Omega_A \left(1 - \left\langle \overline{x}_A \right\rangle \right) - U_A \left\langle \overline{x}_A \right\rangle f_C \tag{43}$$

where $f_C$ denotes the frequency of exocytosis-triggering astrocytic $Ca^{2+}$ oscillations. Similarly, we can do averaging of equation (18) to obtain a mean-field description of glutamate concentration in the ESS, that is

$$\left\langle \dot{\overline{G}}_A \right\rangle = -\Omega_c \left\langle \overline{G}_A \right\rangle + \beta U_A \left\langle \overline{x}_A \right\rangle f_C \tag{44}$$

where $\beta = \rho_A n_v G_v$. Since the time course of glutamate in the ESS can be assumed to be much faster than the duration of astrocyte effect on synaptic release [31, 47] we can take into account only those processes of glutamate time course that are slower than the overall clearance rate $\Omega_c$. Equation (44) above, thus simplifies to

$$\left\langle \overline{G}_A \right\rangle = \frac{\beta}{\Omega_c} U_A \left\langle \overline{x}_A \right\rangle f_C \tag{45}$$

Averaging of equation (19) then provides

$$\left\langle \dot{\overline{\Gamma}} \right\rangle = -\Omega_G \left\langle \overline{\Gamma} \right\rangle + O_G \left\langle \overline{G}_A \right\rangle \left(1 - \left\langle \overline{\Gamma} \right\rangle \right) \tag{46}$$

where we have made the approximation that $\left\langle \overline{G_A(1-\Gamma)} \right\rangle = \left\langle \overline{G}_A \right\rangle \left(1 - \left\langle \overline{\Gamma} \right\rangle \right)$. Substituting $\left\langle \overline{G}_A \right\rangle$ in equation (45) into equation (46), allows to express the fraction of bound receptors $\Gamma$ as a function of the astrocyte-released glutamate, which is ultimately dependent on the exocytosis frequency $f_C$ (equation 43). Therefore

$$\left\langle \dot{\overline{\Gamma}} \right\rangle = -\Omega_G \left\langle \overline{\Gamma} \right\rangle + \beta \frac{O_G}{\Omega_c} U_A \left\langle \overline{x}_A \right\rangle \left(1 - \left\langle \overline{\Gamma} \right\rangle \right) f_C \tag{47}$$

Hence, at steady-state

$$X_{A\infty} = \frac{\Omega_A}{\Omega_A + U_A f_C} \tag{48}$$

$$\Gamma_\infty = \frac{\beta O_G U_A X_{A\infty} f_C}{\Omega_c \Omega_G + \beta O_G U_A X_{A\infty} f_C} = \frac{\beta \Omega_A O_G U_A f_C}{\Omega_A \Omega_c \Omega_G + (\Omega_c \Omega_G + \beta \Omega_A O_G) U_A f_C} \tag{49}$$

We can eventually substitute equation (49) in (42) to obtain an analytical expression of synaptic basal release probability as function of the frequency of astrocyte exocytosis. That is

$$U_{0\infty} = \frac{\Omega_A \Omega_c \Omega_G U_0^* + (\Omega_c \Omega_G U_0^* + \alpha \beta \Omega_A O_G) U_A f_C}{\Omega_A \Omega_c \Omega_G + (\Omega_c \Omega_G + \beta \Omega_A O_G) U_A f_C} \tag{50}$$

The relative error of this approximation can be estimated by the Cauchy-Schwarz inequality of the probability theory as pointed out in Section II.3 (equation 33):

$$\frac{\left| \langle x_A \Gamma \rangle - \langle x_A \rangle \langle \Gamma \rangle \right|}{\langle x_A \rangle \langle \Gamma \rangle} \leq c_{x_A} c_\Gamma \tag{51}$$

where $c_{x_A}$ and $c_\Gamma$ stand for the coefficients of variation of the random variables $x_A$ and $\Gamma$ respectively and satisfy the following:

$$c_{x_A}^2 = \frac{U_A^2 f_C}{2(\Omega_A + f_C(1 - U_A)U_A/2)} \tag{52}$$

$$c_\Gamma^2 \approx \frac{(1 - \vartheta)(2\Omega_G + f_C)\left((1 - \vartheta)\Omega_G + f_C(1 + (2\tilde{\Gamma} - 1)\vartheta)\right)}{(\Omega_G + f_C)(2\Omega_G + f_C(1 - \vartheta^2))\tilde{\Gamma}^2} - 1 \tag{53}$$

with $\tilde{\Gamma} = 1 + \vartheta \Omega_G / ((\vartheta - 1)f_C - \Omega_G)$ and $\vartheta = \exp(-O_G \beta U_A^2 / \Omega_c)$ (Appendix B). For the cases considered in our study, the error does not exceed ~7% (Figure S7B).

# Appendix A: Model equations

## 1.  Astrocyte Ca²⁺ dynamics

### 1.1.  Conditioning IP₃ signal

$$m(t) = m_0 + k\sqrt{IP_3(t) - I_b}$$

with:

−  $m_0$, constant component of the IP₃ signal (in normalized units);

−  $k$, scaling factor;

−  $IP_3(t)$, externally driven IP₃ signal (in normalized units);

−  $I_b$, IP₃ threshold for CICR (in normalized units).

### 1.2.  Calcium signal:

$$C(t) = C_0 + \lambda_{AM} m_{AM}(t) \sin^w \left( 2\pi \cdot \lambda_{FM} m_{FM}(t) \cdot f_C t + \varphi_C \right)$$

with:

−  $C_0$, constant component of the Ca²⁺ signal (in normalized units);

−  $\lambda_{AM}, \lambda_{FM}$, binary parameters that equal to 1 if AM and/or FM encoding features are taken into account in the astrocyte Ca²⁺ dynamics;

−  $f_C$, frequency of Ca²⁺ oscillations;

−  $\varphi_C$, phase of Ca²⁺ oscillations;

−  $w$, shape factor (it must be a positive even integer).

## 2.  Astrocytic glutamate release

### 2.1.  Glutamate exocytosis

$$\dot{x}_A = \Omega_A (1 - x_A) - U_A \sum_i x_A \, \delta(t - \tau_i)$$

with:

−  $\tau_i$,  instants  of  glutamate  release  from  the  astrocyte,  such that $\exists \tau_i : C(\tau_i) = C_{thr} \wedge \dot{C}(\tau_i) > 0$;

−  $x_A$, fraction of astrocytic glutamate vesicles available for release;

−  $U_A$, basal release probability of astrocytic glutamate vesicles;

−  $\Omega_A$, recovery rate of released astrocyte glutamate vesicles.

## 2.2.    Glutamate time course in the extracellular space

$$G_{\text{rel}}(\tau_i) = \rho_A n_v G_v \cdot U_A x_A(\tau_i^-)$$

$$\dot{G}_A = \sum_i G_{\text{rel}}(\tau_i) - v_u \frac{G_A}{G_A + K_u} - r_d G_A$$

This latter equation is implemented as

$$\dot{G}_A = \sum_i G_{\text{rel}}(\tau_i) - \Omega_c G_A$$

where $\Omega_c = r_d + v_u/K_u$.

In the above, it is:

– $G_{\text{rel}}$, glutamate released from the astrocyte into the ESS;

– $G_A$, glutamate concentration in the ESS;

– $G_v$, glutamate concentration within astrocytic vesicles;

– $n_v$, number of ready-releasable astrocytic vesicles;

– $\rho_A$, ratio between the average volume of astrocytic vesicles and the ESS volume;

– $r_d$, glutamate clearance rate in the ESS by diffusion;

– $v_u$, maximal glutamate uptake rate by transporters;

– $K_u$, transporters' affinity for glutamate;

– $\Omega_c$, glutamate clearance rate in the ESS.

## 3.    Presynaptic receptors

$$\dot{\Gamma} = O_G G_A (1 - \Gamma) - \Omega_G \Gamma$$

with:

– $\Gamma$, fraction of activated presynaptic receptors;

– $O_G$, onset rate of astrocyte modulation of synaptic release probability;

– $\Omega_G$, recovery rate of astrocyte modulation of synaptic release probability.

## 4.    Synaptic release

$$U_0 = (1 - \Gamma) U_0^* + \alpha \Gamma$$

$$\dot{x} = \Omega_d (1 - x) - \sum_i u x \delta(t - t_i)$$

$$\dot{u} = \Omega_f (U_0 - u) + U_0 \sum_i (1 - u) \delta(t - t_i)$$

with:

- $U_0$, basal synaptic release probability function;

- $x$, fraction of synaptic glutamate vesicles available for release;

- $u$, per-spike usage of available glutamate vesicles;

- $t_i$, instant of synaptic release upon arrival of the $i$-th action potential;

- $U_0^*$, basal synaptic release probability (i.e. without astrocyte);

- $\alpha$, "effect parameter" of astrocyte regulation of synaptic release;

- $\Omega_d$, recovery rate of released synaptic glutamate vesicles;

- $\Omega_f$, rate of synaptic facilitation.

## Appendix B: Estimation of the coefficient of variation of Γ

In order to estimate the coefficient of variation of Γ we need a recursive expression of the peak value $\Gamma_n$ associated to *n*-th glutamate release event from the astrocyte. This can be done by solving equation (19) for Γ(t), nonetheless some approximations can be done in our case to make the solution analytically tractable.

We start from the observation that the frequency of Ca²⁺ oscillations in astrocytes can be assumed to be much lower than the rate of replenishment of astrocytic glutamate resources, i.e. $f_C << \Omega_A$ [44, 66]. As discussed in the section "Persistent Ca²⁺ oscillations in astrocytes can regulate presynaptic short-term plasticity" of "Results", it follows that astrocytic glutamate exocytosis can be described by quantal release events of almost identical magnitude roughly equal to $RR_{An} = U_A x_{An} \approx U_A^2$. Accordingly the time course of glutamate released from the astrocyte by the *n*-th exocytotic event occurring at $t = \tau_n$, is (equation 18):

$$G_A\left(t \geq \tau_n\right) = \beta \cdot RR_{An} \exp\left(-\Omega_c t\right) = \beta U_A^2 \exp\left(-\Omega_c t\right) \tag{A1}$$

Experimental evidences also suggest that the onset of astrocyte effect on synaptic release and the rate of glutamate degradation in the extrasynaptic space are much faster than the recovery rate from astrocyte modulation, i.e. $\Omega_G << \Omega_c, O_G G_A$ (see Appendix C). Thus we can assume that at its onset till it reaches its peak value $\Gamma_n$ at $t = \hat{t}_n$, astrocyte modulation is set mainly by the time course of release glutamate and the binding of this latter to such receptors. Accordingly, for $\tau_n \leq t \leq \hat{t}_n$, equation (19) can be simplified into

$$\dot{\Gamma} \cong O_G G_A\left(1 - \Gamma\right) \tag{A2}$$

where $G_A$ is given by equation (A1). On the other hand, once glutamate in the extrasynaptic space is cleared, astrocyte modulation monoexponentially decays from its peak value at rate $\Omega_G$ till the next glutamate release from the astrocyte (assumed to occur at $t = \tau_{n+1} = \tau_n + \Delta t$). Namely for $\hat{t}_n \leq t \leq \tau_n + \Delta t$ it is

$$\Gamma\left(t\right) = \Gamma_n \exp\left(-\Omega_G t\right) \tag{A3}$$

Since $f_C << \Omega_A << \Omega_c, O_G G_A$ (Appendix C) we can assume that the whole glutamate released by the $n$-th release event at $t = \tau_n$ is cleared before the following exocytotic event. Thus solving equation (A2) with the initial condition $\Gamma(t = \tau_{n+1}) = \Gamma_n \exp(-\Omega_G \Delta t)$, provides an iterative expression for $\Gamma_n$ such as

$$\Gamma_{n+1} = \vartheta \cdot \Gamma_n \exp(-\Omega_G \Delta t) + 1 - \vartheta \qquad (A4)$$

with $\vartheta = \exp(-O_G \beta U_A^2 / \Omega_c)$. Assuming steady-state conditions, i.e. $\Gamma_{n+1} \cong \Gamma_n$, equation (A4) can be used to estimate $\langle \Gamma \rangle$ and $\langle \Gamma^2 \rangle$ that are needed to eventually compute $c_r$, the coefficient of variation of $\Gamma$.

The average value of the exponential decay factor in equation (A4) is the integral over all positive $\Delta t$ values of $\exp(-\Omega_G \Delta t)$ times the probability density for a Poisson train of glutamate exocytotic events occurring at rate $f_C$ and that produce an inter-event interval of duration $\Delta t$ (see "Persistent Ca$^{2+}$ oscillations in astrocytes can regulate presynaptic short-term plasticity" of "Results" in the main text). Recall that inter-event intervals of a Poisson distribution are exponentially distributed so that the probability of occurrence of a inter-event interval of duration $\Delta t$ is $f_C \exp(-f_C \Delta t)$. Thus, the average exponential decrement is

$$\langle \exp(-\Omega_G \Delta t) \rangle = f_C \int_0^\infty \exp(-\Omega_G \Delta t) \exp(-f_C \Delta t) d\Delta t = \frac{f_C}{\Omega_G + f_C} \qquad (A5)$$

In order for $\Gamma$ to return on average to its steady-state value between glutamate release events, we must therefore require that

$$\langle \Gamma \rangle = \tilde{\Gamma} = \frac{(\vartheta - 1)(\Omega_G + f_C)}{f_C(\vartheta - 1) + \Omega_G} \qquad (A6)$$

Averaging over the square of $\Gamma_n$ as given by equation (A4) provides $\langle \Gamma^2 \rangle$ and $c_r$ can be thus computed accordingly. Comparison of equation (A6) with equation (49) (Figure S8) shows that the error introduced by the above rationale in the computation of $c_r$ is roughly up to ~10% within the frequency range, i.e. 0.01–1 Hz, of Ca$^{2+}$ oscillations considered in this study.

## Appendix C: Parameter estimation

**Synaptic parameters**. Single hippocampal boutons normally release at most a single quantum of neurotransmitter [98, 99]. Accordingly, reported release probabilities ($U_0$*) for these synapses are small, generally comprised between ~0.09 [52] and ~0.6 [98] with average values between ~$0.3 - 0.55$ [30]. Notwithstanding there could also be specific synapses that exhibit probabilities ranging <$0.05 - 0.9$ [100]. In general, facilitating hippocampal synapses are found with lower (basal) release probability [100].

Vesicles in the readily releasable pool preferentially undergo rapid endocytosis, typically occurring within $1 - 2$ s (i.e. $\Omega_d = 0.5 - 1$ Hz) [101]. However, vesicle recycling could be as fast as $10 - 20$ ms [98, 102], implying a maximum recovery rate of $\Omega_d = 50 - 100$ Hz. Facilitation rates ($\Omega_f$) can be estimated by the decay time of intracellular $Ca^{2+}$ increases at presynaptic terminals following action potential arrival [103, 104]. Accordingly, typical decay times for $Ca^{2+}$ transients are reported to be <500 ms [103] with an upper bound between $0.65 - 2$ s [104]. Such $Ca^{2+}$ transients though shall be taken as upper limit of $Ca^{2+}$ level decay due to the high affinity of the $Ca^{2+}$ indicator used to image them [105, 106]. Therefore, estimated facilitation rates can be as low as $\Omega_f = 0.5$ Hz and range up to 2 Hz [14] or beyond [104].

**Astrocytic calcium dynamics**. For the purposes of our study both $Ca^{2+}$ and $IP_3$ signals in equations (4-6) can be assumed to be normalized with respect to their maxima. Furthermore, because glutamate exocytosis from astrocytes likely occurs in concomitance only with $Ca^{2+}$ increases above basal $Ca^{2+}$ concentration [32, 63, 80], we can take $C_0$ and $I_b$ equal to 0. In this fashion both $C(t)$ and $IP_3(t)$ in equations (4-6) vary within 0 and 1.

The threshold $Ca^{2+}$ concentration ($C_{thr}$) of glutamate exocytosis in astrocytes is estimated to be between ~125 nM [63] and ~850 nM [59]. Given that in stimulated astrocytes, peak $Ca^{2+}$ concentration could reach 1 μM or beyond [63], these values suggest at most the range of ~$0.13 - 0.8$ for $C_{thr}$ in our model. Finally, reported values for the frequency of evoked $Ca^{2+}$ oscillations ($f_C$) in astrocytes can be as low as ~0.01 Hz [47, 67] and range up to 0.1 Hz [18]. In our description we assume that maximal amplitude or frequency of $Ca^{2+}$ oscillations correspond to maximal stimulus, i.e. $IP_3 = 1$. Accordingly, we take $k = 1$.

**Astrocytic glutamate exocytosis**. Exocytosis of glutamate from astrocytes is seen to occur more readily at processes than at cell bodies [39, 109]. Vesicles observed in astrocytic processes have regular (spherical) shape with typical diameters ($d_v$) between 27.6 ± 12.3 nm [30] and 110 nm [39]. Accordingly, vesicular volume $V_v$ ranges between ~$2 − 700 \cdot 10^{-21}$ dm$^3$. Vesicular glutamate content is approximately the same or at most as low as one third of synaptic vesicles at adjacent nerve terminals [31, 44]. Given that glutamate concentration in synaptic vesicles is estimated between ~$60 − 150$ mM [58], then astrocytic vesicular glutamate ($G_v$) likely is in the range of ~$20 − 150$ mM.

The majority of glutamate vesicles at astrocytic processes clusters in close proximity to the plasma membrane, i.e. <100 nm, but about half of them is found within a distance of $40 −$ 60 nm, suggesting the presence of 'docked' vesicles in the astrocytic process [31]. Borrowing the synaptic rationale that docked vesicles corresponds approximately to readily releasable ones [52], then the average number of glutamate vesicles available for release ($n_v$) could be between ~$1 − 6$ [31]. Furthermore, because release probability is proportional to the number of docked vesicles [52], and such docked vesicles approximately correspond to 13% of vesicles at astrocytic processes [31] we can estimate that (basal) release probability of astrocytic exocytosis $U_A$ is < 0.13. On the other hand, single Ca$^{2+}$-increases can decrease the number of vesicles in the process up to 18 ± 14% its original value [39]. In the approximation of a single exocytotic event, this sets the upper limit of $U_A$ as high as 0.82 ± 0.14. In reality multiple releases from the same process likely occur when Ca$^{2+}$-dependent glutamate exocytosis from astrocytes is observed [30, 45, 110] hinting that $U_A$ could be smaller than this limit.

Rate of vesicle recycling is dictated by the exocytosis mode. Both full-fusion of vesicles and kiss-and-run events have been observed at astrocytic processes [30] with the latter likely to occur more often [30, 40]. The most rapid recycling pathway corresponds to kiss-and-run fusion, where the rate is mainly limited by vesicle fusion with plasma membrane and subsequent pore opening [111]. Indeed reported pore-open times can be as short as 2.0 ± 0.3 ms [40]. This value corresponds to a maximal rate of vesicle recycling $\Omega_A$ of approximately $\Omega_A < 450$ ± 80 Hz. Actual astrocytic vesicle recycling rates could be though much slower than this value if recycling could depend on timing of calcium oscillations [36]. In this latter case, Ca$^{2+}$ oscillations at single astrocytic processes could be as slow as ~0.010 Hz [31]. Notwithstanding, for fast release events confined within 100 nm from the astrocyte plasma membrane, the re-acidification time course of a vesicle, could be as long as ~1.5 s [66] hinting an average recycling rate for astrocyte exocytosis of $\Omega_A \approx 0.6$ s$^{-1}$, yet slower than that measured for hippocampal neurons.

**Glutamate time course**. An astrocytic vesicle of 50 nm diameter (i.e. $V_v \approx 65 \cdot 10^{-21}$ dm$^3$) filled with 50 mM glutamate could release into the ESS up to (equation 12) $M = (50 \cdot 10^{-3}$ M$)(65 \cdot 10^{-21}$ dm$^3)N_A \approx 2000$ molecules, roughly one third of those estimated in synaptic vesicles [44]. The average distance from release site $\ell$ travelled by a glutamate molecule during the release time $t_{rel}$ can be estimated by the Einstein-Smoluchowski relationship [52] as $\ell = \sqrt{2D^* t_{rel}}$ where $D^*$ is the glutamate diffusion coefficient in the ESS (treated as an isotropic porous medium [112]). With $D^* \approx 0.2$ [113] and $t_{rel} \approx 1$ ms [40], it is $\ell \approx 0.63$ μm. If we take as mixing volume for the released vesicle the diffusion volume within $\ell$ distance from the release site, then [53] $V_e = 4\pi\zeta \, \ell^3/3$, where $\zeta$ is the volume fraction [112]. With $\zeta = 0.1$ [78] then it is $V_e \approx 10^{-16}$ dm$^3$ and the corresponding contribution to glutamate concentration in the ESS space given by a released vesicle is (equation 14) $G_A = (2000$ molecules$)/(N_A \cdot 10^{-21}$ dm$^3) \approx 30$ μM, that is $\rho_A = V_v/V_e = (65 \cdot 10^{-21})/10^{-16}$ dm$^3 \approx 65 \cdot 10^{-5}$. Assuming an astrocytic pool of $n_v = 4$ docked vesicles, and an average release probability of $U_A = 0.5$, our estimations suggest a peak glutamate concentration immediately after exocytosis of (equation 14) $G_A = 0.5 \cdot 4 \cdot 30$ μM $= 60$ μM, which is indeed in the experimentally-measured range of $1 - 100$ μM [114].

Glutamate transporters are likely not saturated by astrocytic glutamate [92]. This is indeed the case for our estimations too, given an effective glutamate binding affinity for the transporters between $K_u \approx 100 - 150$ mM [115]. Therefore, we approximate glutamate time course by a monoexponentially decaying term as in equation (18). Imaging of extrasynaptic glutamate dynamics in hippocampal slices hints that the decay is fast, with glutamate clearance that is mainly carried out within ~100 ms from peak concentration [56]. Indeed, under the hypothesis of sole diffusion, we can estimate that the concentration in the mixing volume $V_e$ after $t^* = 50$ ms is [53] $G_A = \left(M/8\zeta \, N_A \left(\pi D^* t^*\right)^{3/2}\right) \exp\left(-\ell^2/4D^* t^*\right) \approx 40$ nM which is indeed close to the suggested extracellular glutamate resting concentration of ~25 nM [116]. Accordingly to equation (18) then: $G_A\left(t = t^*\right) = G_A\left(t = 0\right)\exp\left(-\Omega_c t^*\right)$

$\Rightarrow \Omega_c = 1/t^* \cdot \ln\left(G_A\left(t = 0\right)/G_A\left(t = t^*\right)\right)$          $= \qquad 1/(50 \cdot 10^{-3}$ s$) \cdot$

·ln(60·10$^{-6}$ M/40·10$^{-9}$ M) ≈ 150 s$^{-1}$. Alternatively the fact that extracellular glutamate concentration decays to 25 nM within 100 ms from its peak [56], leads to an estimation of $\Omega_c$ ≈ 80 s$^{-1}$. The effective clearance rate is expected to be larger for the existence of uptake [49], which is not explicitly included in our estimation.

**Astrocyte modulation of synaptic release**. Presynaptic depression observed following activation of presynaptic mGluRs by astrocytic glutamate could lasts from tens of seconds [33] to ~2 − 3 min [74]. Similarly group I mGluR-mediated facilitation by a single Ca$^{2+}$ increase in an astrocytic process, may affect synaptic release at adjacent synapses for as long as ~50 − 60 s [32, 30]. Values within ~1 − 2 min have been also reported in the case of an involvement of NMDA receptors [31, 34]. Accordingly for astrocyte-mediated facilitation we can estimate the rate of recovery from astrocyte modulation $\Omega_G$ to be <0.5 − 1.2 min$^{-1}$.

According to our description (equation 19), the rising time of astrocyte effect depends on a multitude of factors of difficult estimation such as the glutamate time course in proximity of presynaptic receptors, the kinetics of these latter as well as their density and the intracellular mechanism that they trigger. Nonetheless, the astrocyte effect on synaptic release usually reaches its maximum within the first 1 − 5 seconds from the rise of Ca$^{2+}$ in the astrocyte [32, 33, 34]. This observation motivated us to assume that, for the purpose of our analysis, the effect of astrocyte modulation of synaptic release could be negligible during its rise with respect to its decay. Accordingly, we consider heuristic values of $O_G$ that could be consistent with such fast onset. In particular, a single Ca$^{2+}$ increase could lead up to ~150 − 200% increase of release from adjacent synapses [31, 49, 30]. However evidence from early studies in vitro hints that the entity of modulation of synaptic release due to the astrocyte could virtually be any up to ~10 times the original [34].

In our model the possible maximal astrocyte-induced facilitation depends on the resting value $U_0$* as well as on the value of the effect parameter a. Indeed from equation (21) in order to have facilitation (i.e for $U_0^* < \alpha \leq 1$), it must be $(1-\Gamma)U_0^* + \alpha\Gamma \geq \eta U_0^*$ for *η > 1* which requires $(\eta-1)U_0^*/(\alpha - U_0^*) \leq \Gamma \leq 1 \Rightarrow 1 < \eta \leq \alpha/U_0^*$. That is, for α = 1, starting from $U_0$* = 0.5, the astrocyte could at most increase the release probability up to two times its original values, since indeed $U_0$* ≤ (2)(0.5) ≤ 1. In the opposite case of astrocyte-induced presynaptic depression, namely for $0 \leq \alpha < U_0^*$, maximal depression sets instead an upper bound for the allowed value of $\Gamma$. From equation (21) indeed it follows that $\left(1-\Gamma\right)U_0^* + \alpha\Gamma \geq \eta U_0^*$ with $0 < \eta < 1$ if and only if $\Gamma \geq \left(1-\eta\right)U_0^*/\left(U_0^* - \alpha\right)$. If we assume maximal depression and minimal facilitation to be respectively ~20% and 120% the resting $U_0^*$ value, it follows that $0.2 < \Gamma < 0.8$. Given that the peak of astrocyte effect can be estimated as $\Gamma_{peak} = O_G U_A G_A/\Omega_c$ (equations 15, 19, 45), it follows that ~$0.4 < O_G < 2$.

# References

1.      Markram H, Wang Y, Tsodyks M (1998) Differential signaling via the same axon of neocortical pyramidal neurons. Proc Natl Acad Sci USA 95: 5323–5328.

2.      Tsodyks MV, Markram H (1997) The neural code between neocortical pyramidal neurons depends on neurotransmitter release probability. Proc Natl Acad Sci USA 94: 719-723.

3.      Zucker RS, Regehr WG (2002) Short-term synaptic plasticity. Annual Rev Physiol 64: 355-405.

4.      Bollmann JH, Sakmann B, Gerard J, Borst G (2000) Calcium sensitivity of glutamate release in a calyx-type terminal. Science 289: 953-957.

5.      Schneggenburger R, Neher E (2000) Intracellular calcium dependence of transmitter release rates at a fast central synapse. Nature 406: 889-893.

6.      Del Castillo J, Katz B (1954) Quantal components of the end-plate potential. J Physiol 124: 560-573.

7.      Fuhrmann G, Segev I, Markram H, Tsodyks M (2002) Coding of temporal information by activity-dependent synapses. J Neurophysiol 87: 140-148.

8.      Tsodyks M, Pawelzik K, Markram H (1998) Neural networks with dynamic synapses. Neural Computation 10: 821-835.

9.      Agulhon C, Petravicz J, McMullen AB, Sweger EJ, Minton SK, et al. (2008) What is the role of astrocyte calcium in neurophysiology? Neuron 59: 932-946.

10.     Berridge MJ, Bootman MD, Roderick HL (2003) Calcium signalling: dynamics, homeostasis and remodelling. Nature Rev 4: 517-529.

11.     Nimmerjahn A (2009) Astrocytes going live: advances and challenges. J Physiol 587: 1639-1647.

12.     Bezprozvanny I, Watras J, Ehrlich BE (1991) Bell-shaped calcium-response curves of Ins(1,4,5)$P_3$- and calcium-gated channels from endoplasmic reticulum of cerebellum. Nature 351: 751-754.

13.     De Pittà M, Volman V, Levine H, Pioggia G, De Rossi D, et al. (2008) Coexistence of amplitude and frequency modulations in intracellular calcium dynamics. Phys Rev E 77: 030903(R).

14.     De Pittà M, Goldberg M, Volman V, Berry H, Ben-Jacob E (2009) Glutamate-dependent intracellular calcium and $IP_3$ oscillating and pulsating dynamics in astrocytes. J Biol Phys 35: 383-411.

15.     De Pittà M, Volman V, Levine H, Ben-Jacob E (2009) Multimodal encoding in a simplified model of intracellular calcium signaling. Cogn Proc 10: 55-70.

16.     Carmignoto G (2000) Reciprocal communication systems between astrocytes and neurones. Prog Neurobiol 62: 561-581.

17.     Berridge MJ (1997) The AM and FM of calcium signaling. Nature 389: 759-760.

18.     Cornell-Bell AH, Finkbeiner SM, Cooper MS, Smith SJ (1990) Glutamate induces calcium waves in cultured astrocytes: long-range glial signaling. Science 247: 470-473.

19.     Berridge MJ, Lipp P, Bootman MD (2000) The versatility and universality of calcium signalling. Nat Rev Mol Cell Biol 1: 11-21.

20.     Falcke M (2004) Reading the patterns in living cells - the physics of $Ca^{2+}$ signaling. Adv Phys 53: 255-440.

21.     Izhikevich EM (2000) Neural excitability, spiking and bursting. Int J Bif Chaos 10: 1171-1266.

22.     Kuznetsov Y (1998) Elements of Applied Bifurcation Theory. New York, U.S.A.: Springer, 2nd edition.

23.     Kuga N, Sasaki T, Takahara Y, Matsuki N, Ikegaya Y (2011) Large-scale calcium waves traveling through astrocytic networks *in vivo*. J Neurosci 31: 2607-2614.

24.     Goldberg M, De Pittà M, Volman V, Berry H, Ben-Jacob E (2010) Nonlinear gap junctions enable long-distance propagation of pulsating calcium waves in astrocyte networks. PLoS Comput Biol 6: e1000909.

25.     Scemes E, Giaume C (2006) Astrocyte calcium waves: What they are and what they do. Glia 54: 716-725.

26.     Fiacco TA, McCarthy KD (2004) Intracellular astrocyte calcium waves *in situ* increase the frequency of spontaneous AMPA receptor currents in CA1 pyramidal neurons. J Neurosci 24: 722-732.

27.     Parpura V, Haydon PG (2000) Physiological astrocytic calcium levels stimulate glutamate release to modulate adjacent neurons. Proc Natl Acad Sci USA 97: 8629-8634.

28.     Pasti L, Zonta M, Pozzan T, Vicini S, Carmignoto G (2001) Cytosolic calcium oscillations in astrocytes may regulate exocytotic release of glutamate. J Neurosci 21: 477-484.

29.     Parpura V, Zorec R (2010) Gliotransmission: exocytotic release from astrocytes. Brain Res Rev 63: 83-92.

30.    Bezzi P, Gundersen V, Galbete JL, Seifert G, Steinhäuser C, et al. (2004) Astrocytes contain a vesicular compartment that is competent for regulated exocytosis of glutamate. Nat Neurosci 7: 613-620.

31.    Hua X, Malarkey EB, Sunjara V, Rosenwald SE, Li WH, et al. (2004) $Ca^{2+}$-dependent glutamate release involves two classes of endoplasmic reticulum $Ca^{2+}$ stores in astrocytes. J Neurosci Res 76: 86-97.

32.    Verkhratsky A, V P (2010) Recent advances in (patho)physiology of astroglia. Acta Pharmacologica Sinica 31: 1044-1054.

33.    Santello M, Volterra A (2009) Synaptic modulation by astrocytes via $Ca^{2+}$-dependent glutamate release. Neuroscience 158: 253-259.

34.    Bergersen LH, Gundersen V (2009) Morphological evidence for vesicular glutamate release from astrocytes. Neuroscience 158: 260-265.

35.    Jourdain P, Bergersen LH, Bhaukaurally K, Bezzi P, Santello M, Domercq M, et al. (2007) Glutamate exocytosis from astrocytes controls synaptic strength. Nature Neurosci 10: 331-339.

36.    Zhang Q, Pangršic T, Kreft M, Kržan M, Li N, et al. (2004) Fusion-related release of glutamate from astrocytes. J Biol Chem 279: 12724-12733.

37.    Montana V, Malarkey EB, Verderio C, Matteoli M, Parpura V (2006) Vesicular transmitter release from astrocytes. Glia 54: 700-715.

38.    Stenovec M, Kreft S M Grilc, Pokotar M, Kreft ME, Pangršic T, et al. (2007) $Ca^{2+}$-dependent mobility of vesicles capturing anti-VGLUT1 antibodies. Exp Cell Res 313: 3809-3818.

39.    Crippa D, Schenk U, Francolini M, Rosa P, Verderio C, et al. (2006) Synaptobrevin2-expressing vesicles in rat astrocytes: insights into molecular characterization, dynamics and exocytosis. J Physiol 570: 567-582.

40.    Chen X, Wang L, Zhou Y, Zheng LH, Zhou Z (2005) "Kiss-and-run" glutamate secretion in cultured and freshly isolated rat hippocampal astrocytes. J Neurosci 25: 9236-9243.

41.    Bowser DN, Khakh BS (2007) Two forms of single-vesicle astrocyte exocytosis imaged with total internal reflection fluorescence microscopy. Proc Natl Acad Sci USA 104: 4212-4217.

42.    Marchaland J, Calì C, Voglmaier SM, Li H, Regazzi R, et al. (2008) Fast subplasma membrane $Ca^{2+}$ transients control exo-endocytosis of synaptic-like microvesicles in astrocytes. J Neurosci 28: 9122-9132.

43.    Südhof TC (2004) The synaptic vesicle cycle. Annu Rev Neurosci 27: 509-547.

44.    Perea G, Araque A (2005b) Properties of synaptically evoked astrocyte calcium signal reveal synaptic information processing by astrocyte. J Neurosci 25: 2192-2203.

45.    Santello M, Bezzi P, Volterra A (2011) TNFα controls glutamatergic gliotransmission in the hippocampal dentate gyrus. Neuron 69: 988-1001.

46.    Diamond JS (2005) Deriving the glutamate clearance time course from transporter currents in CA1 hippocampal astrocytes: transmitter uptake gets faster during development. J Neurosci 25: 2906 –2916.

47.    Diamond JS, Jahr CE (1997) Transporters buffer synaptically released glutamate on a submillisecond time scale. J Neurosci 17: 4672– 4687.

48.    Clements JD (1996) Transmitter timecourse in the synaptic cleft: its role in central synaptic function. Trends Neurosci 19: 163-171.

49.    Tzingounis AV, Wadiche JI (2007) Glutamate transporters: confining runaway excitation by shaping synaptic transmission. Nature Rev 8: 935-947.

50.    Aguado F, Espinosa-Parrilla JF, Carmona MA, Soriano E (2002) Neuronal activity regulates correlated network properties of spontaneous calcium transients in astrocytes *in situ*. J Neurosci 22: 9430-9444.

51.    Venance L, Stella N, Glowinski J, Giaume C (1997) Mechanism involved in initiation and propagation of receptor-induced intercellular calcium signaling in cultured rat astrocytes. J Neurosci 17: 1981-1992.

52.    Schikorski T, Stevens CF (1997) Quantitative ultrastructural analysis of hippocampal excitatory synapses. J Neurosci 17: 5858-5867.

53.    Barbour B, Häusser M (1997) Intersynaptic diffusion of neurotransmitter. Trends Neurosci 20: 377-384.

54.    Rusakov DA, Kullmann DM (1998) Extrasynaptic glutamate diffusion in the hippocampus: ultrastructural constraints, uptake, and receptor activation. J Neurosci 18: 3158-3170.

55.    Diamond JS, Jahr CE (2000) Synaptically released glutamate does not overwhelm transporters on hippocampal astrocytes during high-frequency stimulation. J Neurophysiol 83: 2835-2843.

56.    Okubo Y, Sekiya H, Namiki S, Sakamoto H, Iinuma S, et al. (2010) Imaging extrasynaptic glutamate dynamics in the brain. Proc Nat Acad Sci USA 107: 6526-6531.

57.     Scimemi A, Tian H, Diamond JS (2009) Neuronal transporters regulate glutamate clearance, NMDA receptor activation, and synaptic plasticity in the hippocampus. J Neurosci 29: 14581-14595.

58.     Danbolt NC (2001) Glutamate uptake. Progress Neurobiol 65: 1-105.

59.     Südhof T (1995) The synaptic vesicle cycle: a cascade of protein-protein interactions. Nature 375: 645-653.

60.     Suzuki A, Stern SA, O B, Huntely G, Wlaker RH, et al. (2011) Astrocyte-neuron lactate transport is required for long-term memory formation. Cell 144: 810-823.

61.     Nedergaard M, Takano T, Hansen A (2002) Beyond the role of glutamate as a neurotransmitter. Nature Reviews Neuroscience 3: 748-755.

62.     Hertz L, Zielke H (2004) Astrocytic control of glutamatergic activity: astrocytes as stars of the show. TRENDS in Neurosciences 27: 735-743.

63.     Tsacopoulos M, Magistretti P (1996) Metabolic coupling between glia and neurons. The Journal of Neuroscience 16: 877-885.

64.     Cruz N, Ball K, Dienel G (2007) Functional imaging of focal brain activation in conscious rats: impact of [$^{14}$C] glucose metabolite spreading and release. Journal of Neuroscience Research 85: 3254-3266.

65.     Daikhin Y, Yudkoff M (2000) Compartmentation of brain glutamate metabolism in neurons and glia. The Journal of Nutrition 130: 1026S-1031S.

66.     Bonansco C, Couve A, Perea G, Ferradas CA, Roncagliolo M, et al. (2011) Glutamate released spontaneously from astrocytes sets the threshold for synaptic plasticity. Eur J Neurosci 33: 1483-1492.

67.     Stenovec M, Kreft M, Grilc S, Pangršic T, Zorec R (2008) EAAT2 density at the astrocyte plasma membrane and $Ca^{2+}$-regulated exocytosis. Mol Membr Biol 25: 203-215.

68.     Andersson M, Hanse E (2010) Astrocytes impose postburst depression of release probability at hippocampal glutamate synapses. J Neurosci 30: 5776-5780.

69.     Andersson M, Blomstrand F, Hanse E (2007) Astrocytes play a critical role in transient heterosynaptic depression in the rat hippocampal CA1 region. J Physiol 585: 843-852.

70.     Araque A, Parpura V, Sanzgiri RP, Haydon PG (1998) Glutamate-dependent astrocyte modulation of synaptic transmission between cultured hippocampal neurons. Eur J Neurosci 10: 2129-2142.

71.    Perea G, Araque A (2007) Astrocytes potentiate transmitter release at single hippocampal synapses. Science 317: 1083-1086.

72.    Araque A, Sanzgiri RP, Parpura V, Haydon PG (1998) Calcium elevation in astrocytes causes an NMDA receptor-dependent increase in the frequency of miniature synaptic currents in cultured hippocampal neurons. J Neurosci 18: 6822-6829.

73.    Pinheiro PS, Mulle C (2008) Presynaptic glutamate receptors: physiological functions and mechanisms of action. Nature Rev 9: 423-436.

74.    Liu QS, Xu Q, Kang J, Nedergaard M (2004) Astrocyte activation of presynaptic metabotropic glutamate receptors modulates hippocampal inhibitory synaptic transmission. Neuron Glia Biol 1: 307-316.

75.    Niswender CM, Conn PJ (2010) Metabotropic glutamate receptors: physiology, pharmacology, and disease. Annu Rev Pharmacol Toxicol 50: 295-322.

76.    Corlew R, Brasier DJ, Feldman DE, Philpot BD (2010) Presynaptic NMDA receptors: newly appreciated roles in cortical synaptic function and plasticity. Neuroscientist 14: 609-625.

77.    Engelman HS, MacDermott AB (2004) Presynaptic ionotropic receptors and control of transmitter release. Nature Rev 5: 135-145.

78.    Cochilla AJ, Alford S (1998) Metabotropic glutamate receptor-mediated control of neurotransmitter release. Neuron 20: 1007-1016.

79.    Worden MK, Bykhovskaia M, Hackett JT (1997) Facilitation at the lobster neuromuscular junction: a stimulus-dependent mobilization model. J Neurophysiol 78: 417-428.

80.    Hori T, Takahashi T (2009) Mechanisms underlying short-term modulation of transmitter release by presynaptic depolarization. J Physiol 587: 2987-3000.

81.    Awatramani GB, Price GD, Trussell LO (2005) Modulation of transmitter release by presynaptic resting potential and background calcium levels. Neuron 48: 109-121.

82.    Volman V, Ben-Jacob E, Levine H (2007) The astrocyte as a gatekeeper of synaptic information transfer. Neur Comput 19: 303-326.

83.    Haber M, Zhou L, Murai KK (2006) Cooperative astrocyte and dendritic spine dynamics at hippocampal excitatory synapses. J Neurosci 26: 8881-8891.

84.    Panatier A, Theodosis DT, Mothet JP, Touquet B, Pollegioni L, et al. (2006) Glia-derived D-serine controls NMDA receptor activity and synaptic memory. Cell 125: 775-784.

85.     Destexhe A, Mainen ZF, Sejnowski TJ (1994) Synthesis of models for excitable membranes, synaptic transmission and neuromodulation using a common kinetic formalism. J Comp Neurosci 1: 195–230.

86.     Jahr CE, Stevens CF (1990) Voltage dependence of NMDA-activated macroscopic conductances predicted by single-channel kinetics. J Neurosci 10: 3178-3182.

87.     Nadkarni S, Jung P (2007) Modeling synaptic transmission of the tripartite synapse. Phys Biol 4: 1-9.

88.     Schneggenburger R, Sakaba T, Neher E (2002) Vesicle pools and short-term synaptic depression: lessons form a large synapse. Trends Neurosci 25: 206-212.

89.     Katz B, Miledi R (1968) The role of calcium in neuromuscular facilitation. J Physiol 195: 481-492.

90.     Debanne D, Guerineau NC, Giihwiler BH, Thompson SM (1996) Paired-pulse facilitation and depression at unitary synapses in rat hippocampus: quantal fluctuation affects subsequent release. J Physiol 491: 163-176.

91.     Dittman JS, Kreitzer AC, Regehr WG (2000) Interplay between facilitation, depression, and residual calcium at three presynaptic terminals. J Neurosci 20: 1374-1385.

92.     Tsodyks M (2005) Activity-dependent transmission in neocortical synapses. In: Chow CC, Gutkin B, D H, C M, Dalibard J, editors, Methods and Models in Neurophysics, Elsevier. pp. 245-266.

93.     Softky W, Koch C (1993) The highly irregular firing pattern of cortical cells is inconsistent with temporal integration of random EPSPs. J Neurosci 13: 334-350.

94.     Amit DJ, Tsodyks MV (1991) Quantitative study of attractor neural network retrieving at low spike rates: I. Substrate-spikes, rates and neuronal gain. Network 2: 259-273.

95.     Skupin A, Falcke M (2010) Statistical analysis of calcium oscillations. Eur Phys J Special Topics 187: 231-240.

96.     Takata N, Hirase H (2008) Cortical layer 1 and 2/3 astrocytes exhibit distinct calcium dynamics. PLoS ONE 3: e2525.

97.     Shigetomi E, Kracun S, Sovfroniew MS, Khakh BS (2010) A genetically targeted optical sensor to monitor calcium signals in astrocyte processes. Nature Neurosci 13: 759-766.

98.     Stevens CF, Wang Y (1995) Facilitation and depression at single central synapses. Neuron 14: 795-802.

99.     Stevens CF, Wang Y (1994) Changes in reliability of synaptic function as a mechanism for plasticity. Nature 371: 704-707.

100.    Murthy VN, Sejnowski TJ, Stevens CF (1997) Heterogeneous release properties of visualized individual hippocampal synapses. Neuron 18: 599-612.

101.    Pyle JL, Kavalali ET, Piedras-Rentería ES, Tsien RW (2000) Rapid reuse of readily releasable pool vesicles at hippocampal synapses. Neuron 28: 221-231.

102.    Brody DL, Yue DT (2000) Release-independent short-term synaptic depression in cultured hippocampal neurons. J Neurosci 20: 2480-2494.

103.    Emptage NJ, Reid CA, Fine A (2001) Calcium stores in hippocampal synaptic boutons mediate short-term plasticity, store-operated $Ca^{2+}$ entry, and spontaneous transmitter release. Neuron 29: 197-208.

104.    Regehr WF, Delaney KR, Tank DW (1994) The role of presynaptic calcium in short-term enhancement at the hippocampal mossy fiber synapse. J Neurosci 14: 523-537.

105.    Helmchen F, Borst JGG, Sakmann B (1997) Calcium-dynamics associated with a single action potential in a CNS presynaptic terminal. Biophys J 72: 1458-1471.

106.    Regehr WG, Atluri PP (1995) Calcium transients in cerebellar granule cell presynaptic terminals. Biophys J 68: 2156-2170.

107.    Fellin T, Pascual O, Gobbo S, Pozzan T, Haydon PG, et al. (2004) Neuronal synchrony mediated by astrocytic glutamate through activation of extrasynaptic NMDA receptors. Neuron 43: 729-743.

108.    Pasti L, Volterra A, Pozzan T, Carmignoto G (1997) Intracellular calcium oscillations in astrocytes: a highly plastic, bidirectional form of communication between neurons and astrocytes *in situ*. J Neurosci 17: 7817-7830.

109.    Anlauf E, Derouiche A (2005) Astrocytic exocytosis vesicles and glutamate: a high-resolution immunofluorescence study. Glia 49: 96-106.

110.    Domercq M, Brambilla L, Pilati E, Marchaland J, Volterra A, et al. (2006) P2Y1 receptor-evoked glutamate exocytosis from astrocytes: control by tumor necrosis factor-α and prostaglandins. J Biol Chem 281: 30684-30696.

111.    Valtorta F, Meldolesi J, Fesce R (2001) Synaptic vesicles: is kissing a matter of competence? Trends Cell Biol 11: 324-328.

112.    Nicholson C, Phillips JM (1981) Ion diffusion modified by tortuosity and volume fraction in the extracellular microenvironment of the rat cerebellum. J Physiol 321: 225-257.

113.     Nielsen TA, DiGregorio DA, Silver RA (2004) Modulation of glutamate mobility reveals the mechanism underlying slow-rising AMPAR EPSCs and the diffusion coefficient in the synaptic cleft. Neuron 42: 757-771.

114.     Innocenti B, Parpura V, Haydon PG (2000) Imaging extracellular waves of glutamate during calcium signaling in cultured astrocytes. J Neurosci 20: 1800-1808.

115.     Bergles DE, Tzingounis AV, Jahr CE (2002) Comparison of coupled and uncoupled currents during glutamate uptake by GLT-1 transporters. J Neurosci 22: 10153-10162.

116.     Herman MA, Jahr CE (2007) Extracellular glutamate concentration in hippocampal slice. J Neurosci 27: 9736-9741.

# Table S1

Model parameters and respective values used in simulations.

| Parameter | Description | Value | Units |
|---|---|---|---|
| $U_0^*$ | Basal probability of synaptic glutamate release | <0.05 − 0.9[a] | − |
| $\Omega_d$ | Rate of recovery of released synaptic vesicles | 0.5 − 100[a] | $s^{-1}$ |
| $\Omega_f$ | Rate of synaptic facilitation | 0.5 − 2[a] | $s^{-1}$ |
| $\alpha$ | Effect parameter of astrocyte regulation of synaptic release | 0 − 1 | − |
| $U_A$ | Basal release probability of astrocytic glutamate vesicles | <0.8 (0.6) | − |
| $\Omega_A$ | Rate of recovery of released astrocytic vesicles | 0.01 − 450 (0.6) | $s^{-1}$ |
| $n_v$ | Number of readily releasable astrocytic vesicles | 1 − 6 (4) | |
| $G_v$ | Glutamate content of astrocytic vesicles | 20 − 150 (50) | mM |
| $V_v$ | Volume of astrocytic vesicles | $2 − 700 \cdot 10^{-21}$ | $dm^3$ |
| $V_e$ | Mixing volume of released astrocytic glutamate | $\sim 10^{-16}$ | $dm^3$ |
| $\rho_A$ | Volume ratio $V_v/V_e$ | $6.5 \cdot 10^{-4}$ | − |
| $\Omega_c$ | Glutamate clearance rate | >50 − 150 (60) | $s^{-1}$ |
| $O_G$ | Onset rate of astrocyte modulation | 0.2 − 2 (1.5) | $\mu M^{-1} s^{-1}$ |
| $\Omega_G$ | Recovery rate of astrocyte modulation | <0.5 − 1.2 (0.5) | $min^{-1}$ |
| $f_C$ | Frequency of $Ca^{2+}$ oscillations in the astrocyte | 0.01 − 1 (0.1) | Hz |
| $C_0$ | Basal $Ca^{2+}$ concentration | 0 | − |
| $I_b$ | $IP_3$ threshold concentration for astrocyte $Ca^{2+}$ dynamics | 0 | − |
| $k$ | Scaling factor for the $IP_3$ signal | 1 | − |
| $w$ | Shape factor | 20[b] | − |
| $C_{thr}$ | $Ca^{2+}$ threshold for astrocyte exocytosis of glutamate | 0.13 − 0.8 (0.4) | − |
| $\varphi_C$ | Phase of $Ca^{2+}$ oscillations | 0 | rad |

[a] In the simulations, a *depressing* synapse was characterized by: $\Omega_d = 2 \ s^{-1}$, $\Omega_f = 3.33 \ s^{-1}$, $U_0^* = 0.5$; whereas a *facilitating* synapse was given by: $\Omega_d = 2 \ s^{-1}$, $\Omega_f = 2 \ s^{-1}$, $U_0^* = 0.15$.

[b] $w$ must be a positive even integer.

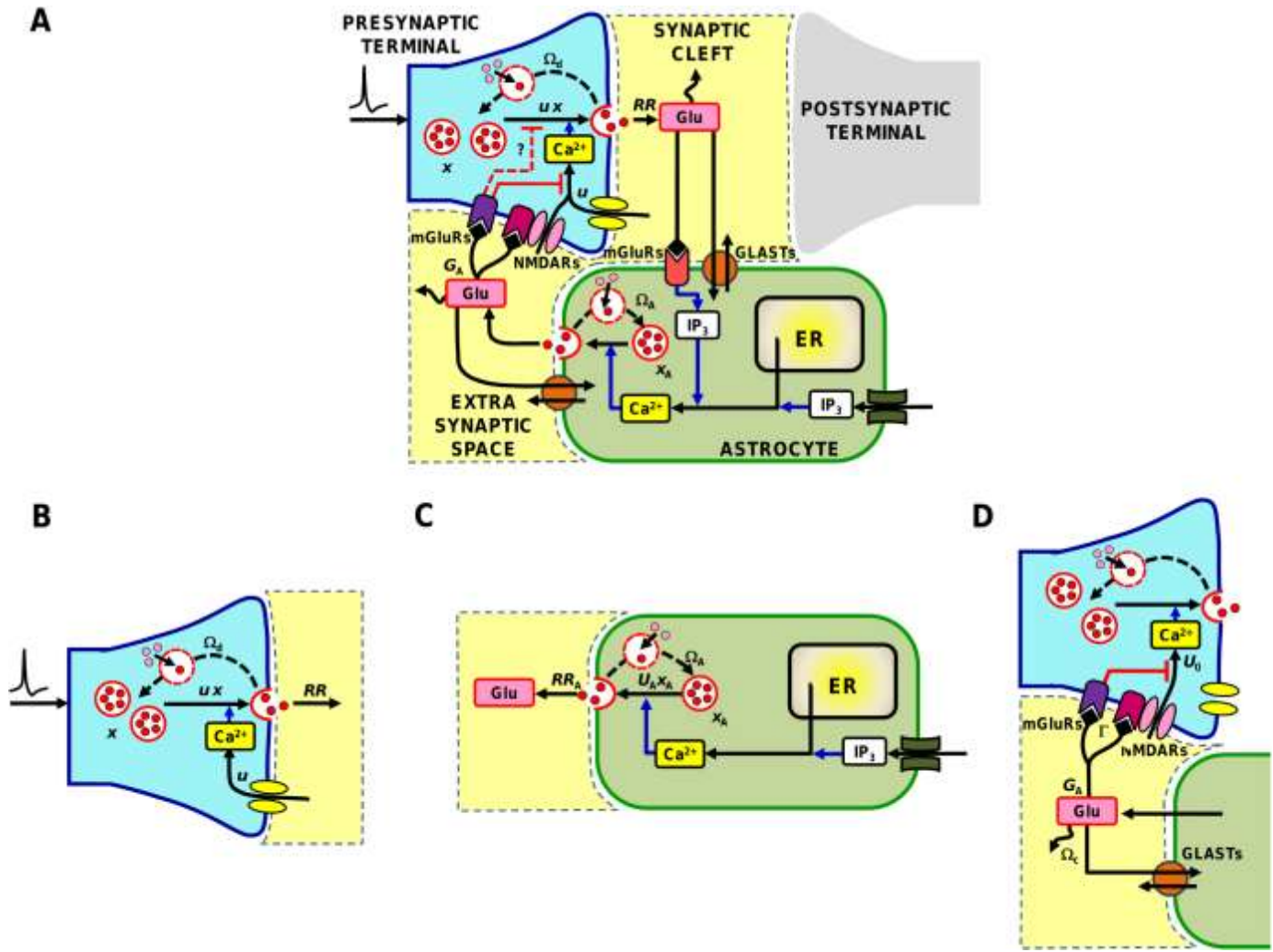

**Figure 1**. Glutamate-mediated astrocyte regulation of synaptic glutamate release in the hippocampus. (**A**) Glutamate exocytosis from synapses is modulated by the amount of available glutamate ($x$) and the fraction ($u$) of resources used by each presynaptic spike, which reflects presynaptic residual Ca$^{2+}$ concentration. Upon an action potential, an amount $RR = ux$ of available glutamate is released to produce a postsynaptic response, and it is later reintegrated into the synapse at rate $\Omega_d$. In the synaptic cleft, released glutamate is cleared by diffusion and uptake by astrocytic glutamate transporters (GLASTs). Part of such glutamate though could also spill out of the cleft and bind to metabotropic glutamate receptors (mGluRs) of neighboring astrocytic processes. The bound receptors then trigger Ca$^{2+}$ release from astrocytic endoplasmic reticulum (ER) stores that is mediated by inositol 1,4,5-trisphospate (IP$_3$). Increasing cytosolic Ca$^{2+}$ levels then triggers glutamate release from the astrocyte by a process similar to synaptic glutamate exocytosis. In turn, released astrocytic glutamate diffuses extrasynaptically and binds to pre-terminal receptors (mGluRs or NMDARs) which can modulate further glutamate release from the synapse by different mechanisms, some of which remain to be elucidated. Calcium dynamics by the astrocyte can also be controlled by other mechanisms, including gap junction-mediated intercellular IP$_3$ diffusion from neighboring astrocytes or external artificial stimulation. The present study takes into account only this scenario. (**B-D**) Building blocks of our model of astrocyte-synapse glutamatergic interactions: (**B**) presynaptic terminal; (**C**) astrocyte; (**D**) glutamate signaling between astrocyte and presynaptic terminal.

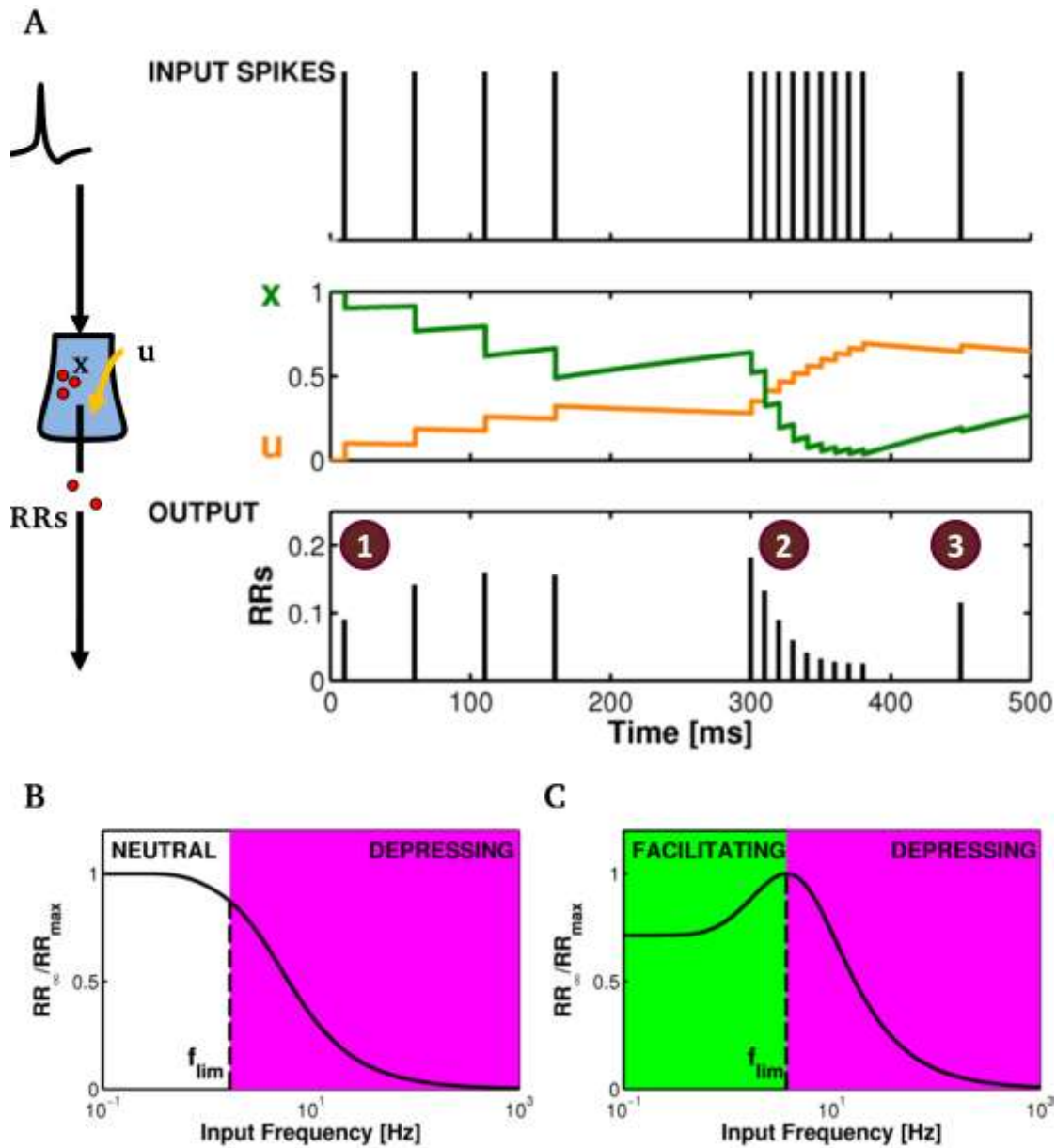

**Figure 2.** Mechanisms of short-term synaptic plasticity in the TM model. (**A**) A train of presynaptic spikes (*top*) can trigger release of synaptic glutamate resources (RRs, *bottom*) in a variegated fashion by different mechanisms of short-term synaptic plasticity. The interplay between dynamics of synaptic variables $u$ and $x$ (*middle*) can bring forth (1) facilitation (STP), (2) short-term depression (STD) or (3) recovery from depression. (**B-C**) Mean-field analysis can be deployed to obtain the (normalized) steady-state frequency response of a synapse (*solid line*). (**B**) The latter monotonically decreases for input frequencies larger than the limiting frequency ($f_{lim}$, *dashed line*) for a depressing synapse (*red shaded area*). (**C**) In the case of facilitating synapses instead, the frequency response is bimodal hinting occurrence of facilitation for input frequencies below $f_{lim}$ (*green shaded area*). Parameters: (**A**) $\Omega_d = 1.67\ s^{-1}$, $\Omega_f = 1.0\ s^{-1}$, $U_0 = 0.5$; (**B**) $\Omega_d = 2\ s^{-1}$, $\Omega_f = 3.3\ s^{-1}$, $U_0 = 0.5$, $RR_{max}\ 0.5$; (**C**) $\Omega_d = 2\ s^{-1}$, $\Omega_f = 2\ s^{-1}$, $U_0 = 0.15$, $RR_{max} = 0.21$.

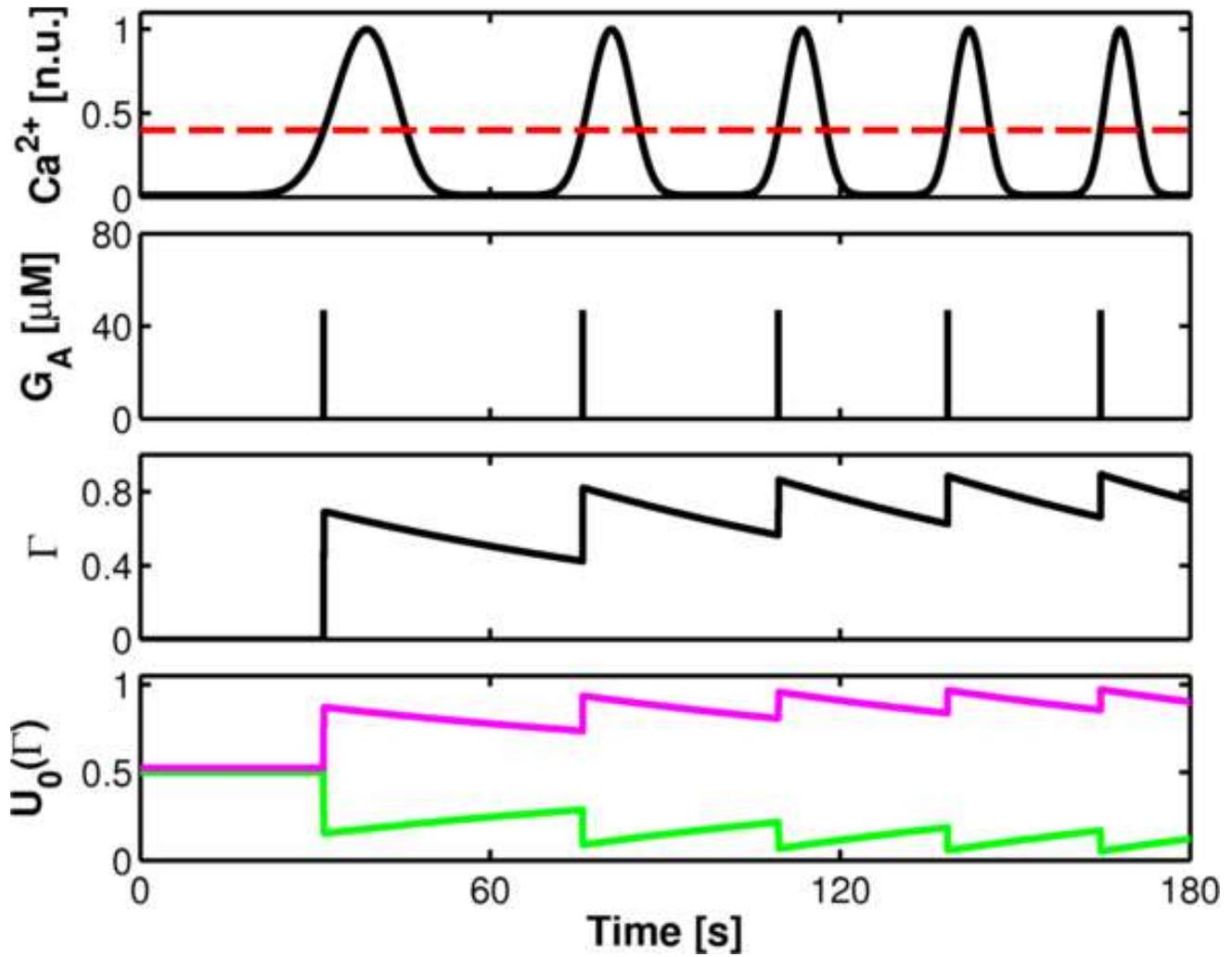

**Figure 3**. A model of astrocyte modulation of synaptic basal release probability. Astrocyte Ca$^{2+}$ oscillations beyond a threshold value ($C_{thr}$, *dashed red line*, *top panel*), trigger transient increases of glutamate ($G_A$) in the extrasynaptic space surrounding presynaptic receptors. The fraction ($\Gamma$) of this latter that bind with astrocytic glutamate modulate synaptic basal release probability ($U_0(\Gamma)$, *bottom panel*). Depending on the nature of presynaptic receptors, lumped in the "effect" parameter $\alpha$, astrocytic glutamate can either decrease (for $0 < \alpha < U_0^*$) or increase for $U_0^* < \alpha < 1$) synaptic release, decreasing or increasing $U_0$ respectively. Here we show the two border cases of $\alpha = 0$ (*green line*) and $\alpha = 1$ (*magenta line*). Parameters: $U_A = 0.3$, $\Omega_G = 1$ min$^{-1}$. Other parameters as in Table S1.

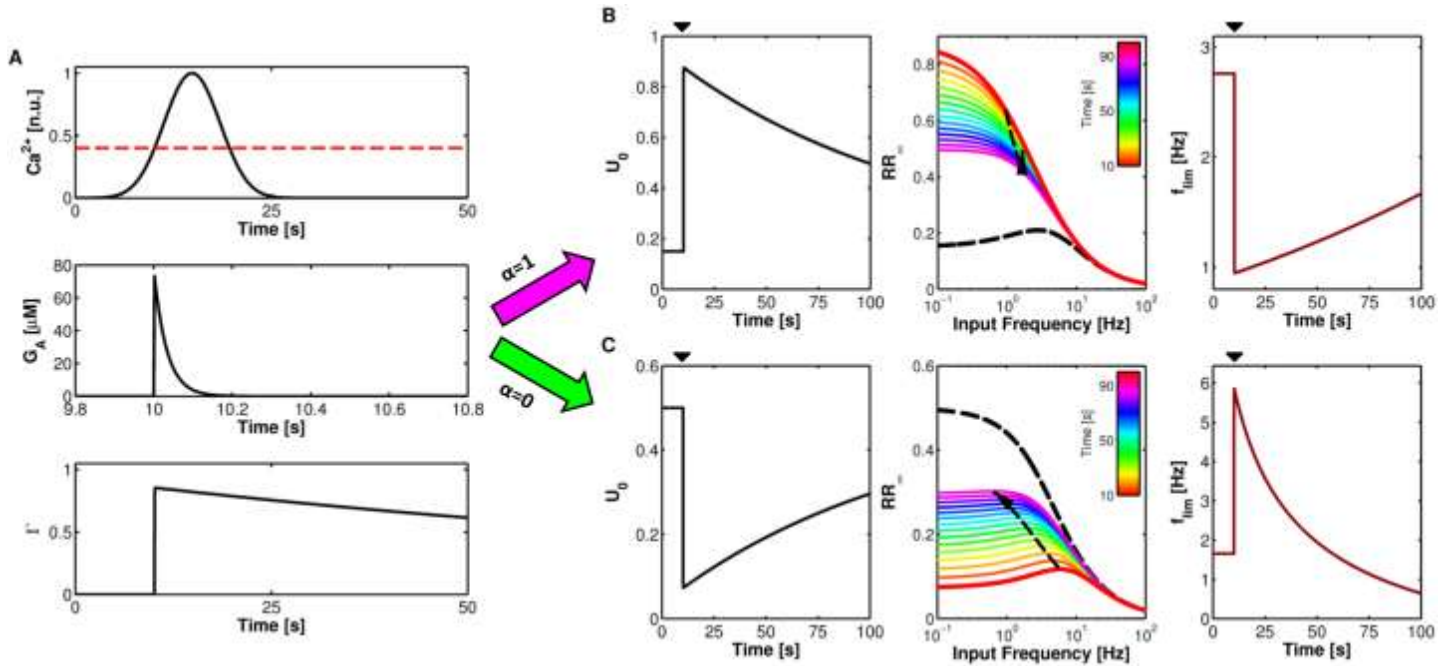

**Figure 4**. Mechanism of astrocyte regulation of synaptic release. (**A**, *top*) When intracellular Ca$^{2+}$ increases beyond the threshold value for exocytosis (*dashed red line*), the astrocyte releases an amount of glutamate into the extrasynaptic space (*black mark*, *middle*). The resulting fast transient increase of extracellular glutamate activates presynaptic receptors which, depending on the "effect" parameter $\alpha$ in our model, can decrease (**C**, $\alpha = 0$) or increase (**B**, $\alpha = 1$) the synaptic basal release probability $U_0$. Mean-field analysis predicts that steady-state evoked synaptic glutamate release ($RR_\infty$) is respectively diminished (**C**, *middle*) or increased (**B**, *middle*) (*colored lines, snapshot color codes* for the time after release of astrocytic glutamate) with respect to the case without astrocytic glutamate (*dashed black line*). (**B,C**, *right*) Equation (4) allows to elucidate how the limiting frequency ($f_{\text{lim}}$) of the synapse changes under astrocyte signaling. Parameters as in Table S1.

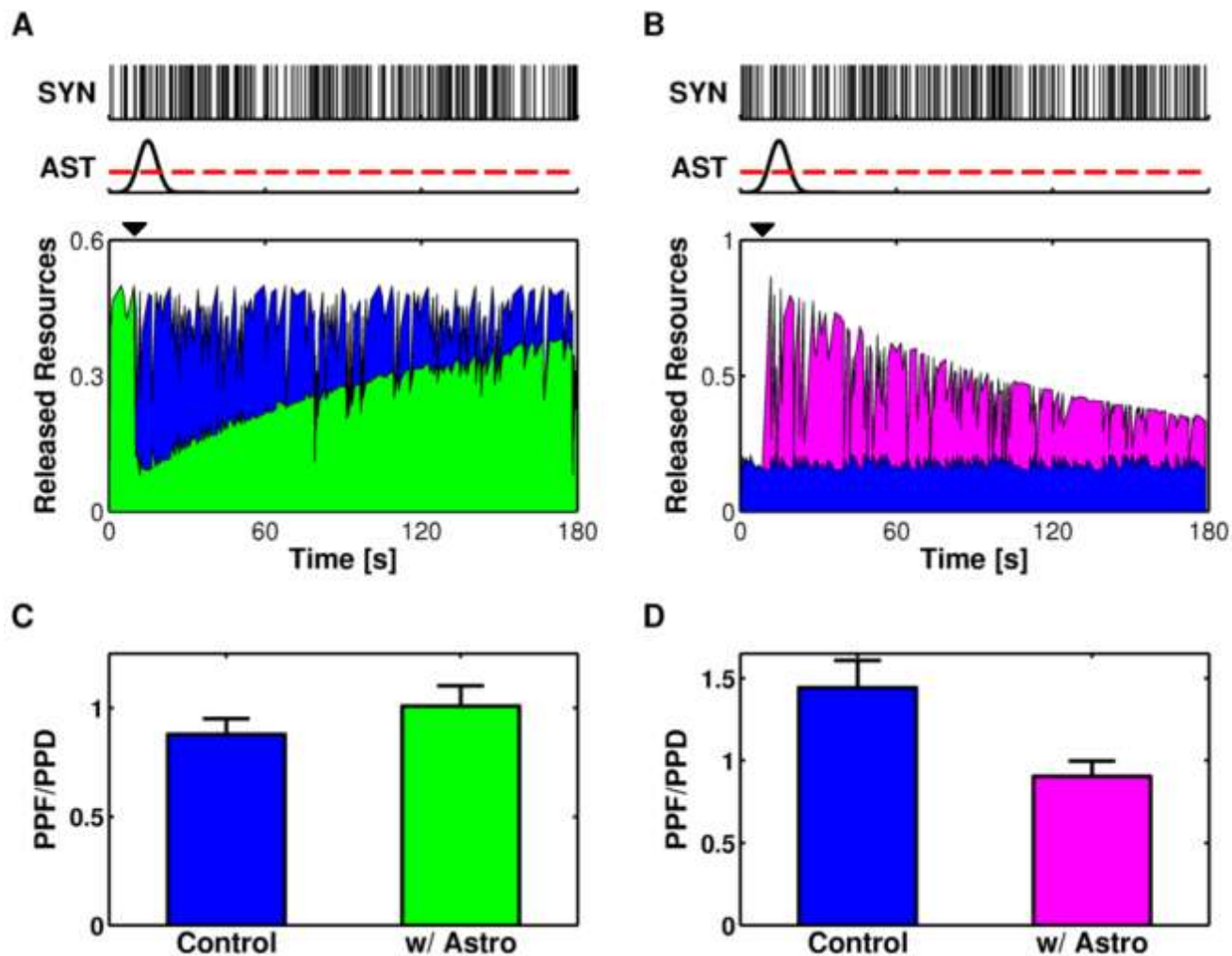

**Figure 5**. Release-decreasing vs. release-increasing astrocytes. The glutamate resources released by two different model synapses (**A**: depressing; **B**: facilitating) in response to a generic Poisson spike train (SYN, *top*), and without astrocytic signaling, are shown in *blue* (*bottom*). When the astrocyte is included, even a single event of glutamate exocytosis from this latter (onset at *t* = 10 s, *black mark*) triggered by a Ca²⁺ increase therein (AST, *top*), can deeply affect the amount of synaptic resources released by the same input. The nature of the change depends on the nature of presynaptic receptors. (**A**) For α = 0, the effect of the astrocytic glutamate is a global decrease of synaptic release, which fades away slowly from its onset at rate $\Omega_G$ (*green area*). (**B**) On the other hand, for α = 1 the effect of astrocytic glutamate is to increase synaptic release (*magenta area*). The global change of the amount of released resources is accompanied also by local changes in terms of paired-pulse plasticity. (**C**) For the depressing synapse with release-decreasing astrocyte in (**A**), the ratio between facilitated (PPF) and depressed (PPD) spike pairs, increases in favor of the former. (**D**) The opposite instead occurs for the case of release-increasing astrocyte with the facilitating synapse in (**B**). Bar + Error bar: Mean + Standard Deviation for *n* = 100 Poisson spike trains with the same average rate. Parameters as in Table S1.

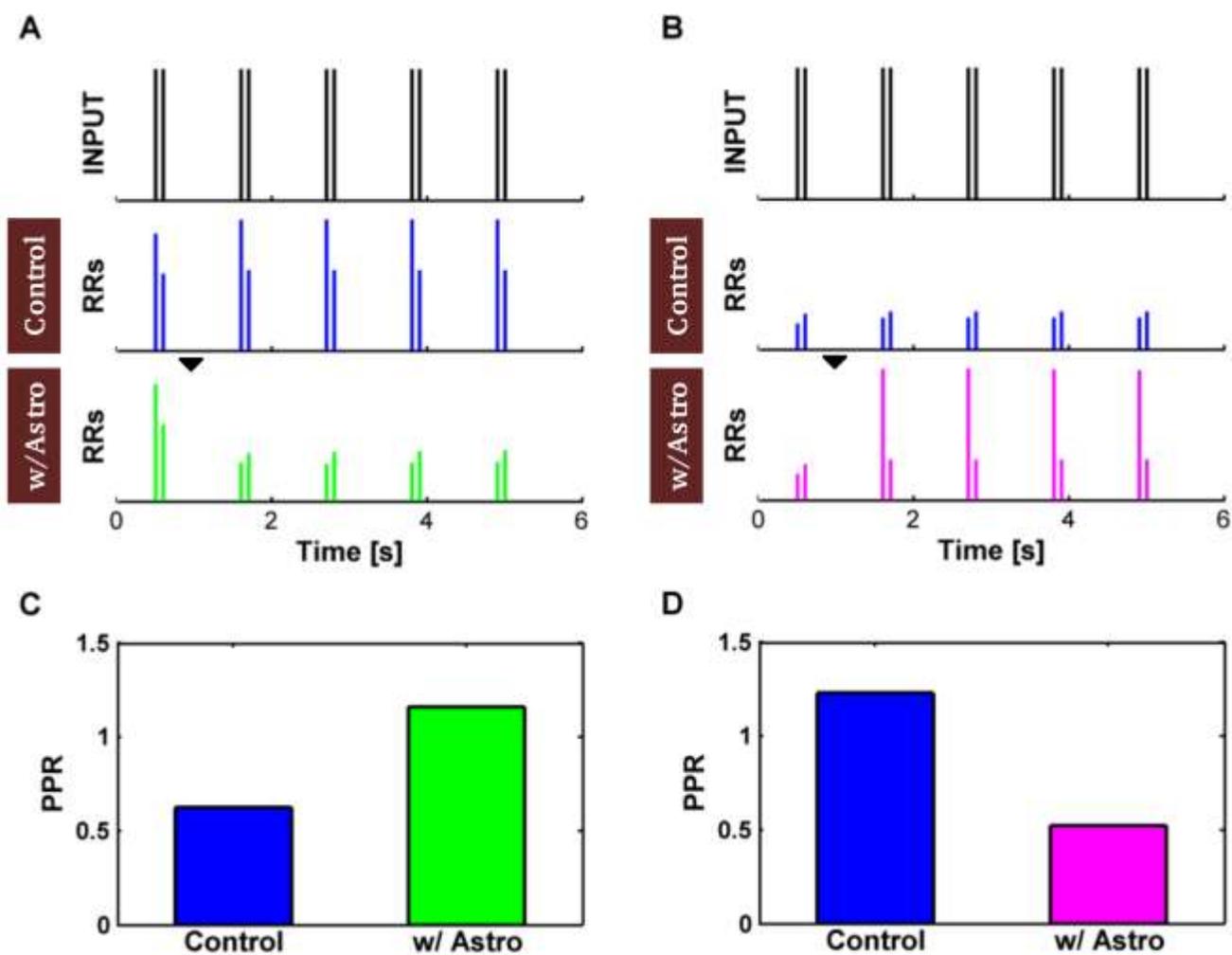

**Figure 6**. Astrocytic glutamate modifies paired-pulse plasticity. (**A**) A depressing and (**B**) a facilitating synapse are stimulated by a sequence of spike pairs (ISI = 100 ms) at 1 Hz (*top*) and the released resources (RRs) in absence ("Control", *middle*) vs. in presence of the astrocyte (*bottom*) are monitored. The amount of resources released by spike pairs dramatically changes in presence of glutamate exocytosis from the astrocyte (*black mark* at $t$ = 1 s, *bottom*). This behavior evidences a change of paired-pulse plasticity at these synapses which is summarized by the histograms in (**C**,**D**). The release-decreasing astrocyte (i.e. α = 0) on the depressing synapse in (**A**) remarkably increases the average synaptic paired-pulse ratio (PPR) (**C**), while the release-increasing astrocyte (i.e. α = 1) on the facilitating synapse in (**B**) decreases the PPR (**D**), which marks the onset of stronger paired-pulse depression. Parameters as in Table S1.

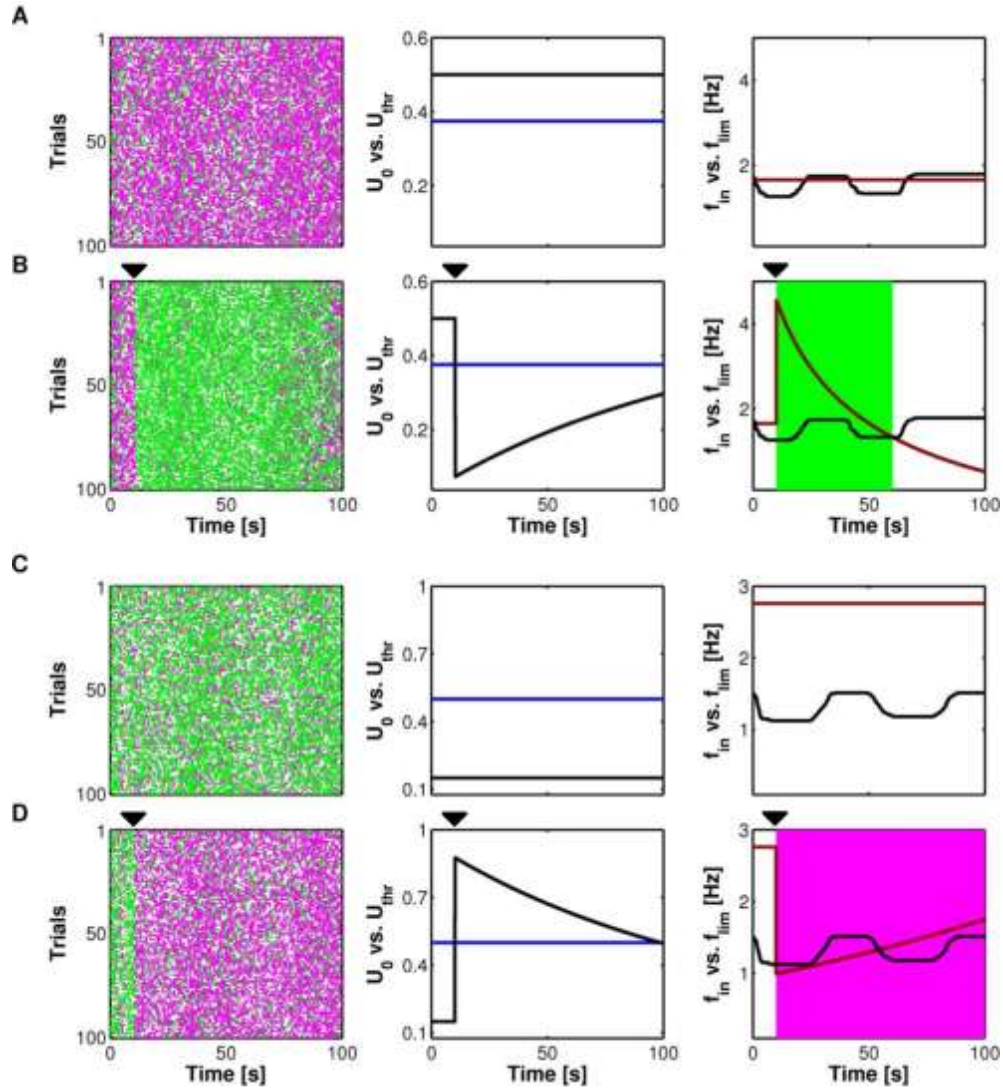

**Figure 7**. Astrocytic glutamate regulates transitions between facilitation and depression at the same synapse. (**A**,**B**, *left*) Raster plots of a depressing synapse, without (**A**) and in presence of (**B**) a single event of glutamate exocytosis from the astrocyte (onset at the *black mark* at $t = 1$ s, $\alpha = 0$) for $n = 100$ Poisson spike trains with the same average frequency $f_{in}$ (*black line* in **A**,**B**, *right*). A *green dot* marks an input spike that released more resources than its preceding one, while a *magenta dot* represents an input spike that released less resources than its previous one. (**A**,**B**, *middle*) The increase of facilitated spike pairs by release-decreasing astrocytic glutamate on the depressing synapse is due to the decrease of synaptic basal release probability $U_0$ (*black line*) beyond the switching threshold $U_{thr}$ (*blue line*) while the limiting frequency ($f_{lim}$, *dark red line*) increase above the average input frequency ($f_{in}$, *black line*). In such situation in fact, both conditions needed for short-term facilitation are fulfilled (see "Mechanisms of short-term presynaptic plasticity" in "Methods"). (**C, D**) The opposite occurs for a facilitating synapse under the effect of a release-increasing astrocyte ($\alpha = 1$). In this case in fact, astrocytic glutamate makes $U_0$ increase beyond $U_{thr}$ (**D**, *middle*) while $f_{lim}$ switches from above to below $f_{in}$, thus marking onset of depression (**D**, *right*). The same results can alternatively be obtained analyzing the slope of the $RR_\infty$ curve (equation 3) for $f_{in}(t)$ (Figure S9). Parameters as in Table S1.

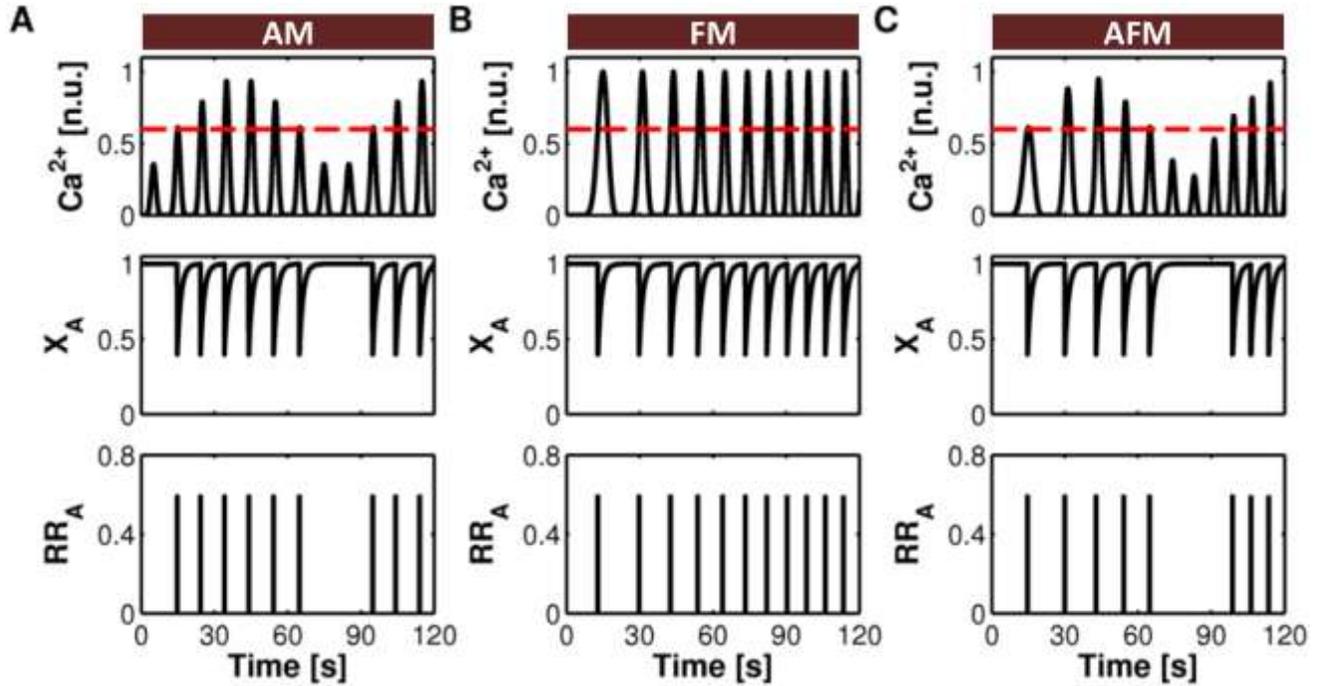

**Figure 8**. Different Ca$^{2+}$ patterns trigger glutamate release from the astrocyte at different frequencies. Fast reintegration of released glutamate and low-frequency Ca$^{2+}$ oscillations translate (**A**) AM, (**B**) FM and (**C**) AFM Ca$^{2+}$ dynamics (*top*) into different frequency-modulated sequences of glutamate release events (GREs) from the astrocyte (*middle*), all of equal magnitude (*bottom*). Parameters as in Table S1.

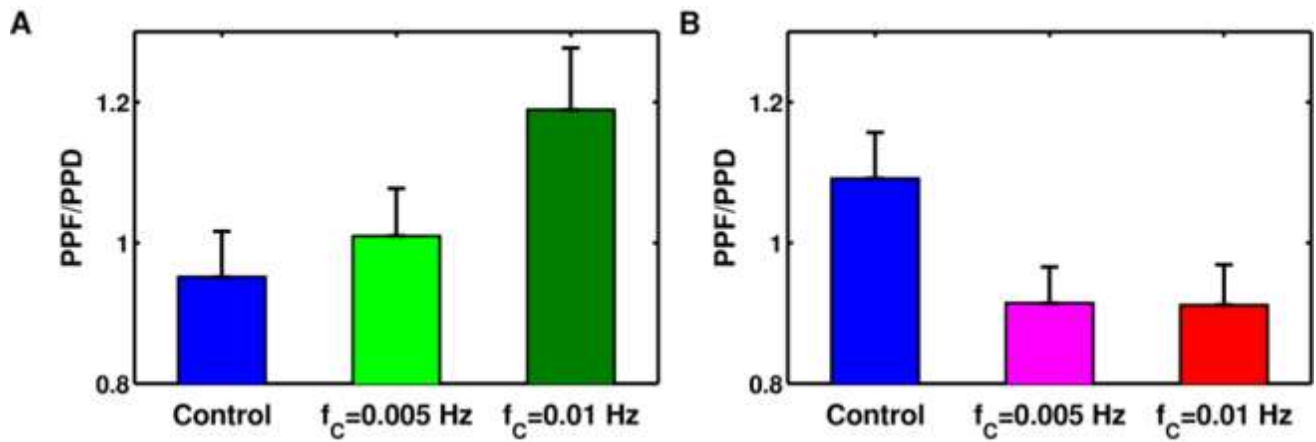

**Figure 9**. The frequency of astrocytic glutamate release controls the transition between depression and facilitation. Paired-pulse plasticity is considered here for $n = 100$ different Poisson spike trains with the same statistics (as in Figure 7) in presence of persistent glutamate release from the astrocyte. A synapse that in the absence of astrocytic glutamate ("Control") is otherwise depressing (**A**), can display increasingly more PPF for increasing GRE frequencies ($f_C$) in presence of release-decreasing astrocyte ($\alpha = 0$). Conversely, a facilitating synapse (**B**) shows increasing PPD for increasing $f_C$, under the influence of release-increasing astrocyte ($\alpha = 1$). Parameters as in Table S1.

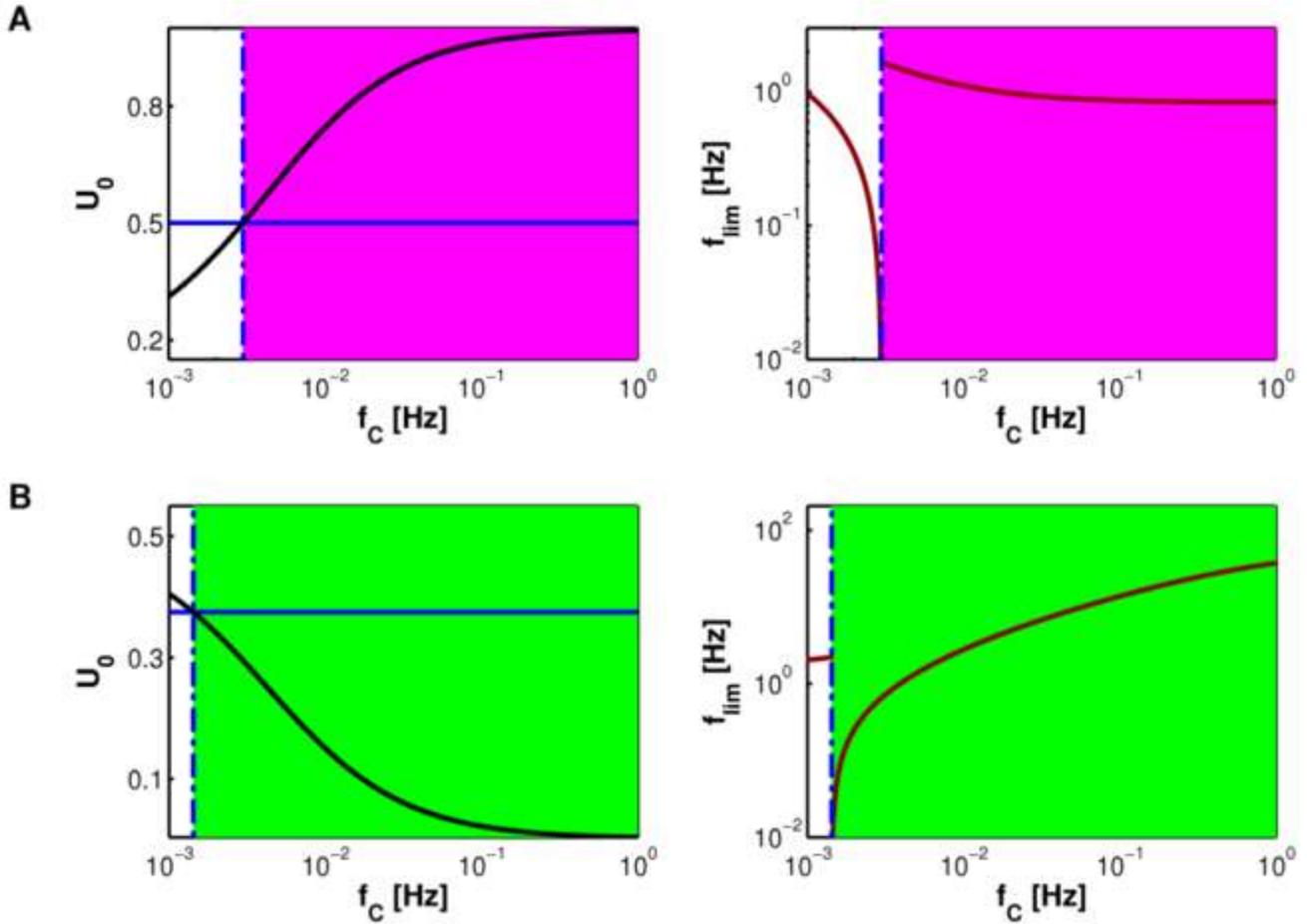

**Figure 10**. Mean-field analysis of astrocyte regulation of presynaptic short-term plasticity. (**A**,**B**, *left*) With increasing GRE frequencies $f_C$, $U_0$ (*solid black line*) crosses the switching threshold $U_{thr}$ (*solid blue line*), setting the conditions for a transition either towards predominant depression, for a facilitating synapse with release-increasing astrocyte (**A**, *magenta-shaded area*) or towards predominant facilitation, for a depressing synapse with release-decreasing astrocyte (**B**, *green-shaded area*). (**A**,**B**, *right*) We can also map $f_{lim}$ as a function of $f_C$ (*dark red line*). The crossing of $U_0$ with $U_{thr}$, coincides with a discontinuity of $f_{lim}$ (equation 4) and sets a threshold frequency ($f_{thr}$) (*dashed blue line*) which marks, for proper input stimuli, the frequency of astrocyte glutamate release that allows switching from facilitation to depression or vice versa. Parameters: (**A**) $\alpha = 0$; (**B**) $\alpha = 1$. Other parameters as in Table S1.

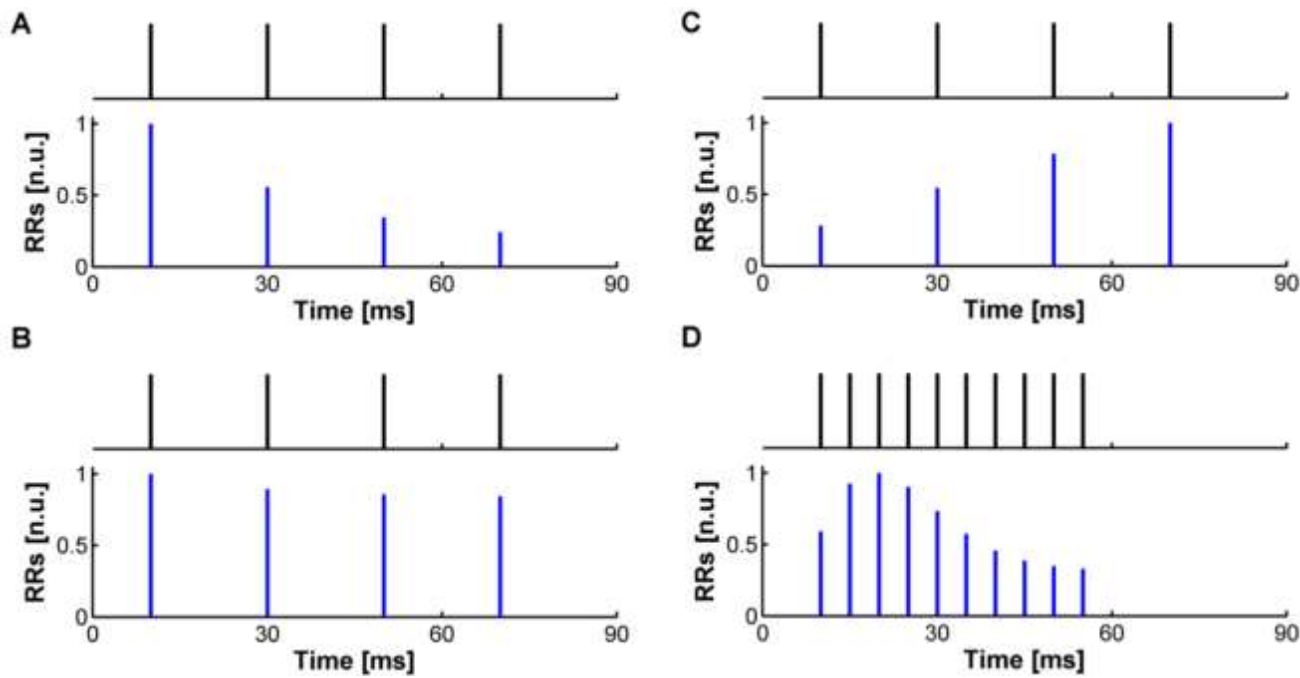

**Figure S1**. Conditions for short-term depression and facilitation in the TM model. Short-term plasticity in the TM model is brought forth by inherent synaptic parameters such as $\Omega_d$, $\Omega_f$ and $U_0$, and the frequency of incoming spikes. (**A,B**) Depressing synapses are generally characterized by $\Omega_f > \Omega_d$. In these latter, input spikes at $f_{in} > \Omega_d$ (**A**) mark the onset of short-term depression (STD) due to fast depletion of the pool of releasable resources. (**B**) Alternatively, STD can also be observed in high-fidelity synapses, namely synapses characterized by high values of $U_0$. (**C,D**) Facilitating synapses instead are characterized by $\Omega_f < \Omega_d$ and low release probability. In these latter (**C**), incoming spikes at $\Omega_f < f_{in} < \Omega_d$ (or $f_{in} > \Omega_d$, $\Omega_f$) build up presynaptic residual Ca$^{2+}$ levels, increasing the synaptic release, thus evidencing facilitation. (**D**) However, the progressive increase of release probability due to facilitation leads to concomitant growing depletion of the releasable pool and STD eventually takes over facilitation. Legend: input presynaptic spikes are in *black*, released resources (RRs, *blue*) are normalized with respect to their maximum (**A,B**: $RR_{max} = 0.5$; **C**: $RR_{max} = 0.18$; **D**: $RR_{max} = 0.2$). Parameters: (**A**) $\Omega_d = 2$ s$^{-1}$, $\Omega_f = 1000$ s$^{-1}$, $U_0{*} = 0.5$, $f_{in} = 50$ Hz; (**B**) $\Omega_d = 20$ s$^{-1}$, $\Omega_f = 1000$ s$^{-1}$, $U_0 = 0.5$, $f_{in} = 50$ Hz; (**C**) $\Omega_d = 100$ s$^{-1}$, $\Omega_f = 1.25$ s$^{-1}$, $U_0 = 0.05$, $f_{in} = 50$ Hz; (**D**) $\Omega_d = 10$ s$^{-1}$, $\Omega_f = 1.25$ s$^{-1}$, $U_0 = 0.1$, $f_{in} = 200$ Hz.

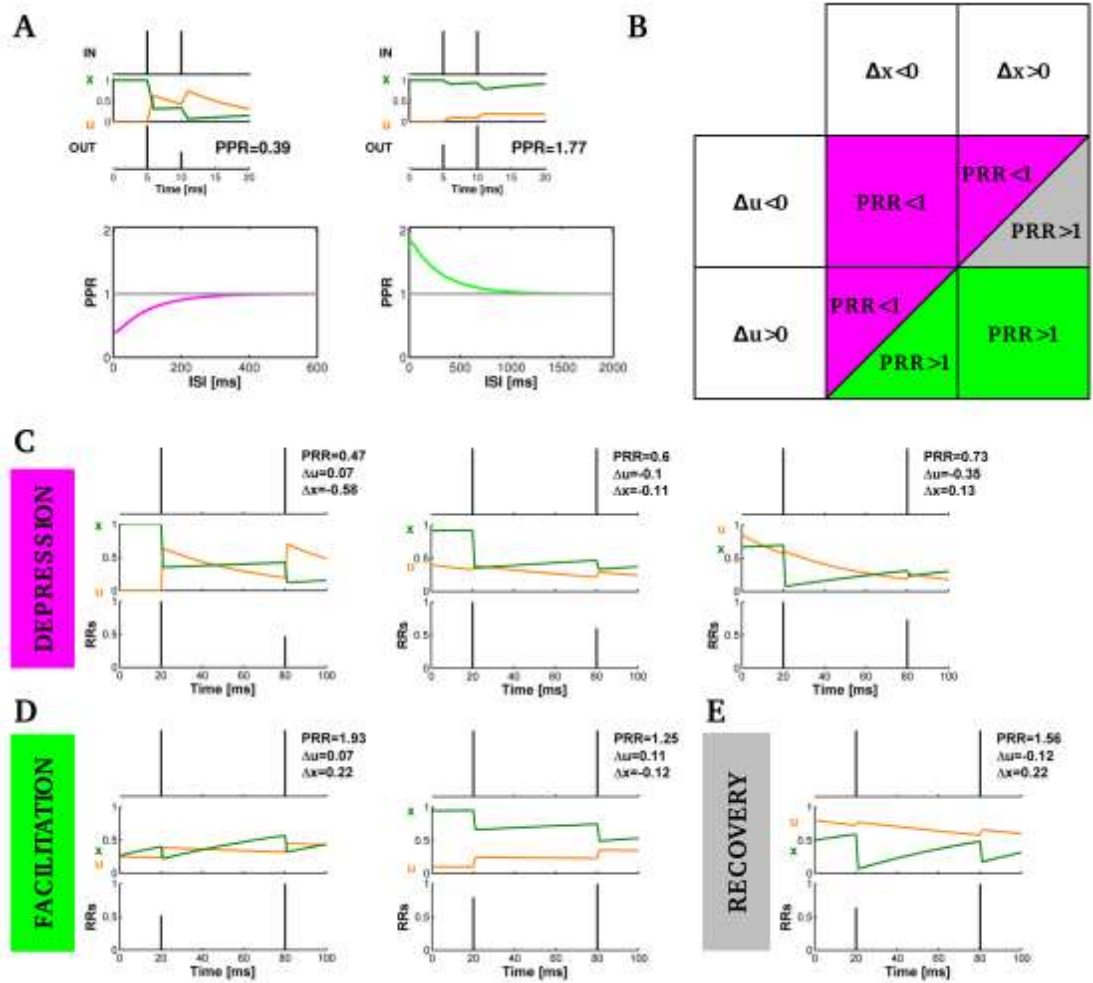

**Figure S2**. Paired-pulse plasticity. (**A**, *top*) In a typical paired-pulse stimulus protocol, a pair of spikes with controlled interspike interval is delivered to the synapse and synaptic response to the second spike ($RR_2$) is compared to synaptic response to the first spike ($RR_1$) by means of paired-pulse ratio, defined as $\mathrm{PPR} = RR_2/RR_1$ . (**A**, *left*) Values of PPR less than 1 mark paired-pulse depression (PPD) as in such conditions $RR_2 < RR_1$. (**A**, *right*) On the contrary, when PPR > 1, then $RR_2 > RR_1$ and paired-pulse facilitation (PPF) is observed. The farther the PPR from unity, the stronger the PPD (or PPF). (**A**, *bottom*) The value of PPR critically depends on the interspike interval (ISI) of spike pairs and approaches zero for very long ISIs reflecting the fact that short-term synaptic plasticity is a transient phenomena. (**B**) For a *generic* input spike trains, the PPR between consecutive spikes in a pair is not sufficient to distinguish between PPD and PPF. Depending on the spike timing and on the past synaptic activity in fact, PPR > 1 could also result from sufficient reintegration of the pool of releasable resources ($\Delta x > 0$), despite a decrease of residual Ca$^{2+}$ between the two spikes in a pair (i.e. $\Delta u < 0$). This situation corresponds to a different form of synaptic plasticity dubbed as "recovery from depression" [13]. (**C-E**) Examples of different short-term plasticity mechanisms listed in the Table (A) displayed by the TM model. Parameters: (**A**, *left*) $\Omega_d = 10$ s$^{-1}$, $\Omega_f = 100$ s$^{-1}$, $U_0 = 0.7$, $RR_{max} = 0.7$; (**A**, *right*) $\Omega_d = 100$ s$^{-1}$, $\Omega_f = 33$ s$^{-1}$, $U_0 = 0.05$; (**C**, *left*) $\Omega_d = 2$ s$^{-1}$, $\Omega_f = 20$ s$^{-1}$, $U_0 = 0.65$; (**C**, *middle*) $\Omega_d = 3.33$ s$^{-1}$, $\Omega_f = 10$ s$^{-1}$, $U_0 = 0.1$; (**C**, *right*) $\Omega_d = 4$ s$^{-1}$, $\Omega_f = 20$ s$^{-1}$, $U_0 = 0.1$; (**D**, *left*) $\Omega_d = 10$ s$^{-1}$, $\Omega_f = 3.33$ s$^{-1}$, $U_0 = 0.2$; (**C**, *right*) $\Omega_d = 5$ s$^{-1}$, $\Omega_f = 1$ s$^{-1}$, $U_0 = 0.16$; (**E**) $\Omega_d = 10$ s$^{-1}$, $\Omega_f = 5$ s$^{-1}$, $U_0 = 0.2$.

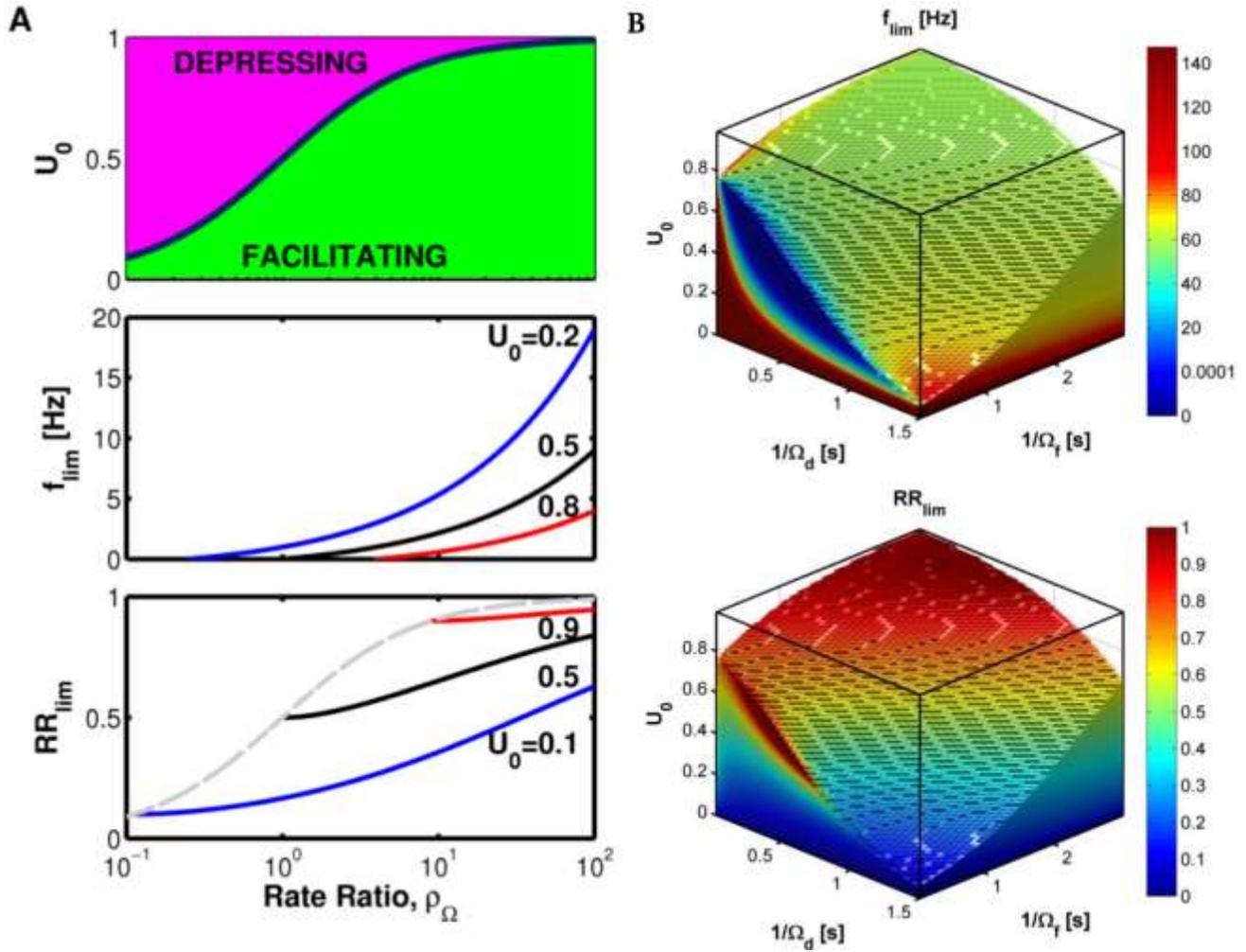

**Figure S3**. The switching threshold in the TM model. (**A**, *top*) Mapping of depressing (*red*) and facilitating (*green*) synapses in the parameter plane $U_0$ vs. $\rho_\Omega = \Omega_d/\Omega_f$. The two types of synapses are separated by the switching threshold (*black line*) given by $U_{thr} = \rho_\Omega/(1 + \rho_\Omega)$ (equation S39). (**A**, *middle*) The limiting frequency $f_{lim}$ of a facilitating synapse coincides with the peak frequency of maximal steady-state release of neurotransmitter is maximal (see also Figure 2D). For fixed facilitation rates (i.e. $\Omega_f$ = const), such limiting frequency increases with $\rho_\Omega$, namely with faster rates ($\Omega_d$) of reintegration of synaptic resources. In such conditions in fact the larger $\Omega_d$, the higher the rate of input spikes before the onset of depression. For the same reason, higher $f_{lim}$ are also found in correspondence of lower values of synaptic basal release probability $U_0$ at given $\rho_\Omega$. (**A**, *bottom*) The peak of released resources at the limiting frequency (equation S41) instead increases with $U_0$ to the detriment of its range of variation (recall in fact, that $0 < RR_{lim} < 1$). (**B**, *top*) Facilitation regions in the parameter space and mapping therein of $f_{lim}$ (equation S40) and (**B**, *bottom*) $RR_{lim}$ (equation S41), show strong nonlinear dependence of both quantities on synaptic parameters.

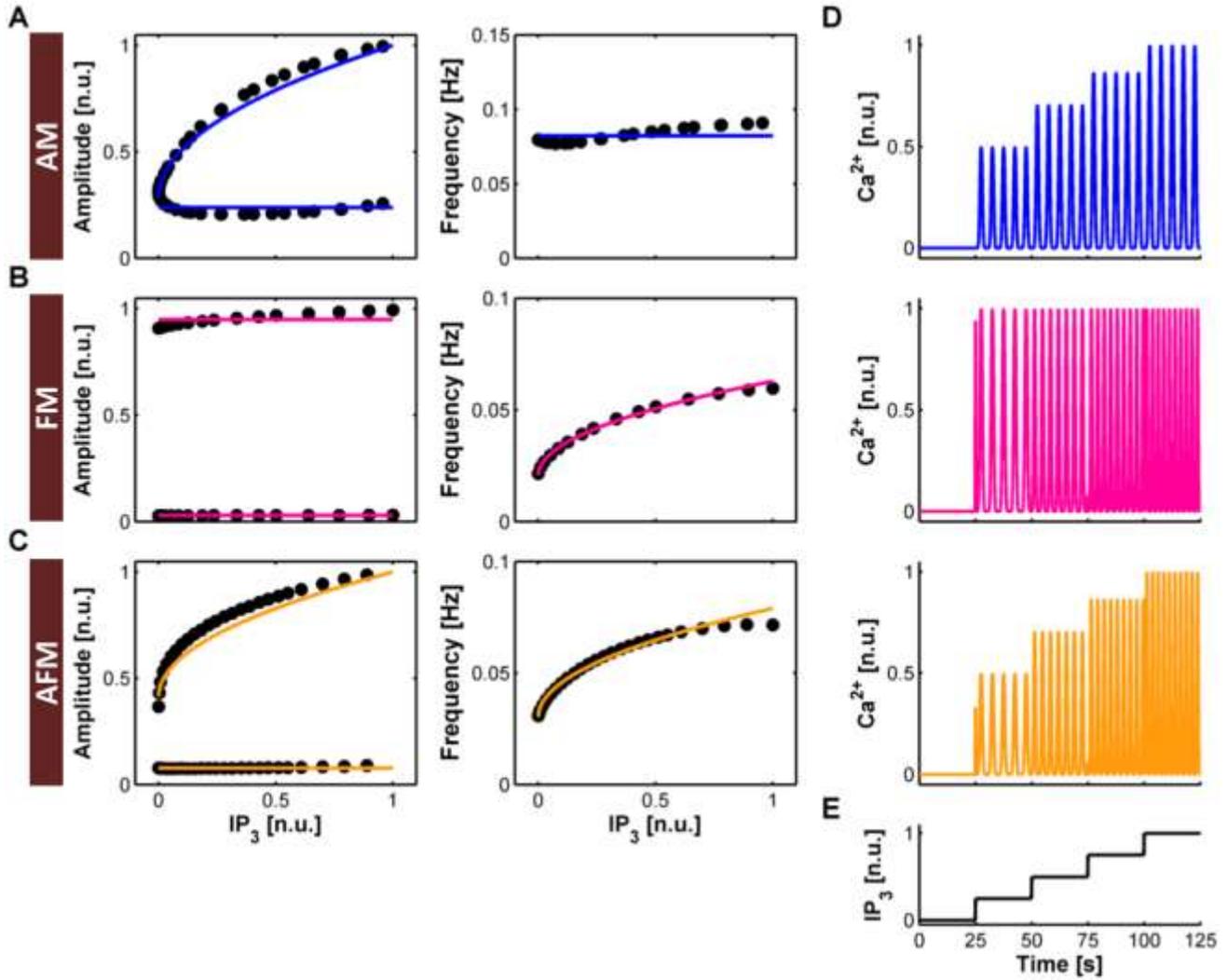

**Figure S4**. Astrocyte calcium dynamics. (**A-C**) Superposition of stereotypical functions (*solid line*) on numerically-solved (*black circles*) amplitude and frequency of (**A**) AM-encoding, (**B**) FM-encoding and (**C**) AFM-encoding Ca²⁺ oscillations as obtained from the Li-Rinzel model of Ca²⁺ dynamics [76, 91] (see Text S1, Section I.2). (**D**) Corresponding Ca²⁺ oscillations pertaining to these three modes for the case of an IP₃ stimulus as in (**E**). Data in (**A-C**, *left* and *middle*) are from [76]. For convenience, only persistent oscillations are considered. The oscillatory range is rescaled between 0 and 1 and amplitude of oscillations is normalized with respect to the maximal Ca²⁺ concentration. Data were fitted by equations (S4-S6) with $m_i(t) = m_0 + k_i \sqrt{(IP_3(t) - I_b)}$ assuming $I_b = 0$. (**A**) $C_0 = 0.239$, $m_0 = 0.256$, $k = 0.750$; (**B**) $C_0 = 0.029$, $C_{max} = 0.939$, $m_0 = 0.210$, $k = 0.470$, $f_C = 0.1$ Hz; (**C**) $C_0 = 0.079$, $m_{0,AM} = 0.449$, $k_{AM} = 0.611$, $m_{0,FM} = 0.310$, $k_{FM} = 0.480$, $f_C = 0.1$ Hz. (**D**) $C_0 = 0$, $m_{0,AM} = 0$, $m_{0,FM} = 0$ Hz, $k_{AM} = 1$, $k_{FM} = 1$, $f_C = 0.1$ Hz, $I_b = 0$.

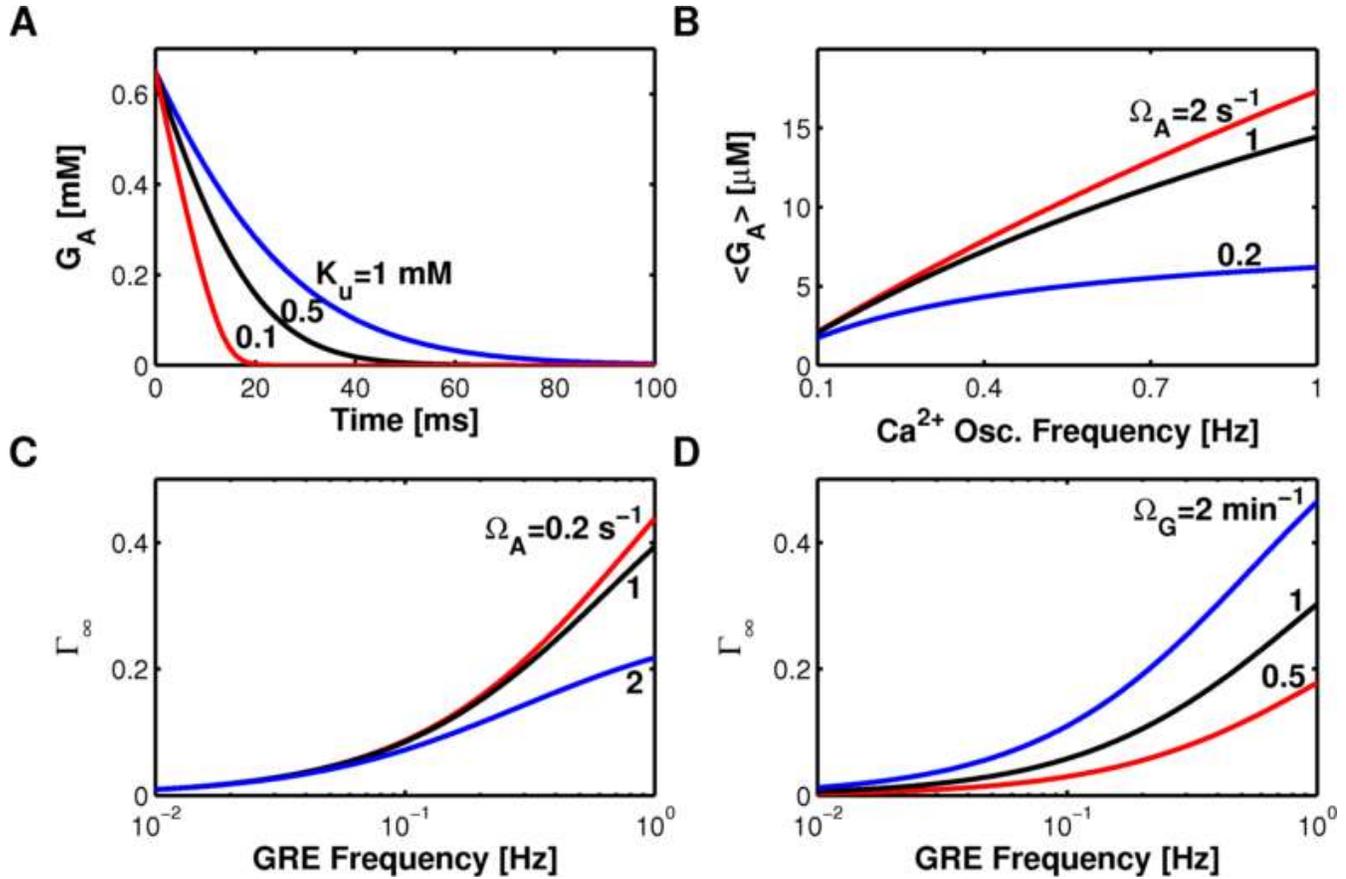

**Figure S5**. Astrocytic glutamate and presynaptic receptor activation. (**A**) Time course of astrocyte-released glutamate ($G_A$) in the extrasynaptic space strongly depends on the affinity of astrocytic glutamate transporters for their substrate, i.e. $K_u$. Several experiments showed that such transporters are not saturated [92] which allows approximating the time course of extrasynaptic glutamate by a single monoexponential decay at rate $\Omega_c$ (Text S1, Section I.4). (**B**) Glutamate concentration in the extrasynaptic space around targeted presynaptic receptors depends on average on $\Omega_A$, that is the rate of reintegration of released glutamate in the astrocyte. On a par with depletion of synaptic resources, for presynaptic spike frequencies larger than $\Omega_d$, the slower $\Omega_A$ the stronger the depletion of the astrocytic pool of releasable glutamate for increasing $Ca^{2+}$ oscillations (assumed suprathreshold in this figure). Accordingly, each $Ca^{2+}$ oscillation releases progressively less glutamate. (**C**) The strength of astrocyte modulation of synaptic release depends among the others, on the time course of astrocyte-released glutamate, thus on both $\Omega_c$ and $\Omega_A$ rates. Accordingly, at steady-state the average peak of astrocyte effect on synaptic release (i.e. $\Gamma_\infty$, equation S49) increases with the GRE frequency and is stronger for faster rates of reintegration of astrocytic glutamate. (**D**) The strength of astrocyte modulation also depends on past activation of pre-terminal receptors. Thus, it is critically regulated by the decay rate $\Omega_G$, which biophysically correlates with inherent cellular properties of presynaptic terminal and/or targeted receptors. Experiments show that astrocyte modulation of synaptic release rises fast after glutamate exocytosis, and decays very slowly [30-33], at rates that could be comparable to typical frequencies of $Ca^{2+}$ oscillations in the astrocyte [44]. This, in turn, accounts for a progressive saturation of receptors by increasing GRE frequencies for small values of $\Omega_G$. Parameters: (**A**) $v_u = 60$ mMs$^{-1}$, $r_d = 0$ s$^{-1}$; (**B**) $G_v = 100$ mM, $C_{thr} = 0$, $\Omega_c = 60$ s$^{-1}$; (**C-D**) $\Omega_c = 60$ s$^{-1}$, $O_G = 1$ μM$^{-1}$s$^{-1}$; $n_v = 4$, $G_v = 50$ mM, $U_A = 0.5$, $\rho_A = 6.5\cdot10^{-4}$, $\Omega_G = 0.67$ min$^{-1}$.

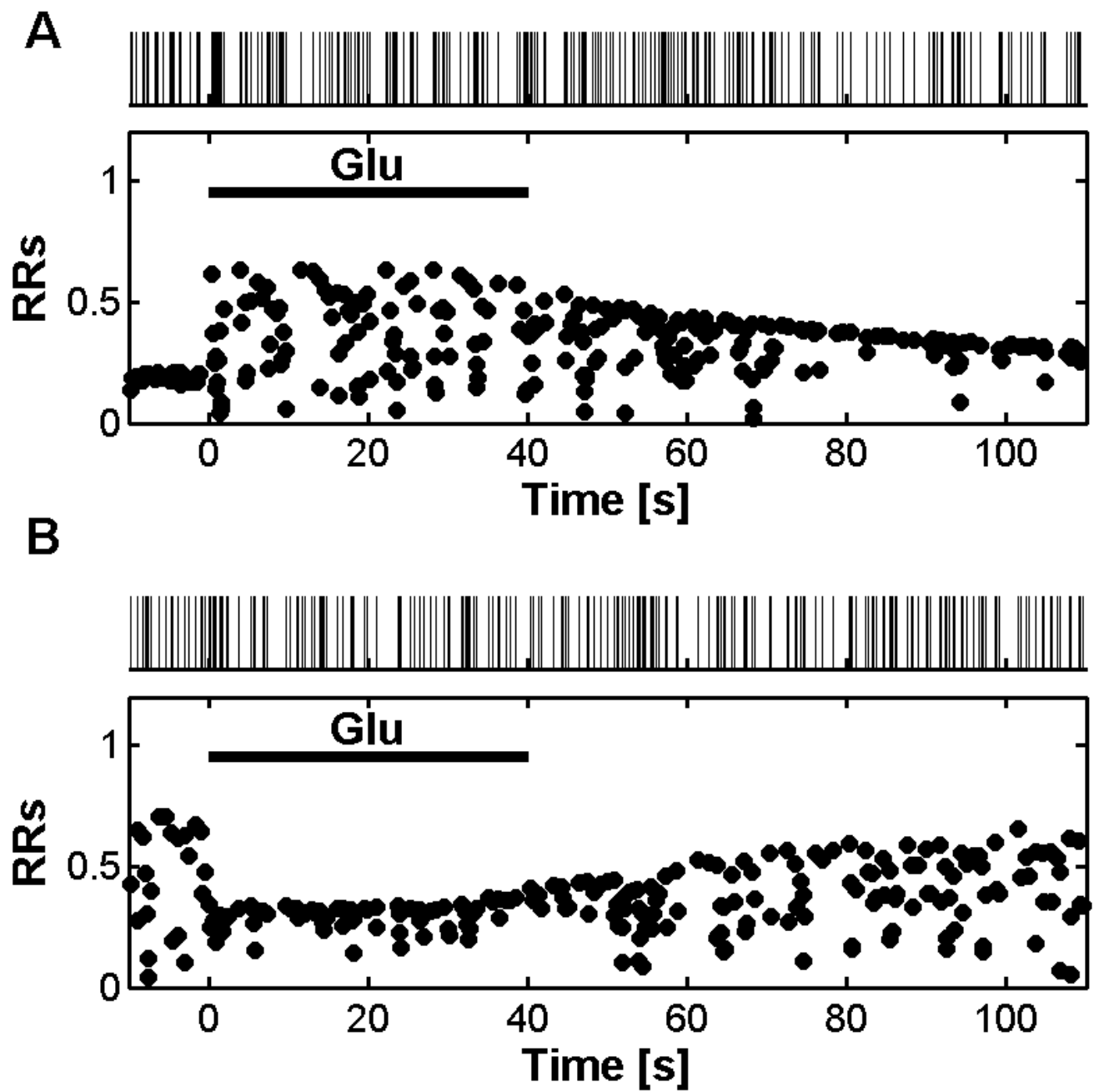

**Figure S6.** Regulation of synaptic release by presynaptic glutamate receptors. Simulated bath perfusion by 100 µM glutamate (Glu) for 20 s on a model synapse, can either increase (**A**) or decrease (**B**) synaptic release (RRs) evoked by a generic stimulus (**A,B**, *top*). These results closely reproduce experimental observations [93-95] and provide our model with general biophysical consistency. Parameters: (**A**) $\Omega_d = 2$ s$^{-1}$, $\Omega_f = 3.3$ s$^{-1}$, $U_0^* = 0.8$, $\alpha = 0$; (**B**) $\Omega_d = 2$ s$^{-1}$, $\Omega_f = 2$ s$^{-1}$, $U_0^* = 0.15$, $\alpha = 1$;       $U_A = 0.4$,       $\Omega_G = 1$ min$^{-1}$,       $O_G = 1$ µM$^{-1}$s$^{-1}$.       Other       parameters       as       in       Table S1.

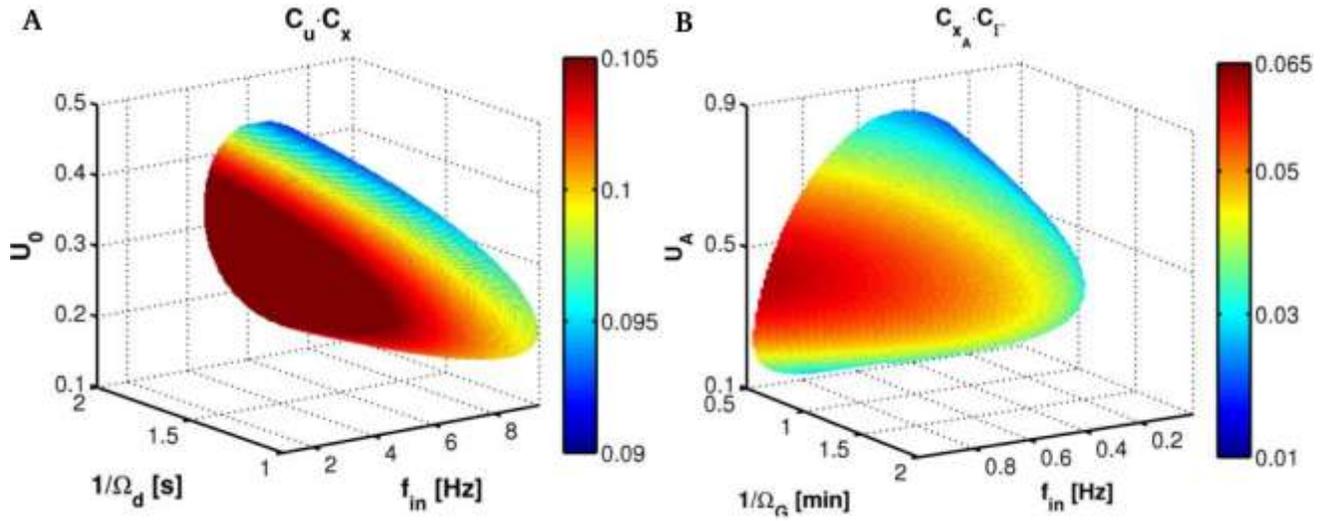

**Figure S7**. Range of validity of the mean-field description. (**A**) Product of coefficients of variations for the two synaptic variables $x$ and $u$ as a function of frequency, allows to estimate the region of validity of the mean-field description (equations S31-S32). In particular, in the domain of the parameter space considered in this study, the error made by averaging exceeds 10% only for a narrow region of such space confined between $4 < f_{in} < 6$ Hz. (**B**) Analogous considerations hold for averaging of equations (S7, S18, S19). Mapping of the product of coefficients of variations of $x_A$ and $\Gamma$ shows that in this case, the error is less than than 7% in the whole parameter space. Parameters: (**A**) $\Omega_f = 2.5$ s$^{-1}$; (**B**) $O_G = 1.5$ μMs$^{-1}$. Other parameters as in Table S1.

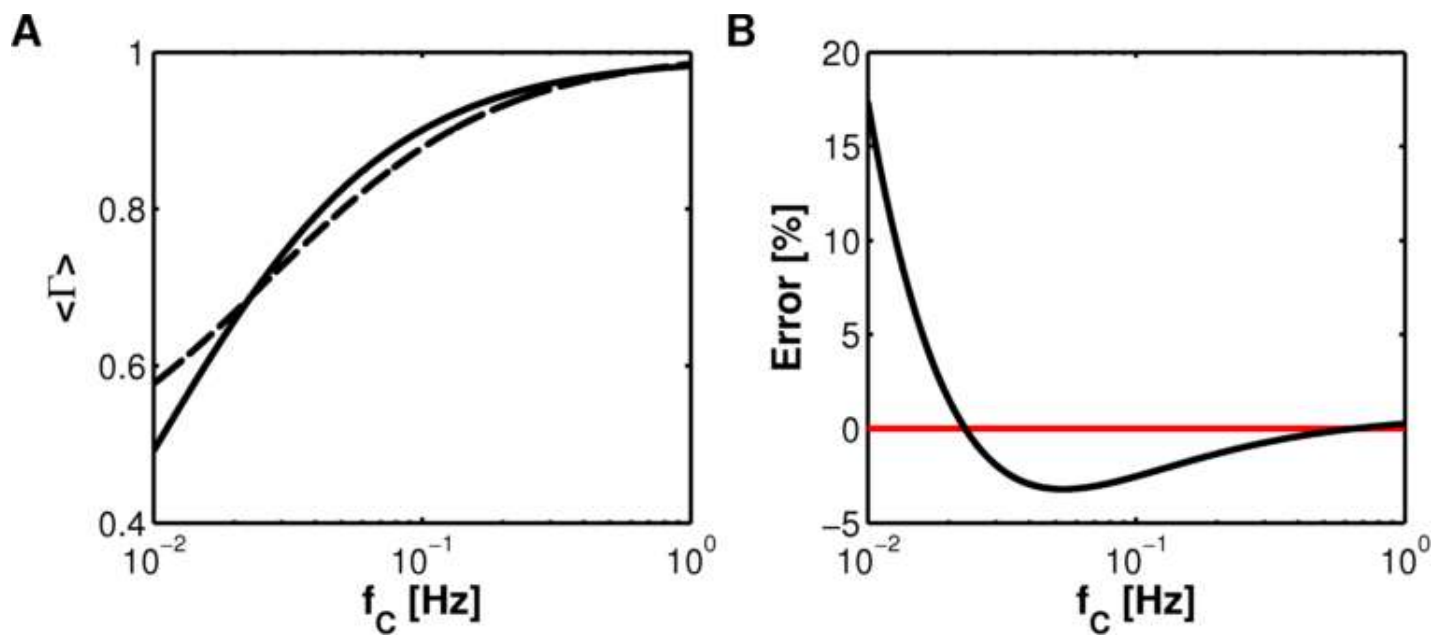

**Figure S8**. Estimation of $\langle \Gamma \rangle$. (**A**) Comparison between the exact analytical solution for $\langle \Gamma \rangle$ (i.e $\Gamma_{\infty}$ from equation S49; *solid line*) and the approximated one (i.e. $\tilde{\Gamma}$ in equation SA6; *dashed line*) used in the computation of the coefficient of variation $c_r$, and (**B**) relative percent error of $\tilde{\Gamma}$ with respect to $\Gamma_{\infty}$. For very low frequencies of Ca²⁺ oscillations ($f_C$), $\tilde{\Gamma}$ diverges from $\Gamma_{\infty}$ as a result of the assumption of $f_C$-independent, constant quantal release from the astrocyte, introduced in equation (SA1). While $\Gamma_{\infty}$ tends to zero as Ca²⁺ oscillations become more and more sporadic because eventually no glutamate is released from the astrocyte, $\tilde{\Gamma}$ instead does not. This ultimately leads to an incorrect estimation of $c_r$ which is not relevant however within the frequency range of Ca²⁺ oscillations considered in this study.

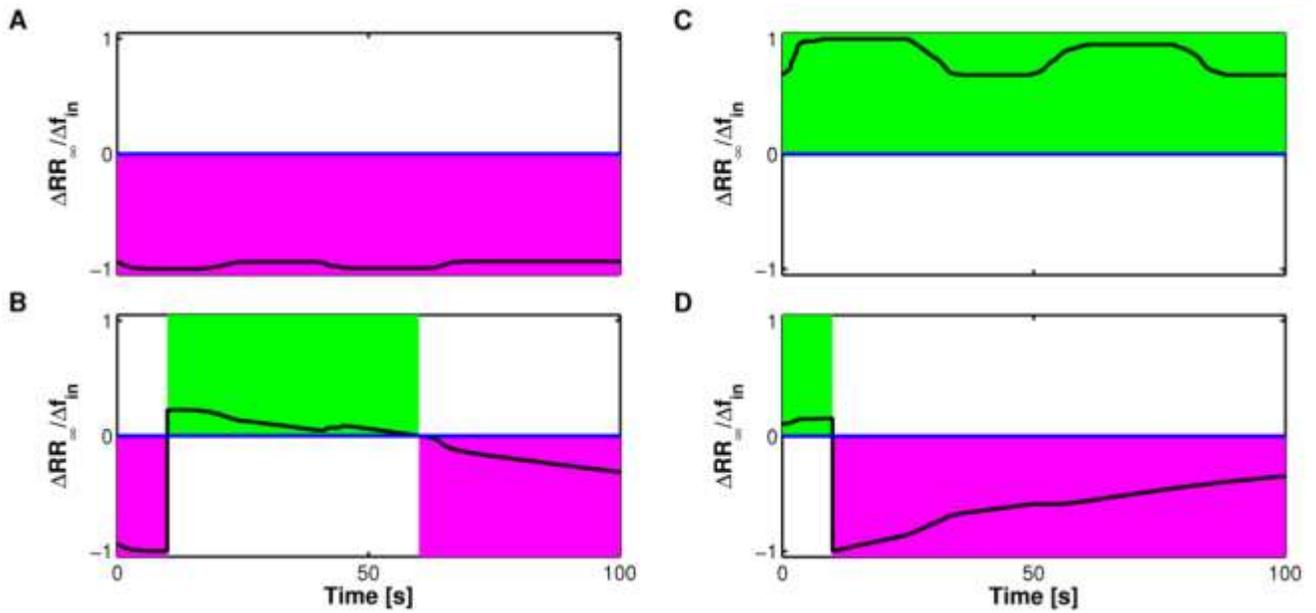

**Figure S9**. Slope analysis. Estimation of the trial-averaged slope $\Delta RR_\infty / \Delta f_{in}$ of the synaptic frequency response curve for any value in time of the input frequency $f_{in}$ (that is the derivative of $RR_\infty$ (equation 3) with respect to $f_{in}$) allows characterization of any transitions of synaptic plasticity. The method is alternative to that outlined in Figure 7, and relies on the observation that in our model of synaptic plasticity, short-term facilitation is likely to occur whenever $\Delta RR_\infty / \Delta f_{in} > 0$ for given input rates, otherwise short-term depression is predominant (see also Text S1, Section II.1). Letters correspond to those in Figure 7, and refer to results of slope analysis for the corresponding cases therein, that is: (**A**) depressing synapse without and (**B**) with release-decreasing astrocyte, and (**C**) facilitating synapse without and (**D**) with release-increasing astrocyte. *Green-shaded* areas denote predominant PPF, *magenta-shaded* areas stand for predominant PPD. Slope values are normalized by their maximum absolute value. Parameters are as in Table S1.

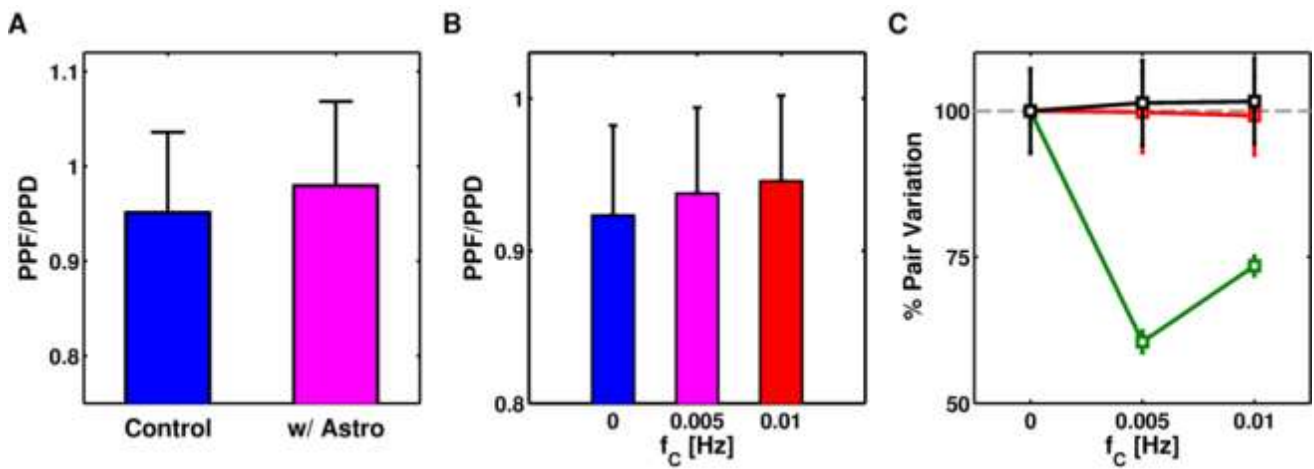

**Figure S10**. Release-decreasing astrocyte on a facilitating synapse. (**A**) Analysis of paired-pulse plasticity in presence of a single glutamate exocytotic event from the astrocyte (same conditions of Figure 5A) shows an increase of the number of facilitated spike pairs (*green bar*) with respect to "Control" simulations (i.e. without astrocyte) (*blue bar*) (bar + error bar: mean + standard deviation). (**B**) Moreover, the larger the frequency of glutamate release from the astrocyte, the stronger the effect. (**C**) Detailed analysis of the different forms of short-term plasticity ongoing within spike pairs – PPF (*dark green*), PPD (*red*) and "recovery from depression" (*black*) – reveals that the increase of the ratio PPF/PPD detected in (**A-B**) is mainly imputable to an increase of PPF accompanied by a reduction of recovery from depression. These results confirm the general notion discussed in the text that the effect of a release-decreasing astrocyte coincides with an increase of paired-pulse facilitation (PPF) (see also Figure 7D). Nonetheless, we note that this effect is less pronounced than in a depressing synapse (compare Figures 5A with S9A and Figure 9A with S9B). Data based on $n = 100$ Poisson input spike trains with average rate as in Figure 7C. Data in (**C**) are normalized with respect to their "Control" value: PPF = 197, PPD = 205, recovery = 27. Parameters as in Table S1 with $\alpha = 0$.

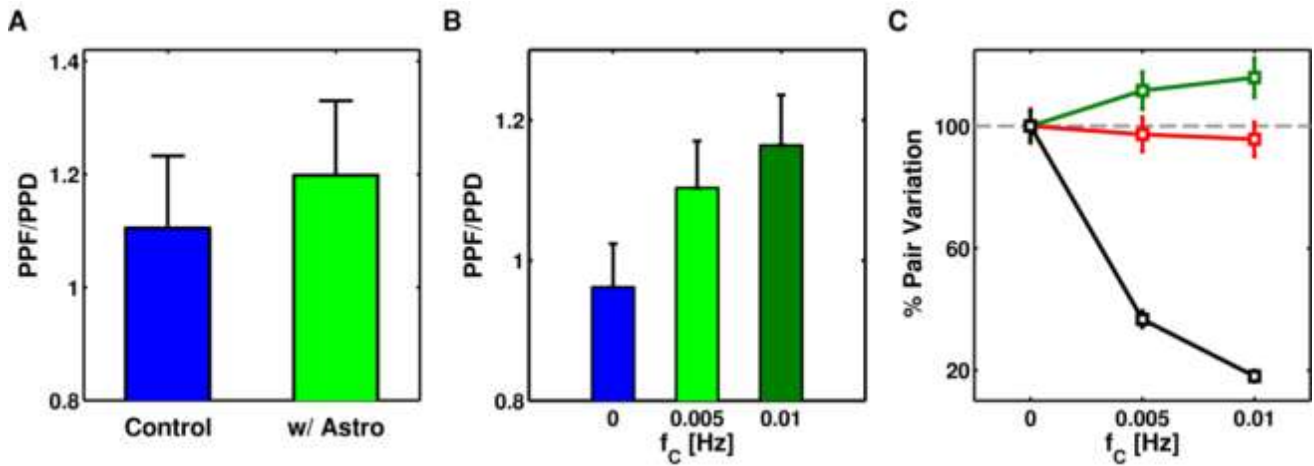

**Figure S11**. Release-increasing astrocyte on a depressing synapse. (**A**) Analysis of paired-pulse plasticity either for a single (same conditions of Figure 5B) and (**B**) for persistent glutamate exocytosis from the astrocyte, shows an increase of facilitated spike pairs (*magenta/red bars*) with respect to the "Control" simulations (i.e. in absence of the astrocyte) (*blue bars*). (**C**) A closer inspection on the nature of ongoing paired-pulse plasticity (PPF: *green*, PPD: *red* and "recovery from depression": *black*) reveals that such increase is actually caused by an increase of recovery from depression (Control: PPF = 4, PPD = 146, recovery = 135). Bar + Error bar: Mean + Standard deviation. Data based on *n* = 100 Poisson input spike trains with average rate as in Figure 7A. Parameters as in Table S1 with α = 1.